# White Paper:
# Verbesserung des Record Linkage für die Gesundheitsforschung in Deutschland

August 2023

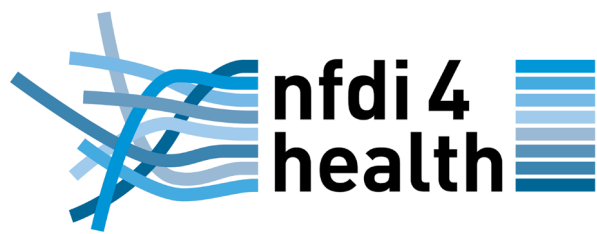

# Inhaltsverzeichnis

# Abkürzungsverzeichnis

| | |
|---|---|
| ACM | Association for Computing Machinery (www.acm.org) |
| ADR | Administrative Data Research |
| ADT | Arbeitsgemeinschaft Deutscher Tumorzentren e.V. (www.tumorzentren.de) |
| AG | Arbeitsgruppe |
| AOK | Allgemeine Ortskrankenkasse (www.aok.de) |
| API | Application Programming Interface: Softwareschnittstelle zu Programmen oder Betriebssystemen |
| ASCII | American Standard Code for Information Interchange; standardisierter Zeichensatz, der von den meisten Computersystemen interpretiert werden kann. |
| AWMF | Arbeitsgemeinschaft der Wissenschaftlichen Medizinischen Fachgesellschaften (www.awmf.org) |
| BDSG | Bundesdatenschutzgesetz |
| BF | Bloomfilter |
| BfArM | Bundesinstitut für Arzneimittel und Medizinprodukte (www.bfarm.de) |
| BFCC | Baltic Fracture Competence Center (https://www.bfcc-project.eu/registry.html) |
| BIPS | Leibniz-Institut für Präventionsforschung und Epidemiologie - BIPS (www.bips-institut.de) |
| BKRG | Bundeskrebsregisterdatengesetz (https://www.gesetze-im-internet.de/bkrg/) |
| BMBF | Bundesministerium für Bildung und Forschung (www.bmbf.de) |
| BMC | BioMed Central; Verlag (www.biomedcentral.com) |
| BMG | Bundesministerium für Gesundheit (www.bundesgesundheitsministerium.de) |

BMJ             Bundesministerium der Justiz (www.bmj.de)

BSG             Bundessozialgericht (www.bundessozialgericht.de)

BSGE            Bundessozialgericht (https://www.bsg.bund.de)

BVerfG          Bundesverfassungsgericht (www.bundesverfassungsgericht.de)

BVerfGE         Entscheidung des BVerfG

BVMI            Berufsverband Medizinischer Informatiker e.V. (www.bvmi.de)

CCC             Comprehensive Cancer Center

CCP             Clinical Communication Platform des DKTK

CDISC           Clinical Data Interchange Standards Consortium (www.cdisc.org)

CIOMS           Council for International Organizations of Medical Sciences (www.cioms.ch)

CIRS            Critical Incident Reporting System

CNIL            Commission Nationale de l'Informatique et des Libertés (www.cnil.fr)

CODEX           COVID-19 Data Exchange Platform; im Rahmen von NUM gefördertes Projekt
                zum Aufbau einer zentralen Dateninfrastruktur mit Anschluss an die MII

CORD            Collaboration on Rare Diseases; Verbundprojekt und konsortienüber
                greifender Use Case im Rahmen der MII

CoV             Coronavirus

CoVerlauf       Studie zum Erkrankungsverlauf bei Personen mit einem positiven Test auf
                SARS-CoV-2 bzw. einer COVID-19-Erkrankung (https://www.bips-institut.de/
                covid-19.html#9176)

COVID           Coronavirus Disease 2019; meldepflichtige Infektionskrankheit in Folge einer
                Infektion mit dem Coronavirus SARS-CoV-2

CSV             Comma Separated Values

| | |
|---|---|
| CTR | Clinical Trials Regulation; EU-Verordnung Nr. 536/2014 über klinische Prüfungen mit Humanarzneimitteln und zur Aufhebung der Richtlinie 2001/20/EG |
| DdGW | Daten des Gesundheitswesens, Jährlicher Überblick über aktuelle Daten des Gesundheitswesens vom Bundesministerium für Gesundheit |
| DEMIS | Deutsches Elektronisches Meldesystem für Infektionsschutz |
| DES | Data Encryption Standard; symmetrischer Verschlüsselungsalgorithmus |
| DESH | Deutscher Elektronischer Sequenzdaten-Hub, technische Plattform des RKI zur Übermittlung von Sequenzdaten (https://www.rki.de/DE/Content/InfAZ/N/Neuartiges_Coronavirus/DESH/DESH.html) |
| DFG | Deutsche Forschungsgemeinschaft (www.dfg.de) |
| DGEpi | Deutsche Gesellschaft für Epidemiologie e. V. (http://dgepi.de) |
| DGSMP | Deutsche Gesellschaft für Sozialmedizin und Prävention e.V. (www.dgsmp.de) |
| DIFUTURE | Data Integration for Future Medicine (https://difuture.de) |
| DIZ | Datenintegrationszentrum im Förderkonzept Medizininformatik des BMBF |
| DKFZ | Deutsches Krebsforschungszentrum (www.dkfz.de) |
| DKH | Deutsche Krebshilfe (www.krebshilfe.de) |
| DKTK | Deutsches Konsortium für Translationale Krebsforschung (https://dktk.dkfz.de) |
| DS | Datenschutz |
| DSGVO | Verordnung des Europäischen Parlaments und des Rates zum Schutz natürlicher Personen bei der Verarbeitung personenbezogener Daten, zum freien Datenverkehr und zur Aufhebung der Richtlinie 95/46/EG Datenschutz-Grundverordnung (Verordnung 2016/679) |
| DuD | Zeitschrift „Datenschutz und Datensicherheit" (www.dud.de) |
| DVMD | Deutsche Verband Medizinischer Dokumentare (www.dvmd.de) |
| DZG | Deutsche Zentren der Gesundheitsforschung |
| DZHK | Deutsches Zentrum für Herz-Kreislauf-Forschung e.V. (http://dzhk.de) |
| EDC | Electronic Data Capturing |
| EG | Europäische Gemeinschaft |
| EHDS | European Health Data Space, Regulierungsentwurf der Europäischen Kommission zu einem europäischen Gesundheitsdatenraum (https://health.ec.europa.eu/ehealth-digital-health-and-care/european-health-data-space_en) |
| EHR4CR | Electronic Health Records for Clinical Research, im Rahmen der IMI gefördertes EU-Projekt (https://www.imi.europa.eu/projects-results/project-factsheets/ehr4cr) |
| eID | eindeutiger Identifier |
| EKN | Epidemiologisches Krebsregister Niedersachsen (https://www.krebsregister-niedersachsen.de/) |
| ePA | elektronische Patientenakte |
| EpiLink | Record Linkage-Software, entwickelt von Contiero et al. (2005) |
| E-PIX | Im Rahmen des GANI_MED-Projekts entwickelte und mit dem PIX-Profile von IHE kompatible MPI-Software |
| FAIR | Findable, Accessible, Interoperable, Reusable |
| FAQ | Frequently asked question(s) |
| FASTA | Textbasiertes Format zur Speicherung von Sequenzdaten |
| FDPG | Deutsches Forschungsdatenportal Gesundheit; Zentrale Antrags- und Registerstelle in der MII |
| FDZ | Forschungsdatenzentrum am BfArM nach § 303d SGB V |
| FHIR | Fast Healthcare Interoperability Resources; HL7-Standard (http://hl7.org/fhir) |
| fTTP | federated Trusted Third Party |
| GDSG | Gesellschaft für zentrales Datenmanagement und Statistik im Gesundheitswesen mbH |
| GECCO | German Corona Consensus Datensatz |
| GEKID | Gesellschaft epidemiologischer Krebsregister in Deutschland e.V. (www.gekid.de) |
| GEKN | Gesetz über das Epidemiologische Krebsregister Niedersachsen |
| GePaRD | German Pharmacoepidemiological Research Database des BIPS (https://www.bips-institut.de/forschung/forschungsinfrastrukturen/gepard.html) |
| GKR | Gemeinsames epidemiologisches Krebsregister |
| GKV | Gesetzliche Krankenversicherung |
| GMDS | Deutsche Gesellschaft für Medizinische Informatik, Biometrie und Epidemiologie e.V. (www.gmds.de) |
| GMS | German Medical Science: Webportal der German Medical Science gGmbH (www.egms.de) |
| GPD | Gute Praxis Datenlinkage (https://doi.org/10.1055/a-0962-9933) |
| GPS | Gute Praxis Sekundärdatenanalyse der DGSMP, DGEpi und GMDS |
| GTH | Gesellschaft für Thrombose- und Hämostaseforschung e.V. (www.gth-online.org) |
| HCHS | Hamburg City Health Study (https://hchs.hamburg/) |
| HiGHmed | Heidelberg - Göttingen - Hannover Medical Informatics (www.highmed.org) |
| HL7 | Health Level Seven; Internationale SDO für den Bereich der Interoperabilität von IT-Systemen im Gesundheitswesen (www.hl7.org) |
| HTTP | Hyper Text Transfer Protocol |

| i2b2 | Informatics for Integrating Biology and the Bedside (www.i2b2.org) |
|------|--------------------------------------------------------------------|
| IAB | Industry Advisory Board |
| ICD | International Statistical Classification of Diseases and Related Health Problems |
| ICDMW | International Conference on Data Mining Workshops des IEEE |
| ID | Identifikationsnummer |
| IDAT | Identifizierende Daten (der Patient:innen) |
| IDNrG | Identifikationsnummerngesetz |
| IEC | International Electrotechnical Commission (www.iec.ch) |
| IEEE | International non-profit organization and professional association for the advancement of technology, ursprgl.: Institute of Electrical and Electronics Engineers (www.ieee.org) |
| IfSG | Gesetz zur Verhütung und Bekämpfung von Infektionskrankheiten beim Menschen - Infektionsschutzgesetz |
| IHE | Integrating the Healthcare Enterprise (www.ihe.net) |
| IK | Institutionskennzeichen, bundesweit eindeutige, neunstellige Zahlen für Abrechnungen und Qualitätssicherungsmaßnahmen zur deutschen Sozialversicherung |
| IMI | Innovative Medicines Initiative (www.imi-europe.org) |
| INDEED | Sektorenübergreifende Inanspruchnahme ambulanter Versorgungsstrukturen von Notaufnahmepatienten in Deutschland; im Rahmen des Innovationsfonds gefördertes Projekt |
| INTERREG | EU-Funding-Programm für die Ostsee-Region (2021-2027, https://interreg-baltic.eu) |
| IP | Internet Protocol |

| ISO | International Organization for Standardization (www.iso.org) |
| JMIR | Journal of Medical Internet Research (www.jmir.org) |
| KFRG | Gesetz zur Weiterentwicklung der Krebsfrüherkennung und zur Qualitätssicherung durch klinische Krebsregister – Krebsfrüherkennungs- und -registergesetz |
| KI | Künstliche Intelligenz |
| KIS | Krankenhausinformationssystem |
| KK | Krankenkasse(n) |
| KKN | Klinisches Krebsregister Niedersachsen, AöR (www.kk-n.de) |
| KKR | Klinisches Krebsregister |
| KORA | Kooperative Gesundheitsforschung in der Region Augsburg |
| KR | Krebsregister |
| KV | Kassenärztliche Vereinigung |
| KVen | Kassenärztliche Vereinigungen |
| KVMV | Kassenärztliche Vereinigung Mecklenburg-Vorpommern (www.kvmv.de) |
| KVNR | Krankenversichertennummer |
| LIFE | Leipziger Interdisziplinärer Forschungskomplex zu molekularen Ursachen umwelt- und lebensstilassoziierter Erkrankungen (https://www.uniklinikum-leipzig.de/einrichtungen/life/life-erwachsenenkohorten/life-adult-studie) |
| LIMS | Laboratory Information Management System |
| lit | Littera / Buchstabe |
| LKHG | Landeskrankenhausgesetz |
| LKR | Landeskrebsregister |
| LKRG | Landeskrebsregistergesetz |

LOINC          Logical Observation Identifiers Names and Codes (www.loinc.org)

MAGIC          Mainzelliste, Samply.Auth und der Generische Informed Consent Service als Open-Source-Werkzeuge für Identitäts-, Einwilligungs- und Rechte-management in der medizinischen Verbundforschung; DFG-gefördertes Verbundprojekt

MAGICPL        MAGIC Pseudonymization Language

MainSEL        Mainzelliste Secure EpiLinker: PPRL-Tool auf Basis der Mainzelliste

Mainzelliste   Open-Source-Software zur Pseudonymisierung, Pseudonymverwaltung und zum Record Linkage

MDAT           Medizinische Daten

mdi            mdi – Forum der Medizin_Dokumentation und Medizin_Informatik; von den beiden Verbänden BVMI und DVMD gemeinsam herausgegebene Fachzeitschrift

MI             Medizininformatik

MII            Medizininformatik-Initiative des BMBF (www.medizininformatik-initiative.de)

MIRACUM        Medizininformatik in Forschung und Versorgung in der Universitätsmedizin | Medical Informatics in Research and Care in University Medicine (www.miracum.org)

MODYS          Modular Control and Documentation System, Software-Tool

MPI            Master Patient Index

MVZ            Medizinisches Versorgungszentrum

MWV            Medizinisch Wissenschaftliche Verlagsgesellschaft OHG, Berlin (www.mwv-berlin.de)

NAKO           NAKO Gesundheitsstudie (www.nako.de)

NAPKON         Nationales Pandemie Kohorten Netz; im Rahmen von NUM gefördertes Projekt zum Aufbau von drei Kohorten-Plattformen zur Covid-19-Forschung

| | |
|---|---|
| NFDI | Nationale Forschungsdateninfrastruktur; Idee und Empfehlung aus dem Positionspapier „Leistung aus Vielfalt" des RfII zum Forschungsdatenmanagement von 2016 und von der DFG 2019 ausgeschriebene Projektförderung |
| NFDI4Health | Nationale Forschungsdateninfrastruktur für personenbezogene Gesundheitsdaten; im Rahmen der NFDI geförderter Infrastrukturaufbau (www.nfdi4health.de) |
| NHS | UK National Health Service (www.nhs.uk) |
| nNGM | Nationales Netzwerk Genomische Medizin Lungenkrebs (https://nngm.de/) |
| NTH | Transfer Hub des Netzwerks der Universitätsmedizin |
| NUM | Netzwerk der Universitätsmedizin zu COVID-19 (www.netzwerk-universitaets-medizin.de) |
| OHG | Offene Handelsgesellschaft |
| PCR | Polymerase-Chain-Reaction (Polymerasekettenreaktion) |
| PID | (pseudonymer) Patientenidentifikator |
| PIMS | Patienten-Identifikatoren-Management-System |
| PIX | Patient Identifier Cross-referencing, IHE-Profil zum domänenübergreifenden Abgleich von Patienten-Identifikatoren |
| PKV | Private Krankenversicherung |
| PLZ | Postleitzahl |
| PPRL | Privacy-Preserving Record Linkage |
| PRIMAT | PRIvate Matching Toolbox, PPRL-Toolbox (https://www.toolpool-gesundheits-forschung.de/produkte/primat-private-matching-toolbox) |
| PSN | Pseudonym |
| RatSWD | Rat für Sozial- und Wirtschaftsdaten (www.ratswd.de) |
| REST | Representational State Transfer; Standard zur einfachen Datenübertragung in Webanwendungen auf Basis des HTTP-Protokolls |
| RDP | Routinedatenplattform-Projekt des Netzwerks Universitätsmedizin (NUM) (https://www.netzwerk-universitaetsmedizin.de/projekte/num-rdp) |
| RfII | Rat für Informationsinfrastrukturen (www.rfii.de) |
| RKI | Robert Koch-Institut (www.rki.de) |
| RL | Record Linkage |
| SAIL | Secure-Anonymised-Information-Linkage-Datenbank (https://saildatabank.com/) |
| SARS | Severe Acute Respiratory Syndrome; virale Infektionskrankheit |
| SDO | Standards Development Organization |
| SDTM | Study Data Tabulation Model (CDISC-Standard) |
| SGB | Sozialgesetzbuch |
| SHIP | Study of Health in Pomerania: (http://ship.community-medicine.de) |
| SMITH | Smart Medical Information Technology for Healthcare (www.smith.care) |
| SOAP | Simple Object Access Protocol; vom W3C empfohlener, XML-basierter Protokoll-Standard zur Kommunikation strukturierter Daten mit Webservices per HTTP |
| SOEMPI | Secure Enterprise Master Patient Index, Record Linkage-Tool (https://github.com/MrCsabaToth/SOEMPI) |
| StGB | Strafgesetzbuch |
| Synonym | Anlage mehrerer Datensätze zu einer Person mit jeweils unterschiedlichem PID |
| THS | Treuhandstelle |

TMF          TMF – Technologie- und Methodenplattform für die vernetzte medizinische
             Forschung e.V. (www.tmf-ev.de)

TORCH        Translational Registry for Cardiomyopathies

TTP          Trusted Third Party

UK           United Kingdom

UMG          Universitätsmedizin Göttingen (www.med.uni-goettingen.de)

UUID         Universally Unique Identifier

W3C          World Wide Web Consortium (www.w3.org)

WHO          World Health Organization (http://www.who.org)

XML          extensible Markup Language

ZD           Zeitschrift für Datenschutz (https://rsw.beck.de)

ZfKD         Zentrum für Krebsregisterdaten (https://www.krebsdaten.de)

# Tabellenverzeichnis

# Abbildungsverzeichnis

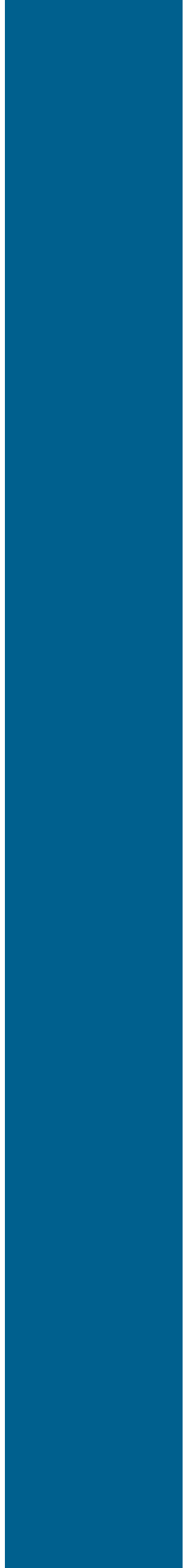

# Executive Summary

Die Verknüpfung von unterschiedlichen Daten auf Personenebene, zum Beispiel von Gesundheitsdaten, mit dem Ziel einen gemeinsamen Datensatz zu erstellen, wird als Record Linkage bezeichnet. Die Daten zu einer Person können dabei bei verschiedenen, voneinander getrennten Datenhaltern wie bspw. Forschungsinstituten, Krebsregistern, Krankenhäusern oder Krankenversicherungen vorliegen. Durch die Verknüpfung lassen sich wissenschaftliche Fragestellungen beantworten, die wegen des beschränkten Variablenumfangs mit einer Datenquelle alleine nicht zu beantworten wären. Diese verknüpften Daten entfalten ein riesiges Potenzial für die Gesundheitsforschung, um die Prävention, Therapie und Versorgung der Bevölkerung zu verbessern. Da es sich bei diesen Gesundheitsdaten um sensible Daten handelt, sind sie durch strenge Rechtsvorschriften gegen potenziellen Missbrauch geschützt.

Die derzeitigen rechtlichen Gegebenheiten (z. B. bundeslandspezifische Gesetzgebungen, aufsichtsbehördenspezifische Auslegungen von Bundesgesetzen) schränken allerdings die Nutzung der wertvollen Gesundheitsdaten für die Forschung so stark ein, dass ihr Potenzial für eine Verbesserung von Prävention und Versorgung bisher nur unzureichend ausgeschöpft werden kann. Das Record Linkage wird in Deutschland insbesondere dadurch erschwert bzw. in vielen Fällen sogar unmöglich gemacht, dass es im Gegensatz zu Ländern wie Dänemark keinen eindeutigen personenbezogenen Identifikator gibt, der eine Zusammenführung von Gesundheitsinformationen über verschiedene Datenkörper hinweg ermöglichen würde. Zudem sind in Deutschland interoperable Lösungen nicht vorhanden, um ein für die Gesundheitsforschung erforderliches, umfassendes studien- und datenkörperübergreifendes Record Linkage in einer gesicherten Umgebung durchführen zu können. Dem berechtigten Interesse auf Schutz der personenbezogenen Daten steht z. B. das Interesse entgegen, die verborgenen Informationen über etwaige Risiken und den möglichen Nutzen von Behandlungen zu erforschen und diese zur Verbesserung der gesundheitlichen Versorgung zu nutzen.

Es werden Lösungen benötigt, die auch in den Fällen eingesetzt werden können, in denen eine individuelle Einwilligung zur Datennutzung für ein spezifisches Forschungsvorhaben nicht eingeholt werden kann, wie z. B. bei der Verknüpfung von Krankenversicherungsdaten mit Krebsregisterdaten, um mögliche Krebsrisiken durch Medikamente bei Millionen von Versicherten aufzudecken. So unterliegt allein schon der Zugang zu einzelnen Datenquellen für Forschende großen Beschränkungen und bedarf einer Genehmigung durch die jeweiligen

Aufsichtsbehörden sowie durch die Datenhalter. Oft wird eine adäquate Datennutzung durch unterschiedliche Auslegung oder Handhabung gesetzlicher Vorgaben durch die zu beteiligenden Einrichtungen behindert oder unmöglich gemacht. Dabei erweist sich auch die föderale Struktur als Hemmschuh, insbesondere, wenn eine Datennutzung, die in einem bestimmten Bundesland erlaubt wird, in einem anderen Bundesland *nicht* erlaubt wird. Eine bestmögliche Nutzung von Gesundheitsdaten für die Forschung ist damit in Deutschland gegenwärtig nicht möglich.

Bei der Durchführung von Record Linkage-Projekten stehen wir vor großen Herausforderungen (siehe Box 1). Oftmals wird von Datenhaltern oder Datenschützern für die Verknüpfung personenbezogener Daten die informierte Einwilligung der einzelnen Studienteilnehmenden gefordert, selbst wenn dies nicht erforderlich ist, z. B. weil klare gesetzliche Regelungen fehlen. Hinzu kommt eine oft unterschiedliche Auslegung der gesetzlichen Rahmenbedingungen durch verschiedene Datenschutzbehörden. Zweitens erlauben die Informationen, die in den zu verknüpfenden Datenquellen enthalten sind, häufig keine exakte Verknüpfung. So ist die Datensatzverknüpfung nicht nur eine rechtliche, sondern auch eine methodische Herausforderung, da in Deutschland keine eindeutige persönliche Kennung in Gesundheitsdaten zur Verfügung steht.

In diesem White Paper schildern wir verschiedene Anwendungsfälle, die die technischen und rechtlichen Rahmenbedingungen veranschaulichen, unter denen Gesundheitsforschung gegenwärtig arbeiten muss, um eine personenbezogene Verknüpfung verschiedener Datenkörper zu realisieren. Insgesamt ist festzuhalten, dass das Record Linkage für die Gesundheitsforschung in Deutschland gegenwärtig weit hinter den Standards anderer europäischer Länder hinterherhinkt. So müssen für jeden Anwendungsfall und jedes Record Linkage-Projekt einzelfallspezifische Lösungen entwickelt, geprüft, ggf. modifiziert und – falls positiv beschieden – umgesetzt werden. Die Limitationen und Möglichkeiten dieser unterschiedlichen und spezifisch auf verschiedene Anwendungsfelder zugeschnittenen Ansätze werden diskutiert, und es werden die Voraussetzungen beschrieben, die erfüllt sein müssen, um einen forschungsfreundlicheren Ansatz für die personenbezogene Datensatzverknüpfung zwischen verschiedenen Datenquellen in Deutschland zu erreichen. Dabei werden auch entsprechende Empfehlungen an den Gesetzgeber formuliert (siehe Box 2). Dieses White Paper soll die Grundlage für eine Verbesserung des Record Linkage für die Gesundheitsforschung in Deutschland schaffen. Es zielt darauf ab, praktikable Lösungen für die personenbezogene

Datensatzverknüpfung von unterschiedlichen Datenquellen anzubieten, die im Einklang mit der europäischen Datenschutz-Grundverordnung stehen.

## Box 1: Herausforderungen beim Record Linkage

→ Je nach Verarbeitungssituation und Datenkategorie unterschiedliche Rechtsgrundlagen

→ Fragmentierte aufsichts- und fachbehördliche Kontrollstrukturen, und damit erheblicher Mehraufwand

→ Föderale Zersplitterung der Datenschutzaufsicht bzw. der zuständigen Aufsichtsbehörden und unterschiedliche Auslegung gesetzlicher Vorgaben

→ Hoher bürokratischer und zeitlicher Aufwand für unterschiedliche Genehmigungsverfahren bei Nutzung von Daten mehrerer Datenhalter aus mehreren Bundesländern

→ Record Linkage ohne Unique Identifier führt zu Verknüpfungsfehlern und Datenverlust

→ **Insgesamt führen diese Beschränkungen dazu, dass nur (Bruch-)Teile der bestehenden Daten verknüpft und für die Gesundheitsforschung genutzt werden können. Das heißt, Forschung wird verhindert oder auf unzureichende Ergebnisse mit eingeschränkter Aussagekraft beschränkt.**

## Box 2: Empfehlungen zur Verbesserung des Record Linkage

→ Einführung eines Unique Identifiers und Etablierung von bereichsspezifischen Pseudonymen

→ Einführung einer federführenden Aufsichtsbehörde inkl. Genehmigungskompetenz zur Schaffung von Rechtssicherheit

→ Etablierung einer dezentral-föderierten Forschungsdateninfrastruktur mit zentralen Komponenten für das Record Linkage u. a. zur

- Vergabe von bundeseinheitlichen Identifikatoren
- Pseudonymisierung und standardisierten Durchführung des Record Linkage
- Zuordnung von bereichsspezifischen Pseudonymen (Vertrauensstelle),
- Verwaltung von Einwilligungen
- Beantwortung und Bearbeitung von Anfragen und Anträgen
- Verzahnung mit dem NFDI4Health Services zur Übersicht verfügbarer Daten und relevanter Metainformationen

→ Schaffung der Voraussetzung für Anknüpfung an europäischen Raum für Gesundheitsdaten

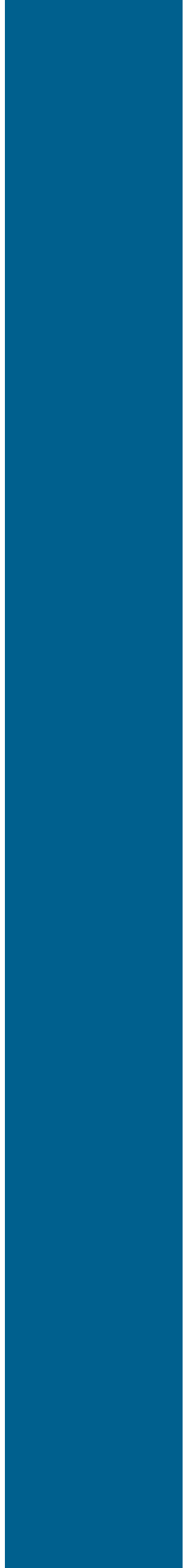

# 1 Einleitung

Neben Primärdaten, die im Rahmen epidemiologischer und klinischer Studien gesammelt werden, gibt es in Deutschland einen zu großen Teilen bisher ungehobenen Schatz an wertvollen gesundheitsbezogenen Informationen auf individueller Ebene, deren Potenzial für die Verbesserung von Prävention, Therapie und Versorgung noch bei Weitem nicht ausgeschöpft wird. Von besonderem Interesse sind hier z. B. Leistungsdaten der gesetzlichen Krankenkassen, Krankenhausdaten, Daten der (elektronischen) Patientenakte, Daten der Rentenversicherungsträger, Berufsverlaufsdaten des Instituts für Arbeitsmarkt- und Berufsforschung (IAB) und Daten aus Krankheitsregistern wie z. B. (bevölkerungsbezogenen) Krebs- oder Mortalitätsregistern. Abgesehen von Krankheitsregistern werden in den meisten sekundären Datenbanken Daten auf individueller Ebene für andere Zwecke als die Forschung gespeichert, z. B. von Krankenversicherungen zu Erstattungszwecken erhobener Leistungsdaten.

Diese Datenkörper enthalten jeweils spezifische auf den jeweiligen Zweck zugeschnittene Informationen, die für sich genommen nur begrenzten Erkenntnisgewinn ermöglichen, weil für die Forschung wichtige Kovariablen fehlen. Erst durch die Kombination verschiedener Datenquellen können wir Fragen beantworten, die große Stichproben erfordern und die die Komplexität des Krankheitsgeschehens umfassend abbilden können. Record Linkage, d. h. die Verknüpfung mit Primärdaten oder die Verknüpfung verschiedener Sekundär- und Registerdatenquellen miteinander auf individueller Ebene, schafft Synergien. Das heißt, eine solche personenbezogene Verknüpfung verschiedener Datenquellen bietet die effizienteste Option für die Gesundheitsforschung, bei der hochinformative, aber noch nicht ausreichend genutzte Daten gemeinsam genutzt und kombiniert werden, um Forschungsfragen zu beantworten, die mit herkömmlichen Ansätzen bzw. mit nur einer Datenquelle nicht beantwortet werden können. So könnten auch klinische Studiendaten mit Daten zur medizinischen Versorgung oder sogar mit anderen Sozialdaten angereichert werden und dadurch einen ganzheitlichen Ansatz ermöglichen, um z. B. die Wirksamkeit und Langzeittoxizität von Arzneimitteln zu ermitteln.

Gerade die genannten Sekundär- und Registerdaten ermöglichen Erkenntnisse mit einem hohen Maß an externer Validität und hoher Relevanz für politische Entscheidungsträger, denn sie enthalten Informationen über Bevölkerungsgruppen, die zwar grundsätzlich keine

Einwände gegen die Nutzung ihrer Daten für die Gesundheitsforschung haben, die aber aus verschiedenen Gründen nur eine geringe Bereitschaft zeigen, sich aktiv an bevölkerungs-bezogenen Primärdatenerhebungen zu beteiligen, z. B. weil ihnen der zeitliche Aufwand zu hoch erscheint.

Aber auch für die Menschen, die sich aktiv an gesundheitsbezogenen Studien beteiligen, bietet die Nutzung von Sekundärdaten viele Vorteile. Typischerweise muss in epidemio-logischen Studien eine große Menge an Informationen über jeden einzelnen Teilnehmenden gesammelt werden, was wiederum zu anspruchsvollen Untersuchungsprotokollen und umfangreichen Fragebögen führt. Die Ergänzung von Untersuchungs- und (subjektiven) Befragungsdaten von Studienteilnehmenden durch administrative (objektive) Daten, die unabhängig von der jeweiligen Studie erhoben wurden, führt nicht nur zu einer Verbesserung der Datenqualität sondern verringert auch den Erhebungsaufwand, weil damit der Umfang der von den Studienteilnehmern zu liefernden Informationen (Primärdaten) auf ein Mindestmaß begrenzt werden kann. Für einige Lebensbereiche, die für die Gesundheitsforschung von höchstem Interesse sind, wie z. B. ihre zurückliegende Medikamenteneinnahme oder ihre medizinische Anamnese, können Studienteilnehmende oft keine zuverlässigen oder ausreichend genauen Angaben machen, so dass entsprechende Forschungsfragen überhaupt nur unter Hinzunahme entsprechender Sekundärdaten beantwortet werden können.

Da es sich bei diesen Sozialversicherungs- und Registerdaten jedoch um hochsensible Daten handelt, sind sie durch strenge Rechtsvorschriften gegen potenziellen Missbrauch geschützt. Die derzeitigen Vorschriften erschweren die Wiederverwendung von Sozialversicherungsdaten für Zwecke der Gesundheitsforschung. Das Record Linkage wird in Deutschland vor allem dadurch erschwert, dass es im Gegensatz zu den nordischen Ländern wie Dänemark keinen eindeutigen personenbezogenen Identifikator gibt. In Abwesenheit eines personen-spezifischen deutschlandweit gültigen und für die medizinische Forschung verwendbaren Identifikators müssen oftmals die personenidentifizierenden Merkmale Namen, Geschlecht, Geburtsdatum etc. verwendet werden. In Deutschland wurden verschiedene alternative Verknüpfungsansätze erprobt, die jeweils ihre spezifischen Einschränkungen (March et al., 2018) und Qualitätsprobleme wie Verknüpfungsfehler (Harron et al., 2017) aufweisen. Viele dieser alternativen Verfahren sind probabilistischer Natur und damit potenziell fehlerbehaftet. Fehler in der Zuordnung, z. B. durch Verwechslungen hervor-gegangen aus Schreibfehlern, führen zu falschen oder fehlenden Verknüpfungen der für die Analyse benötigten Daten einer Person und können damit einen immensen Einfluss auf die Analyse und die Interpretierbarkeit der Ergebnisse haben. Dem entgegen würde ein einheitlicher personenspezifischer Identifikator Zuordnungsfehler minimieren und damit sowohl die Qualität des Record Linkage verbessern als auch – bei Schaffung entsprechender Strukturen – den Schutz der Daten vor unbefugtem Zugriff verbessern. Bislang stehen dem aber gesetzliche Vorschriften und ethische sowie Datenschutzbedenken bezüglich des Missbrauchs des Identifikators durch Dritte (z. B. Cyber-Security Angriffe) entgegen. Interoperable Lösungen in festen Strukturen und Institutionen, die ein für die Gesundheitsforschung erforderliches studien- und datenkörperübergreifendes Record Linkage in einer gesicherten Umgebung durchführen können, sind bislang in Deutschland nicht vorhanden.

Hier muss eine wichtige Abwägung getroffen werden: Dem berechtigten Interesse auf Schutz der personenbezogenen Gesundheitsdaten steht z. B. das Interesse der Krankenversicherten entgegen, dass die in ihren Versichertendaten verborgenen Informationen über etwaige Risiken und den möglichen Nutzen von Behandlungen und Versorgungsleistungen erforscht und zur Verbesserung der gesundheitlichen Versorgung genutzt werden. Eine spezifische Einwilligung des Einzelnen hierfür kann in vielen Fällen nicht eingeholt werden. Millionen von Studienteilnehmenden, deren Daten auf individueller Ebene in sekundären Datenbanken gesammelt wurden, um eine informierte Zustimmung zu bitten, ist i.d.R. weder durchführbar noch wissenschaftlich akzeptabel: Die Bitte um eine informierte Zustimmung würde zu einer stark selektierten Stichprobe aufgrund von Non-Responder Bias führen, was wiederum die Gültigkeit und Verallgemeinerbarkeit der Forschungsergebnisse einschränken würde. In Deutschland sind beispielsweise Sozialversicherungsdaten durch § 75 Sozialgesetzbuch (SGB) X geschützt. Dieser Paragraph erlaubt die Weitergabe solcher Daten zu Forschungszwecken ohne informierte Einwilligung nur auf Antrag für ein bestimmtes Forschungsprojekt oder Forschungsgebiet und nur für begrenzte Zeiträume, sofern die legitimen Rechte auf individuelle Vertraulichkeit der Betroffenen nicht beeinträchtigt werden oder das öffentliche Interesse an dieser Forschung das Recht auf Vertraulichkeit der Betroffenen bei weitem überwiegt.

So unterliegt allein schon der Zugang zu einer einzelnen sekundären Datenquelle für Forschende großen Beschränkungen und bedarf einer ausführlichen Begründung und Genehmigung durch die jeweilige Aufsichtsbehörde sowie durch den Datenhalter. Darüber hinaus ist die Verknüpfung von Sekundärdatenquellen untereinander oder mit

Primärdatenbanken in Deutschland extrem schwierig und zeitaufwendig – und daher oft unmöglich. Eine bestmögliche Nutzung von Sekundärdaten für die Gesundheitsforschung ist damit in Deutschland – im Gegensatz zu anderen europäischen Ländern – gegenwärtig nicht möglich.

Hier stehen wir vor zwei großen Herausforderungen. Erstens erfordert die Verknüpfung personenbezogener Daten die informierte Einwilligung der Studienteilnehmenden, sofern es keine spezielle gesetzliche Regelung gibt. Hinzu kommt eine oft unterschiedliche Auslegung der gesetzlichen Rahmenbedingungen durch die verschiedenen (Landes-) Datenschutz-behörden. Ohne klare und überregional gültige gesetzliche Vorgaben werden wichtige Studien, die z. B. die Patientensicherheit verbessern könnten, nicht durchführbar sein. Zweitens erlauben die Informationen, die in den zu verknüpfenden Datenquellen enthalten sind, oft keine exakte deterministische Verknüpfung. So ist die Datensatzverknüpfung nicht nur ein rechtliches, sondern auch ein technisches Problem, da in Deutschland, anders als in vielen anderen Ländern, gesundheitsbezogene Daten ohne eindeutige Kennung gespeichert werden. Das bedeutet, dass zum Beispiel ein Datensatz aus einem Krebsregister nicht direkt mit dem Leistungsdatensatz derselben Person bei einer gesetzlichen Krankenkasse verknüpft werden kann. Diese Datenquellen verwenden unterschiedliche Algorithmen zur Generierung und Verschlüsselung der personenbezogenen Informationen, was zu unterschiedlichen Pseudonymen führt. Je nach angewendetem Verfahren und zur Verfügung stehenden Identifikatoren bergen Datensatzverknüpfungsansätze jedoch häufig das Risiko, Verknüpfungsfehler einzuführen (Harron et al., 2017).

Wir schildern verschiedene Anwendungsfälle, die die technischen und rechtlichen Rahmen-bedingungen veranschaulichen, unter denen Gesundheitsforschung gegenwärtig arbeiten muss, um eine personenbezogene Verknüpfung verschiedener Datenkörper zu erreichen. Je nach Anwendungsgebiet werden die verschiedenen Verknüpfungsmethoden für die datenschutzfreundliche Datensatzverknüpfung (Privacy-Preserving Record Linkage, PPRL) bewertet, die in Deutschland und anderen europäischen Ländern mehr oder weniger erfolgreich umgesetzt wurden. Hier sind insbesondere probabilistische und deterministische Verknüpfungsmethoden zu unterscheiden, die abhängig von dem Forschungszweck, der Struktur und Quelle der zu verknüpfenden Daten sowie dem rechtlichen (insbesondere den EU-Datenschutzbestimmungen), organisatorischen und technischen Umfeld infrage kommen. Aufbauend auf March et al. (2018) werden die Erfahrungen, die bei der Verknüpfung verschiedener gesundheitsbezogener Datenquellen in Deutschland in früheren Studien gemacht wurden, beschrieben und ausgewertet. Basierend auf dieser Auswertung werden die im deutschen rechtlichen und administrativen Kontext entwickelten Lösungen mit den identifizierten Best-Practice-Verknüpfungsansätzen verglichen, um Lösungen vorzuschlagen und die Best-Practice-Ansätze im Hinblick auf ihre Vereinbarkeit mit der Datenschutz-Grundverordnung (DSGVO) und unter Berücksichtigung der wissenschaftlichen Erfordernisse zu bewerten. Die Limitationen und Möglichkeiten dieser unterschiedlichen und spezifisch auf verschiedene Anwendungsfelder zugeschnittenen Ansätze werden diskutiert, und es werden die Voraussetzungen beschrieben, die erfüllt sein müssen, um einen forschungsfreund-licheren Ansatz für die personenbezogene Datensatzverknüpfung zwischen verschiedenen Datenquellen (Record Linkage) in Deutschland zu erreichen. Dabei werden auch ent-sprechende Forderungen an den Gesetzgeber formuliert.

Dieses White Paper soll die Grundlage für eine Verbesserung der Gesundheitsforschung in Deutschland schaffen. Es zielt darauf ab, die Auffindbarkeit und Zugänglichkeit von Sekundär- und Registerdaten zu verbessern und praktikable Lösungen für die personenbezogene Datensatzverknüpfung (Record Linkage) von Primär- und Sekundärdaten anzubieten.

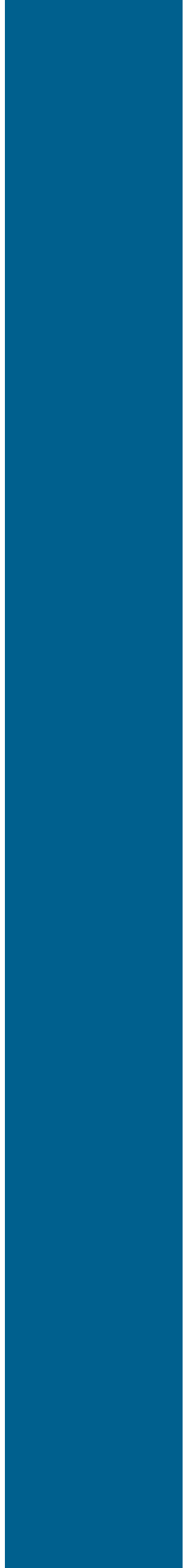

# 2 Grundlagen des Record Linkage in der Gesundheitsforschung

In diesem Kapitel wird die grundlegende Terminologie für das Record Linkage dargestellt, auf die sich die darauffolgenden Use Cases beziehen. Die Begriffe basieren hauptsächlich auf den entsprechenden Definitionen des Artikels „Quo vadis Datenlinkage in Deutschland? Eine erste Bestandsaufnahme" (March et al., 2018) und sind im Glossar (Tabelle Anhang 1) aufgeführt. Im Unterschied zu dem Papier von March et al. wird Record Linkage in diesem White Paper nicht nur als rein technischer Vorgang aufgefasst, sondern schließt darüber hinaus auch die Betrachtung der organisatorischen und datenschutzrechtlichen Aspekte mit ein. Außerdem wird in diesem Kapitel auf einzelne zentrale Begriffe eingegangen, um erstens eine systematische Einordnung hinsichtlich Datenquellen, Verfahren des Record Linkage und Privacy-Preserving Record Linkage (PPRL) vornehmen zu können und zweitens die besonderen Aspekte der Einbindung einer Treuhandstelle / Vertrauensstelle im Kontext der Gesundheitsforschung hervorzuheben. Im Anhang ist zudem eine Übersicht über die zur Verfügung stehenden Software-Tools fürs Record Linkage zu finden (Tabelle Anhang 2). Für eine ausführliche Einführung in Record Linkage sei auf Christen et al. (2020) verwiesen.

## 2.1 Datenquellen

Weltweit werden Daten zielgerichtet in speziellen Datenquellen erfasst und verwaltet. Datenquellen können unterschiedlich technisch organisiert und verfügbar sein, z. B. in Dateien und Datenbanken. In der Medizin sind insbesondere solche Datenquellen von Interesse, die personenbezogene Daten von Patient:innen oder Studienteilnehmenden enthalten. Die Speicherung, Verwaltung, Bearbeitung und Analyse dieser Daten orientiert sich an einem speziellen Zweck, sei es in der Patientenversorgung zur Dokumentation von diagnostischen Untersuchungsergebnissen oder in klinischen / epidemiologischen Studien zur Untersuchung medizinisch-wissenschaftlicher Fragestellungen. Beispielsweise wird die Dokumentation der Patientenversorgung in speziellen Krankenhausinformationssystemen (KIS) vorgenommen; für die medizinische Forschung werden oftmals spezielle Studiendatenbanken angelegt oder auf bestehende Register, wie die Krebsregister, zurückgegriffen; Abrechnungsdaten, die Informationen zu Diagnosen oder Prozeduren enthalten, sind unter anderem bei den Krankenkassen und beim Bundesamt für Arzneimittelsicherheit (BfArM) zu finden und Meldeämter verwalten mit zeitlichem Bezug die Adressdaten der Einwohner einer Region und deren Vitalstatus (lebend / verstorben).

Die Datenquellen lassen sich aus Sichtweise der Datennutzenden in primäre und sekundäre Datenquellen einteilen, wobei die Datennutzung primärer Datenquellen auf deren ursprünglichen Zweck, z. B. die Dokumentation in der Patientenversorgung, ausgerichtet ist. Dagegen hat die sekundäre Datennutzung die Wiederverwendung der Daten zum Ziel, d. h. die Nutzung der gesammelten Daten für einen anderen Zweck. Dies ist zum Beispiel der Fall, wenn Abrechnungsdaten der Krankenkassen für wissenschaftliche Untersuchungen verwendet werden.

Jede Datenquelle zeichnet sich durch Stärken und insbesondere Schwächen aus, wie beispielsweise fehlende und fehlerhafte Werte sowie nicht enthaltende Informationen, die zur Beantwortung einer Fragestellung benötigt werden. Die Schwächen können idealerweise durch eine Verknüpfung (Record Linkage) mit anderen Datenquellen ausgeglichen werden. Ein solches Record Linkage, d. h. die Verknüpfung von personenbezogenen Daten auf Individualniveau, kann sowohl von Primärdaten mit Sekundär- und / oder Registerdaten, aber auch von Sekundär- und / oder Registerdaten untereinander erfolgen (March et al., 2019).

Des Weiteren können die zu verknüpfenden Daten vertikal oder horizontal verteilt vorliegen. In der vertikalen Datenpartitionierung verfügen verschiedene Datenquellen über unterschiedliche Daten (Variablen) zu derselben Personengruppe. Dies betrifft beispielsweise Patient:innen mit seltenen Erkrankungen, deren Diagnostik- und Behandlungs-Odyssee sich oftmals über eine Vielzahl von medizinischen Einrichtungen erstreckt, so dass sich medizinisch-wissenschaftliche Fragestellungen (z. B. zur frühzeitigen Diagnose, Erkrankungsstadien und Therapieoptionen) vor allem mit der Analyse dieser chronologisch geordneten Daten beantworten lassen. Um die entsprechenden Fragestellungen umfassend bearbeiten zu können, bedarf es eines vorangestellten Record Linkage. Bei der horizontalen Datenpartitionierung liegen für unterschiedliche Personen in verschiedenen Datenquellen die gleichen Variablen vor (Gaye et al., 2014; Kohlmayer et al., 2014). Die datenschutzkonforme Nutzung von horizontal verteilten Daten wird beispielsweise in standortübergreifenden medizinischen Forschungsvorhaben und -initiativen benötigt, wie z. B. in den Deutschen Zentren der Gesundheitsforschung (DZG), im Netzwerk Universitätsmedizin (NUM) oder in der Medizininformatik-Initiative (MII).

Ausgehend von der Zielstellung des Forschungsvorhabens muss recherchiert werden, ob sich diese mit bereits existierenden, oftmals für andere Zwecke erhobenen Daten wie z. B. Sekundärdaten beantworten lässt. Zudem muss recherchiert werden, ob geeignete Datenquellen für das Vorhaben nutzbar sind. Hierfür ist zu prüfen, (a) ob die Sekundär- bzw. Registerdaten für das Vorhaben geeignet sind bzw. ob sie für die Beantwortung der Forschungsfrage zweckmäßig und ausreichend sind und (b) ob die Daten für die Forschung verwendet werden dürfen. Sind keine geeigneten Datenquellen zur Beantwortung der Fragestellung vorhanden oder dürfen diese nicht für die Forschung verwendet werden, müssen neue Daten für den Forschungszweck - also Primärdaten - erhoben werden (March et al., 2019).

## 2.2 Das Identifikator-Dilemma

Datenquellen können einen Identifikator enthalten, mit dem die einzelnen Daten einer Person zugeordnet werden können. Ein direkter Identifikator identifiziert eindeutig eine Person; solche Identifikatoren sind einrichtungsgebunden und nicht übergreifend verfügbar. Einrichtungsspezifische Identifikatoren werden beispielsweise in der Patientenversorgung sowie in Studien und Registern verwendet; das Vorhandensein solcher Identifikatoren bleibt für die Öffentlichkeit zumeist unsichtbar. Dagegen ist jede sozialversicherungspflichtige Person mit einer Sozialversicherungsnummer, jede krankenversicherungspflichtige Person mit einer Versichertennummer und beinah jede Person in Deutschland mit einer Steuernummer assoziiert, die zwar übergreifend genutzt werden kann – allerdings nur für bestimmte vorgesehene Zwecke. Mit der Nutzung dieser Identifikatoren entsprechend ihrer Bestimmung (z. B. Abrechnung von medizinischen Leistungen oder steuerrelevanten Vorgängen) sind sie zudem lediglich in ausgewählten Datenquellen verfügbar. **Damit gibt es keinen einheitlichen, eineindeutigen personenspezifischen Identifikator, auf dem ein Record Linkage zwischen beliebigen Datenquellen mit gesundheitsbezogenen Daten basieren kann.**

## 2.3 Verfahren und Arten des Record Linkage

Die Identitätsdaten einer Person (IDAT wie z. B. Name, Vorname, Geburtsdatum, Geschlecht und Adresse) sind in der Praxis oftmals Veränderungen unterworfen, sodass z. B. unterschiedliche Schreibweisen eines Namens auftreten und dadurch verschiedene digitale Identitäten – auch Personenidentitäten genannt – einer tatsächlichen Person entstehen. Andersherum können auch unterschiedliche Personen scheinbar gleiche verfügbare IDAT aufweisen und somit fälschlicherweise dieselbe Personenidentität zugewiesen bekommen. Dies ist insbesondere der Fall, wenn nur wenige oder grobe Informationen über die IDAT vorliegen. Die möglichst fehlerfreie und vollständige Zuweisung der zuzuordnenden Person auf Grundlage der verschiedenen digitalen Identitäten ist das Ziel bei Record Linkage-Prozessen. Record Linkage beschreibt in diesem Zusammenhang die Verknüpfung von Datensätzen in Hinblick auf die Zugehörigkeit zu einer Person (Hampf et al., 2019). Dabei werden direkte oder indirekte Identifikatoren (siehe Tabelle Anhang 1: Glossar) untereinander abgeglichen, um festzustellen, welche Datensätze zur gleichen Person gehören und verknüpft werden können. Beim Linkage mit direkten Identifikatoren kann die Verknüpfung der Datensätze über eindeutig identifizierende Attribute (z. B. Krankenversichertennummer, Name, Adresse, Geburtsdatum) erfolgen. Zum Schutz dieser Identitätsdaten kommen in der Gesundheitsforschung häufig sogenannte vertrauenswürdige Dritte oder Vertrauensstellen bzw. Treuhandstellen zum Einsatz, wenn es darum geht, Record Linkage unter Verwendung von direkten Identifikatoren zu realisieren (Boyd et al., 2015) (mehr dazu in den Abschnitten 2.4 und 3.8).

Bei Linkageverfahren basierend auf indirekten Identifikatoren kommen keine eindeutig identifizierenden Merkmale zum Einsatz. Es werden vielmehr Attribute in den zu verknüpfenden Datensätzen festgelegt, die in Kombination dazu führen können, dass die verknüpften Daten mit hoher Wahrscheinlichkeit ein- und derselben Person zuzuordnen sind (March et al., 2018). Beispielsweise wurden in einer Studie zur Evaluation der Qualität der stationären Versorgung von Herzinfarktpatient:innen in Berlin Krankenkassendaten der AOK Nordost und Daten des Berliner Herzinfarktregisters auf Basis von indirekten Identifikatoren (Alter, Geschlecht, Datum und Uhrzeit der Krankenhausaufnahme) verknüpft (Maier et al., 2015).

Die Record Linkage-Verfahren lassen sich in exakte und fehlertolerante Verfahren einteilen. Bei Ersterem werden nur bei genauer Übereinstimmung eines eindeutigen Identifikators (z. B. Krankenversichertennummer) oder mehrerer Identifikatoren die Daten verschiedener Datenquellen zusammengebracht. Zu den fehlertoleranten Verfahren gehört das regelbasierte Linkage, das distanzbasierte Linkage und das probabilistische Linkage (vgl. Abbildung 1). Während bei dem regelbasierten Linkage für eine erfolgreiche Verknüpfung der Datensätze festgelegt wird, welche Identifikatoren komplett übereinstimmen müssen und bei welchen eine teilweise Übereinstimmung ausreichend ist, werden beim distanzbasierten Linkage die Identifikatoren in ihrer Ausprägung auf Ähnlichkeit hin überprüft.

In diesem Sinne wird mittels sogenannten String-Metriken ermittelt, wie viele Änderungen in der Schreibweise eines Namens nötig sind, um ihn in den zu vergleichenden Namen umzuwandeln. Dabei gilt, je weniger Änderungen dafür nötig sind, um so ähnlicher sind sich diese Namen. So ist z. B. für das Vergleichspaar „Maier" und „Mayer" nur die Änderung eines Buchstabens nötig, um die Namen ineinander umzuformen. Fellegi & Sunter (1969) entwickelten auf Basis der Idee von Newcombe et al. (1959) die Theorie des probabilistischen Record Linkage – also der Verwendung von Wahrscheinlichkeiten im Linkageverfahren. Dabei wird beispielsweise davon ausgegangen, dass es bei dem Abgleich von zwei Datensätzen mit einem landestypisch verbreiteten Namen (z. B. Müller) weniger wahrscheinlich ist, dass diese Datensätze zusammengehören, als wenn es sich um einen selteneren Namen (z. B. Aschowski) handelt (Stegmaier et al., 2019). Ebenso wird die Aussagekraft von unterschiedlichen Identifikatoren – wie z. B. Name und Geschlecht – unterschiedlich bewertet. Nachnamen weisen beispielsweise deutlich mehr Ausprägungen auf als das Geschlecht. Es besteht somit eine geringere Wahrscheinlichkeit, dass bei zwei unterschiedlichen Identitäten der Nachname übereinstimmt als das Geschlecht. Somit wird die Übereinstimmung beim Geschlecht geringer gewichtet als die Übereinstimmung beim Namen. Darüber hinaus erhöht die Übereinstimmung von weiteren Identifikatoren die Wahrscheinlichkeit der Übereinstimmung der Identität. Zusätzlich kann es aus Gründen des Datenschutzes erforderlich sein, die Identifikatoren zu verschlüsseln. In dem Fall müssen die verschlüsselten Identifikatoren untereinander vergleichbar und somit für den Abgleich geeignet sein. Oft werden dafür Kontrollnummern verwendet, die aus der Verschlüsselung von persönlichen Daten wie Name, Vorname, Geburtsname und Geburtsdatum resultieren. Es werden jedoch nur dieselben Kontrollnummern erzeugt, wenn deren Ausgangsdaten in exakt gleicher Schreibweise vorliegen. In einem weiteren Verfahren zur Verschlüsselung von Identifikatoren, das zudem ein fehlertolerantes Record Linkage ermöglicht, werden mit einer geeigneten Zahl von Hash-Funktionen sogenannte Bloomfilter erzeugt. Im Abschnitt 2.4 wird auf diese Verschlüsselungsverfahren zum Zwecke des Record Linkage näher eingegangen (siehe auch Tabelle Anhang 1: Glossar).

Um beim Abgleich von großen Datensätzen die Laufzeit der Linkageverfahren zu verkürzen (Christen, 2012b), wird das sogenannte Blocking verwendet. Dabei werden die Datensätze in Teilmengen oder Blocks anhand von bestimmten Regeln aufgeteilt. Nur die in den entsprechenden Block selektierten Datenmengen werden untereinander verglichen (March et al., 2018). Beispielsweise kann ein Blocking nach Postleitzahl vorgenommen werden, sodass nur Daten mit gleicher Postleitzahl untereinander abgeglichen werden.

Die genannten Linkageverfahren können auch kombiniert werden und verschiedene Identifikatoren bzw. Instrumente zur Anwendung kommen (March et al., 2018):

- Direkte vs. indirekte Identifikatoren

- Klartextangaben vs. verschlüsselte Identifikatoren

- Unterschiedliche Formen des Blockings

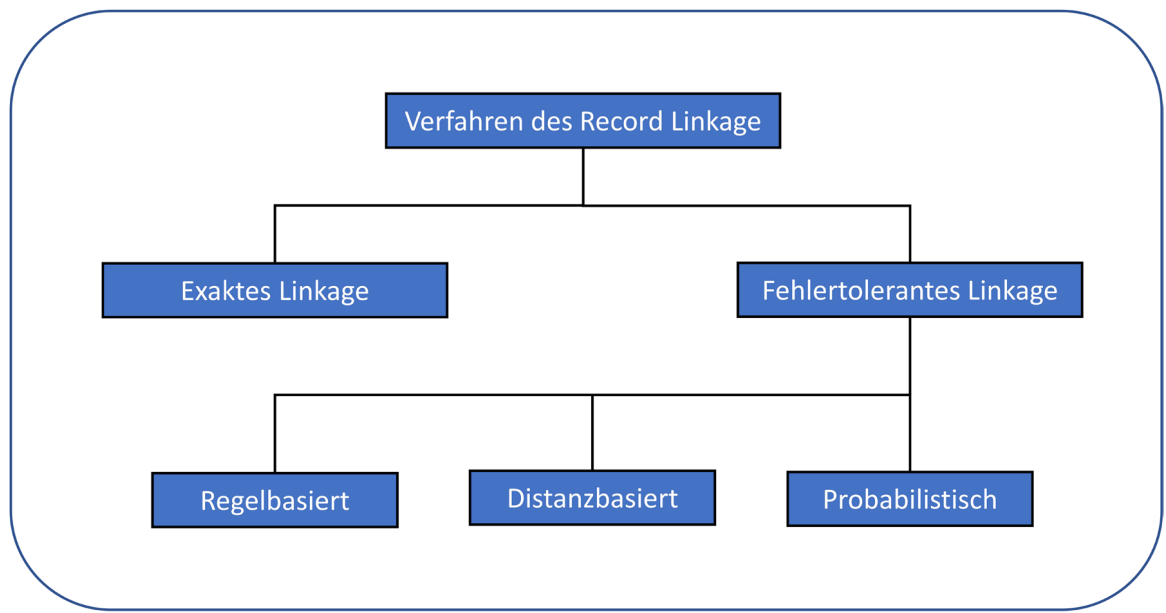

Abbildung 1: Verfahren des Record Linkage auf Basis der Publikation „Quo vadis Datenlinkage in Deutschland? Eine erste Bestandsaufnahme" (nach March et al., 2018)

## 2.4    Privacy-Preserving Record Linkage

Wenn eine unabhängige Vertrauensstelle oder Treuhandstelle genutzt wird, können für das Record Linkage identifizierende Daten wie Name und Geburtsdatum im Klartext zugänglich gemacht werden. Wenn dies aus Datenschutzgründen nicht möglich ist, können Methoden des Privacy-Preserving Record Linkage (PPRL) eingesetzt werden. Hierbei werden keine Klartext-IDAT ausgetauscht. Die Dateneigner transformieren die identifizierenden Daten in eine kodierte Version, aus der keine Rückschlüsse auf die ursprünglichen Daten gezogen werden können. Nur diese kodierten identifizierenden Daten werden an die Linkage-Stelle geschickt, die mit diesen den Abgleich vornimmt, ohne Personen re-identifizieren zu können.

Die Kodierungen basieren in der Regel auf kryptografischen Hashfunktionen, die Klartext-IDAT deterministisch und nur erschwert unumkehrbar transformieren. Einfache Hashfunktionen sollten nur bei exaktem Linkage verwendet werden, da schon leicht abweichende Eingabewerte, z. B. durch Tippfehler, zu völlig unterschiedlichen kodierten Ausgaben führen. Für fehlertolerantes Record Linkage werden daher häufig Kodierungen auf Basis sogenannter ähnlichkeitserhaltender Bloomfilter verwendet. Bloomfilter sind Datenstrukturen, die aus einer Bit-Folge fester Länge bestehen. Zur Überführung in einen Bloomfilter werden die unverschlüsselten (Klartext-)IDAT zunächst in Teile zerlegt z. B. Ma-ai-ie-er für Maier. Diese Teile werden mit Hashfunktionen auf Positionen im Bloomfilter abgebildet und die jeweiligen Bits auf 1 gesetzt. Ähnliche Klartext-IDAT (z. B. Mayer) bestehen aus ähnlichen Teilen (Ma-ay-ye-er), die zu ähnlichen Bloomfiltern führen.

Durch diese ähnlichkeitserhaltende Kodierung werden jedoch Muster aus den Klartext-IDAT (z. B. häufige Namen) in den kodierten IDAT widergespiegelt, wodurch potentiell durch Häufigkeitsanalysen kodierten IDAT ihr Klartext zugeordnet werden kann. Aufgrund dieser Anfälligkeit der Kodierung wurden verschiedene Techniken entwickelt, die darauf abzielen, Bloomfilter zu härten, um sie gegen Re-Identifizierungsangriffe zu wappnen (Franke et al., 2021). Bei Härtungstechniken können beispielsweise die Bitmuster in Bloomfiltern so modifiziert werden, dass zwar die bei Angriffen benötigten Informationen zur Häufigkeitsverteilung der Hamming-Gewichtung (Anzahl der 1-Bits) reduziert werden, aber ein Abgleich mittels Record Linkage-Verfahren möglich bleibt. Secure Multi-Party Computation (SMPC)-basierte Verfahren erreichen durch Nutzung kryptographischer Protokolle wesentlich höhere, mathematisch beweisbare Sicherheitsgarantien auf Kosten der Geschwindigkeit (Stammler et al., 2022).

Für einen Überblick zum Ablauf und den Komponenten von PPRL, insbesondere verschiedener Kodierungstechniken und ihrer Vor- und Nachteile, siehe Christen et al. (2020) und Gkoulalas-Divanis et al. (2021).

Die Ergebnisqualität eines Record Linkage-Vorhabens hängt von einer Vielzahl von Parametern ab, z. B. welche Attribute mit welchem Gewicht im Abgleich berücksichtigt werden sollen. Im Record Linkage auf Basis von unkodierten IDAT können diese Variablen in der Linkage-Einrichtung (z. B. die Treuhandstelle oder Vertrauensstelle) in Abhängigkeit aller vorliegenden Klartext-IDAT angepasst werden. Im PPRL-Szenario hingegen müssen viele dieser Parameter vorab unter den Dateneignern festgelegt werden, da sie bereits zum Zeitpunkt der Kodierung benötigt werden. Dies bringt folgende Herausforderungen mit sich: Die für das jeweilige Record Linkage-Vorhaben geeigneten Parameter müssen ohne gegenseitigen Einblick in die Daten bestimmt werden. Dies ist problematisch, weil die vom Record Linkage-Algorithmus tolerierbaren Fehlerarten zwischen Duplikaten (z. B. Tippfehler, vertauschte Felder) von der Kodierungstechnik abhängen, einige Fehlerarten aber nur bei vorheriger Einsicht der Datensätze verschiedener Dateneigner ermittelt werden können. Eine Möglichkeit dieser Herausforderung zu begegnen, ist die datensatzunabhängige Vorgabe eines Kodierungsschemas, inklusive erforderlicher Attribute, basierend auf Erfahrungswerten und Anforderungen bzgl. der Datensatzeigenschaften. Die Dateneigner sind dann dafür verantwortlich, die Daten in der geforderten Form und Qualität bereitzustellen.

Des Weiteren müssen die Dateneigner das abgestimmte Kodierungsschema auch technisch einheitlich umsetzen. Dies bedarf einer einheitlichen Implementierung oder der Verwendung mehrerer, aber kompatibler Lösungen. Beispielsweise werden bei der Kodierung in der Regel Pseudozufallszahlengeneratoren benötigt, deren Kompatibilität bei der Verwendung unterschiedlicher Implementierungen besonderes Augenmerk erfordert.

Entscheidend für den PPRL-Prozess ist somit, dass vom Dateneigner vergleichbare kodierte Identifikatoren der Linkage-Einrichtung zur Verfügung gestellt werden.

## 2.5 Das Konzept der Vertrauensstelle / Treuhandstelle zur Unterstützung einer datenschutzkonformen Forschung

Die Etablierung datenschutzgerechter Verfahren zur Speicherung und Nutzung personenbezogener Daten ist eine Voraussetzung für Gesundheitsforschung in Deutschland. Basierend auf den gesetzlichen Bestimmungen haben sich in der Praxis Verfahren zur Umsetzung etabliert, so werden u. a. durch den TMF e.V. konkrete Modelle zur Verteilung der Rollen bei der Datenverarbeitung vorgeschlagen (Pommerening et al., 2014).

Eine wichtige Rolle kann dabei ein sogenannter Treuhänder einnehmen. Im Datenschutzleitfaden der TMF wird der Treuhänder als Institution oder Person beschrieben, die zwischen Forschungsdatengebern und der Nachfrageseite von Forschungsdaten als eine unabhängige und neutrale Vertrauensinstanz sicher und gesetzeskonform agiert (Pommerening et al., 2014). Der Treuhänder fungiert in diesem Kontext direkt bei der Umsetzung des Prinzips der informativen Gewaltenteilung. Diese sollte über entsprechende Zugangsregularien auf verschiedenen Ebenen der Datenbestandsmanagements sichergestellt werden. Dabei sollten besonders die für die Forschung essentiellen medizi-nischen Daten (MDAT) von den die Patient:innen oder Proband:innen direkt identifizierenden Daten (IDAT) separat und unabhängig voneinander verwaltet werden. Für die Verwaltung der IDAT wird die Einrichtung einer Treuhandstelle vorgeschlagen, die in diesem Zusammenhang den Schutz der personenidentifizierenden Daten sicherstellt.

Zur Umsetzung des Datenschutzes können durch die Treuhandstelle u. a. folgende Aufgaben übernommen werden (Bialke et al., 2015):

- Führung von Patientenlisten und somit Verwaltung von **direkten Identifikatoren** unter Einbeziehung technischer Verfahren zur eindeutigen Identifizierung, Dopplerausschluss und **Record Linkage**

- Verwahrung und Prüfung von Einwilligungserklärungen

- Widerrufsmanagement von Einwilligungen

- Pseudonymisierung / Anonymisierung von personenbezogenen Daten z. B. für die Speicherung im Forschungskontext oder die Weitergabe an Dritte und Verwaltung der Pseudonyme

- Depseudonymisierung oder Re-Identifikation
  - z. B. bei Rückmeldungen von Forschungsdaten in die Versorgung

- Abgleich von Personendaten mit externen Datenquellen, bspw. Melderegistern

Ergänzend hierzu ist in Abschnitt 3.8 am Beispiel der Unabhängigen Treuhandstelle der Universitätsmedizin Greifswald das Leistungsspektrum einer solchen Vertrauensstelle im Kontext der medizinischen Forschung geschildert.

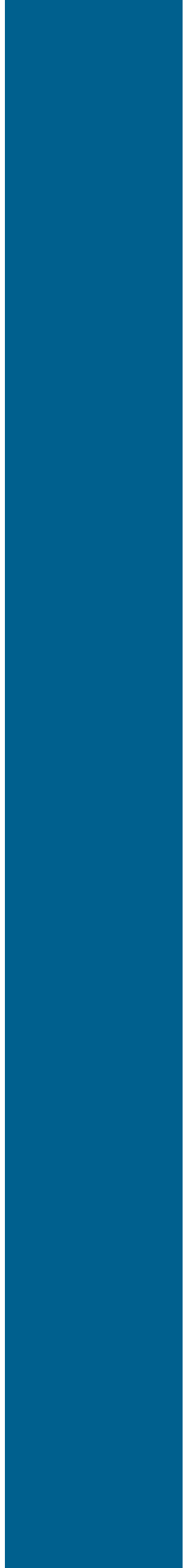

# 3    Use Cases in der Gesundheitsforschung

Im Folgenden werden verschiedene Anwendungsfälle der Verknüpfung von personenbezogenen Gesundheitsdaten in der Gesundheitsforschung, sogenannte Use Cases, beschrieben, um einen Überblick über die praktische Umsetzung von Record Linkage-Projekten in Deutschland zu geben. Dieser Überblick erhebt keinerlei Anspruch auf Vollständigkeit, sondern soll vielmehr einen Einblick in Record Linkage-Projekte aus unterschiedlichen Bereichen der Gesundheitsforschung geben und insbesondere den Zweck, die technische Umsetzung und die Herausforderungen des jeweiligen Vorhabens darstellen. Ein besonderes Augenmerk wurde auf den jeweils verwendeten Datenfluss gelegt – falls nötig veranschaulicht durch sogenannte Datenflussdiagramme. Um ein möglichst breites Spektrum an Beispielen abzubilden, wurden Use Cases sowohl aus dem Bereich der Epidemiologie als auch aus der Surveillance und dem klinischen Bereich ausgewählt. Dementsprechend unterscheiden sich die Use Cases hinsichtlich der zu verknüpfenden Datenquellen und der durchführenden Einrichtung (bzw. der Studie oder des Projekts), was sich im Titel des jeweiliges Use Cases widerspiegelt:

1. Use Case 1: Linkage von Primärdaten und Krankenkassendaten (Studien CoVerlauf und SHIP)
2. Use Case 2: Linkage von Krankenkassen- und Krebsregisterdaten (Projekt DFG-Linkage)
3. Use Case 3: Linkage der Landeskrebsregister am Beispiel von Nordrhein-Westfalen und Niedersachsen
4. Use Case 4: Linkage am Beispiel der molekularen Sars-CoV-2 Surveillance (DESH)
5. Use Case 5: Linkage von Daten aus Routinedaten und -proben und klinischen Erhebungen innerhalb eines rechtlichen Trägers
6. Use Case 6: Linkage von klinischen Routine- und Studiendaten über mehrere rechtliche Träger hinweg am Beispiel des nationalen Netzwerks Genomische Medizin Lungenkrebs
7. Use Case 7: Standortübergreifendes Record Linkage im Deutschen Konsortium für Translationale Krebsforschung (DKTK)
8. Use Case 8: Spezialfall Treuhandstelle – die Unabhängige Treuhandstelle der Unversitätsmedizin Greifswald

9. Use Case 9: Zukünftig – standortübergreifendes Record Linkage in der Medizininformatik-Initiative

10. Use Case 10: Standortübergreifendes Record Linkage im Netzwerk Universitätsmedizin (NUM-RDP)

Die Diversität der Use Cases soll außerdem dem Umstand Rechnung tragen, dass es sich beim Record Linkage von personenbezogenen Gesundheitsdaten um ein Problem handelt, an dem Forschende aus verschiedenen Fachrichtungen arbeiten. Die daraus entstehenden unterschiedlichen Perspektiven sollen so ebenfalls beispielhaft abgebildet werden (Für eine genaue Auflistung der jeweiligen beteiligten Forschenden sei auf das Kapitel „Beiträge der Autoren und Autorinnen" verwiesen). Ein systematischer Vergleich der Use Cases hinsichtlich rechtlicher Grundlage, Umsetzung des Datenschutzes, verwendeter Identifikatoren, Record Linkage-Verfahren und Herausforderungen wird in Kapitel 4 „Allgemeine Vorgehensweise beim Record Linkage" dargestellt. Weitere relevante Anwendungsfälle der Verknüpfung von personenbezogenen Gesundheitsdaten in der Gesundheitsforschung finden sich beispielsweise bei

- Ohlmeier et al. (2015) (Krankenkassendaten/Daten eines Krankenhausinformationssystems),

- Gramlich (2014) (Akutbehandlungsdaten von Schlaganfallpatient:innen/Daten von Frührehabilitationsprogramme und Rehabilitationsbehandlung),

- Schnell & Borgs (2015) (Aufbau einer nationalen Perinataldatenbank),

- Langner et al. (2019a) (Krankenkassendaten/administrative Mortalitätsdaten),

- Dreger et al. (2020) (Strahlenschutzregister/Flugpersonaldatenbanken),

- Fischer-Rosinský et al. (2022) (Notaufnahmedaten/Daten der kassenärztlichen Versorgung),

- Bobeth et al. (2023) (Krankenkassendaten/regionale Krebsregister) sowie

- Trocchi et al. (2022) (Krankenkassendaten/Kinderkrebsregister).

## 3.1 Use Case 1: Linkage von Primärdaten und Krankenkassendaten (Studien CoVerlauf und SHIP)

### 3.1.1 Hintergrund und Ziel

Hochwertige Primärdaten sind ein elementarer Bestandteil der epidemiologischen Forschung (Ahrens et al., 2014). Als Primärdaten werden in diesem Kontext Daten bezeichnet, die direkt zu Studienzwecken erhoben werden. Zu ihrer Erhebung werden unter anderem Fragebögen und medizinische Untersuchungen eingesetzt, was für Studienteilnehmende sowie für Forschende zu hohen zeitlichen Aufwänden und Studienkosten führen kann (Fink, 2014). Weniger zeit- und kostenaufwendig kann es sein, wenn man auf geeignete Sekundärdaten wie z. B. Abrechnungsdaten gesetzlicher Krankenversicherungen (GKV) (Stallmann et al., 2015) zurückgreift (siehe auch Kapitel 2). So enthalten z. B. die zu Abrechnungszwecken erhobenen Krankenkassendaten wertvolle, personenbezogene Informationen wie z. B. zur Demografie, Arzneimittelverordnungen, ambulanten und stationären sowie rehabilitativen Leistungen, Heil- und Hilfsmitteln sowie zu ärztlichen Diagnosen (Pigeot & Ahrens, 2008). Aufgrund von Erinnerungs- oder Wissenslücken können solche Informationen häufig in Primärdatenerhebungen nicht präzise erfasst werden (Ahrens et al., 2020).

Kassendaten sind somit ein wahrer Datenschatz für die epidemiologische Forschung. Allerdings verteilt sich dieser Schatz in Deutschland auf 96 gesetzliche und mehr als 46 private Krankenversicherungen (GKV / PKV) (Mai 2022), wobei erstere mit circa 73 Mio. Versicherten den Großteil der deutschen Bevölkerung abdecken[1,2]. Zudem muss zur Verknüpfung der Krankenkassendaten mit Primärdaten in epidemiologischen Studien neben der Zustimmung der Versicherungen auch eine informierte Einwilligungserklärung (ein sogenannter Informed Consent) und ggf. eine von der PKV benötigte Schweigepflichtentbindungserklärung der Studienteilnehmenden eingeholt werden. Daneben werden Abrechnungsdaten der ambulanten kassenärztlichen Versorgung sowie Arzneimitteldaten von kassenärztlichen Vereinigungen (KV) vorgehalten. Diese stehen vor allem aufgrund datenschutzrechtlicher Anforderungen nicht ohne Weiteres für die Forschung und somit für ein Record Linkage zur Verfügung. Beispielsweise sind separate Einwilligungen für die Nutzung von

---

[1] https://gkv-spitzenverband.de/krankenversicherung/kv_grundprinzipien/alle_gesetzlichen_krankenkassen/alle_gesetzlichen_krankenkassen.jsp, Zugriff 20.05.2022

[2] https://portal.mvp.bafin.de/database/InstInfo/, Zugriff 20.05.2022, Suchkategorie: Krankenversicherer

Medikamentendaten einzuholen. Ein großer Vorteil der Nutzung der KVen als Datenquelle liegt darin, dass aus einer Hand Daten aller gesetzlich Versicherten eines Bundeslandes verfügbar sind. Eine Ausnahme ist Nordrhein-Westfalen bei der sich die Daten auf zwei KVen verteilen. Der Nachteil der KV-Daten liegt darin, dass nur ambulante Versorgungsdaten zur Verfügung stehen. Weder sind Versichertenstammdaten, Daten zur stationären Versorgung noch zu Heil- und Hilfsmitteln oder rehabilitativen Maßnahmen oder zu verschriebenen Medikamenten verfügbar.

Seit 25 Jahren wurde und wird aus den vorgenannten Gründen in Deutschland eine Verknüpfung von Kassendaten mit Primärdaten in einzelnen epidemiologischen Studien durchgeführt (Tabelle 1). Technische Verfahren zur Umsetzung solcher Verknüpfungen sowie damit einhergehende Möglichkeiten und Herausforderungen werden nachfolgend am Beispiel von CoVerlauf und SHIP dargestellt.

Tabelle 1: Übersicht über Studien mit Record Linkage zwischen Primärdaten und Krankenkassendaten.

| Studie | Referenz / Webseite | Daten | Erhebungs-zeitraum | Stichproben-umfang | Einwilligungs-/ Einverständnis-quote |
|---|---|---|---|---|---|
| KORA[1] | John & Krauth (2005) | GKV[9] | 1997-1999; 2005 | 796; 313 | 64 %; 78 % |
| Heinz-Nixdorf-Recall-Studie | Swart et al. (2011) | GKV | Bis 2010 | 4.814 | 90 % |
| AGiL[2] | Swart et al. (2011) | AOK BW[10] | Bis 2011 | 361 | 100 %[13] |
| lida[3] | March (2017) http://www.lida-studie.de | GKV | 2011; 2014 | 6.265; 4.244 | 55 %; 63 % |
| SHIP[4] | Schmidt et al. (2015), Schmidt et al. (2022) | KV Meck.-Vorp.[11] | Ab 2008 | >5000 | 95 % |
| NAKO[5] | Schipf et al. (2020) | GKV, PKV[12] | Ab 2014 | ~200.000 | 94 % |
| HCHS[6] | Jagodzinski et al. (2020), https://hchs.hamburg/ | GKV, PKV | Ab 2016 | ~45.000 | Noch nicht ermittelt |
| LIFE[7] (Follow-Up) | https://www.uniklinikum-leipzig.de/einrichtungen/life/life-er-wachsenenkohorten/life-adult-studie | GKV, PKV | Ab 2017 | 5.665 | 88 % |
| CoVerlauf[8] | https://www.bips-institut.de/covid-19.html#9176 | GKV | 2021 | 1.477 | 85 % |

[1] Kooperative Gesundheitsforschung in der Region Augsburg; [2] Aktive Gesundheitsförderung im höheren Lebensalter; [3] Leben in der Arbeit; [4] Study of Health in Pomerania; [5] NAKO-Gesundheitsstudie; [6] Hamburg City Health Study; [7] Leipziger Forschungszentrum für Zivilisationserkrankungen; [8] Studie zum Erkrankungsverlauf bei Personen mit einem positiven Test auf SARS-CoV-2 bzw. einer COVID-19-Erkrankung; [9] gesetzliche Krankenversicherung; [10] Allgemeinen Ortskrankenkassen für Baden-Württemberg; [11] Kassenärztliche Vereinigung Mecklenburg-Vorpommern; [12] private Krankenversicherung; [13] Einschlusskriterium

### 3.1.2 Technische Umsetzung

Anhand von zwei Beispielen erläutern wir im Folgenden die technische Umsetzung der beschriebenen Record Linkage-Verfahren in der Praxis.

#### 3.1.2.1 CoVerlauf

Bei CoVerlauf handelt es sich um eine „Studie zum Erkrankungsverlauf bei Personen mit einem positiven Test auf SARS-CoV-2 bzw. einer COVID-19-Erkrankung". Mit dieser Studie wurde das Leibniz-Institut für Präventionsforschung und Epidemiologie – BIPS von der Senatorischen Behörde für Gesundheit, Frauen und Verbraucherschutz des Landes Bremen betraut, um folgende drei Ziele zu verfolgen: 1. Schätzung der Häufigkeit von schweren Erkrankungsverläufen nach einer SARS-CoV-2-Infektion bei unterschiedlichen Personengruppen; 2. Ermittlung von Determinanten, die das Auftreten eines schweren Verlaufs beeinflussen; 3. Evaluation des Nutzens von Krankenkassendaten zur Untersuchung von Krankheitsverläufen.

Eine Besonderheit an der CoVerlauf-Studie stellt die Einbindung des Gesundheitsamts Bremen (GA) dar, die nötig ist, da ausschließlich das Gesundheitsamt über die Kontaktdaten der auf SARS-CoV-2 positiv getesteten Personen verfügt und diese auch nicht weitergeben darf. Das bedeutet, dass die erste Kontaktaufnahme über das Gesundheitsamt erfolgen muss. Dazu hat das BIPS ein Schreiben vorbereitet, das das Gesundheitsamt an alle bekannten positiv getesteten Personen verschickt hat. Dieses Schreiben enthält Informationen zur Studie, ein Kontaktformular, auf dem unter anderem die Krankenversichertennummer (KVNR) und die Krankenversicherung angegeben werden konnten, sowie eine zu unterschreibende Einverständniserklärung zur Studienteilnahme und zur Übermittlung der Versichertendaten sowie zur Verknüpfung dieser mit den Primärdaten (Das Formular des Informed Consents und eine allgemeine Vorlage eines solchen Dokuments sind in Intemann et al. (2023) zu finden.). Um an der Studie teilnehmen zu können, mussten diese Unterlagen an das Studienzentrum des BIPS gesendet werden. Anschließend konnte die Befragung (unter anderem zu Soziodemografie, Vorerkrankungen, Medikamenteneinnahme und Lebensstil) telefonisch oder online durchgeführt werden. Die so gewonnenen Primärdaten können dann bei vorliegendem Einverständnis mit den Versichertendaten verknüpft werden. Immerhin stimmten 1617 von 1908 Teilnehmenden (85%) einer Verknüpfung mit den Versichertendaten zu. Zur Verknüpfung ist ein vergleichbares Konzept wie bei der NAKO-Gesundheitsstudie vorgesehen, bei der bereits GKV- und Primärdaten verknüpft werden.

Als direkter Identifikator dient dabei die KVNR. Der Ablauf bei der NAKO-Gesundheitsstudie kann wie folgt zusammenfasst werden (siehe Abbildung 2): Die KVNR wird im Studienzentrum erfasst, mit einer studienspezifischen Identifikationsnummer (ID-S) verknüpft und an eine eigens eingerichtete Treuhandstelle (THS) übermittelt. Die Treuhandstelle generiert weitere Identifikationsnummern (IDs), bspw. für die GKV (ID-GKV) und spätere Forschungsdaten (ID-F). Die ID-GKV wird mit der KVNR und geeigneten Kontrolldaten (Name, Vorname und Geburtsdatum) an die jeweilige GKV weitergeleitet, wo die Verknüpfung zwischen GKV-Daten und ID-GKV stattfindet. Ausschließlich diese beiden Informationen leitet die GKV an ein Integrationszentrum weiter. Dort werden die GKV-Daten schließlich mit der ID-F, die direkt von der THS bezogen wird, als einzige ID in der Forschungsdatenbank für die Auswertung bereitgestellt. Je nach Forschungsfrage können über diese ID GKV- und Primärdaten verknüpft werden. Für zusätzliche Schritte in einer gesonderten Instanz zur Datenaufbereitung und Qualitätssicherung wird eine weitere ID-Nummer verwendet. Alle Daten werden stets verschlüsselt übermittelt.

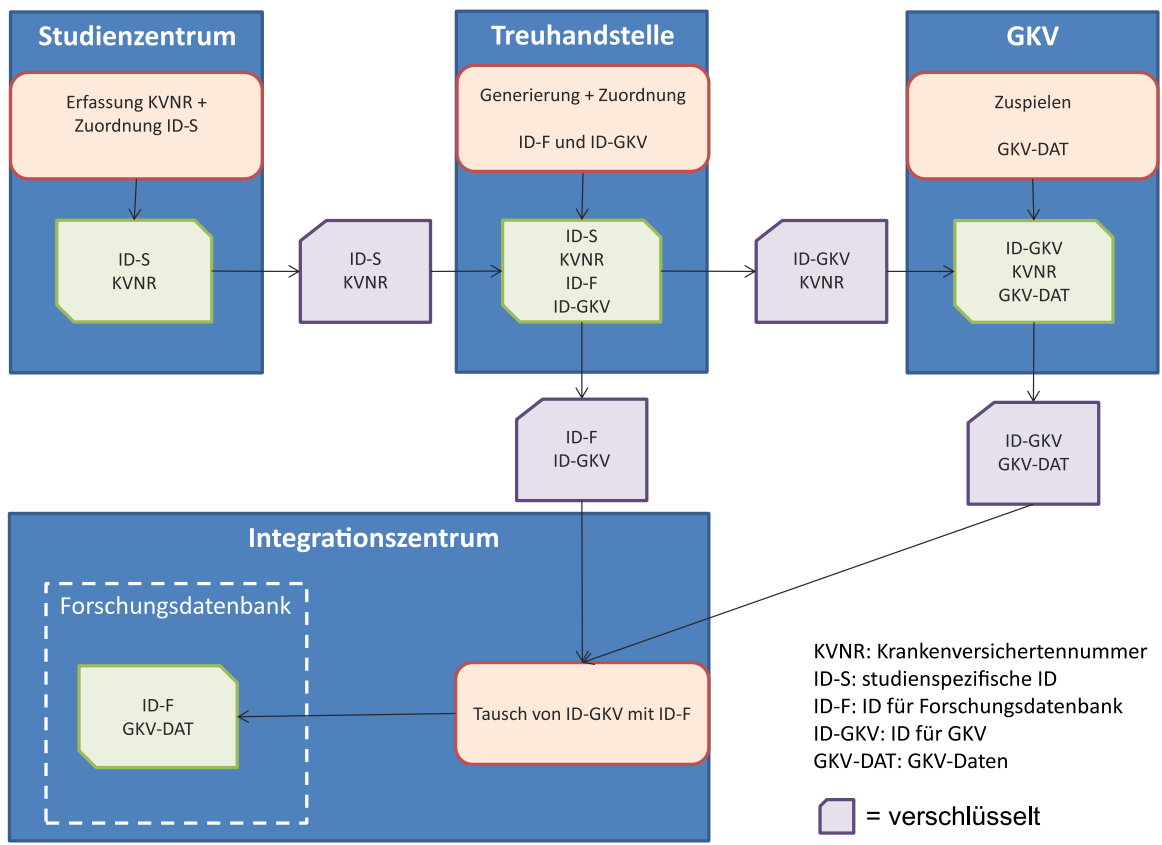

Abbildung 2: Vereinfachter schematischer Datenfluss der GKV-Daten für das Record Linkage bei der NAKO-Gesundheitsstudie

### 3.1.2.2    SHIP

Die Study of Health in Pomerania (SHIP) ist eine bevölkerungsbasierte Studie, die inzwischen drei laufende Kohorten umfasst (Völzke, 2012; Völzke et al., 2011): SHIP-Start (seit 1998, 5 Erhebungswellen), SHIP-Trend (seit 2008, 2 Erhebungswellen) und SHIP-Next (seit 2021, Basiserhebung laufend). Alle drei Kohorten werden in der Region Vorpommern rekrutiert. Eine Besonderheit ist das umfassende Untersuchungsprogramm (Völzke et al., 2011). Während lange Zeit der Fokus allein auf den Primärdaten lag, wurden ab 2008 die Primärdaten mit Abrechnungsdaten der Kassenärztlichen Vereinigung Mecklenburg-Vorpommern (KVMV) verknüpft.

Grundlage für die Verknüpfung ist die Einwilligung der Teilnehmenden zur Nutzung von Daten der ambulanten Versorgung bzw. der KV. Die große Mehrheit stimmte dem zu. Das Linkage mit KV-Daten wird in zwei Schritten umgesetzt, die mit den beteiligten Datenschutzbeauftragten und zuständigen Behörden abgestimmt sind: erstens die Identifikation von SHIP-Proband:innen mit Abrechnungsdaten bei der KVMV und zweitens die Übergabe von Abrechnungsdaten an SHIP. Der erste Schritt wird vom SHIP-Probandenmanagement angestoßen, der zweite Schritt beim SHIP-Datenmanagement abgeschlossen. Beide Schritte sind auf Seiten von SHIP infrastrukturell und personell getrennt. Zudem befindet sich das SHIP-Probandenmanagement in einem vom Internet abgeschirmten Intranet, um für personenidentifizierende Daten die höchstmögliche Sicherheit zu bieten. Im ersten Schrittes werden personenidentifizierende Merkmale von SHIP-Proband:innen (Vorname, Nachname, Geburtsdatum, PLZ und, falls vorhanden, KVNR) an die KVMV über ein verschlüsseltes Netzlaufwerk zur Verfügung gestellt. Genaue Adressangaben werden nicht übermittelt. An die genannten Merkmale ist nur eine projektspezifische ID gekoppelt. Mittels eines für diese Studie entwickelten Programms werden in der KVMV mit einem deterministischen Record Linkage-Algorithmus Übereinstimmungen zwischen den KVMV-Daten und SHIP-Probandendaten gesucht. Namen werden bei diesem Prozess phonetisch standardisiert. Akzeptiert werden nur exakte Übereinstimmungen von allen oben genannten personenidentifizierenden Merkmalen, was das Verfahren technisch sehr einfach und gut nachvollziehbar macht. Nach dem Abschluss von Schritt 1 werden die ausgewählten Abrechnungsdaten in der KV mit der projektspezifischen ID verknüpft und auf einem eigenen Netzlaufwerk an das SHIP-Datenmanagement übergeben. Die verwendeten Netzlaufwerke werden nach Abschluss der jeweiligen Schritte wieder gelöscht. Zugang wird nur zu projektspezifischen Untermengen der Daten nach erfolgreichen Datennutzungsanträgen im Forschungsverbund für Community

Medicine und nach Information der KVMV für Analysen innerhalb der Universitätsmedizin Greifswald gewährt.

### 3.1.3 Möglichkeiten und Herausforderungen

Technisch ist eine Verknüpfung von Primärdaten mit Krankenkassendaten für wissenschaftliche Zwecke also grundsätzlich möglich. Da hierfür in aller Regel das Einverständnis bzw. idealerweise die Einwilligung der Teilnehmenden einzuholen ist (Buchner et al., 2021) und das Studienzentrum den Kontakt mit den Teilnehmenden sowieso handhabt, stellt die zusätzliche Erhebung der KVNR keine Hürde dar. Auf diese Weise steht zudem ein direkter Identifikator zur Verfügung, der in Kombination mit weiteren IDs das Linkage unter Einhaltung des Datenschutzes vereinfacht. Mögliche Probleme und Herausforderungen ergeben sich allerdings an anderen Stellen.

Der Stichprobenumfang der verknüpften Daten hängt somit wesentlich von der erreichten Einwilligungsquote für das Linkage ab. Wie die NAKO-Gesundheitsstudie oder SHIP zeigen, scheint es für die Verknüpfung von Primärdaten mit Sekundärdaten grundsätzlich eine hohe Bereitschaft in der Bevölkerung zu geben (siehe Tabelle 1). Möglicherweise können studienspezifische Merkmale (z. B. persönliches Gespräch, Brief- oder Telefonkontakt) diese Quote beeinflussen, weswegen dies bei der Studienplanung berücksichtigt werden sollte.

Eine zweite Herausforderung stellt der organisatorische Aufwand für die Kooperation mit den Krankenkassen dar. Bei CoVerlauf wurden unter den 1617 eingewilligten Personen 101 unterschiedliche GKVen, PKVen und sonstige vergleichbare Institutionen (z. B. Heilfürsorge oder Beamtenkrankenkassen) registriert. In der NAKO-Gesundheitsstudie wurden nur Kooperationsverhandlungen mit einer Krankenkasse aufgenommen, wenn mindestens 20 Versicherte ihre Einwilligung für die Verknüpfung mit ihren Daten dieser Krankenkasse gegeben haben. Wendet man diese Regel auf CoVerlauf an, so verringert sich der Stichprobenumfang auf 1352 (71%) und die Anzahl der GKVen und PKVen auf 14. Insbesondere für kleine Studien muss somit im Einzelfall überlegt werden, ob der Aufwand in einem vertretbaren Verhältnis zum Nutzen steht. Beispielsweise erscheint es wenig lohnend, mit den 17 bei CoVerlauf registrierten GKVen zu kooperieren, die lediglich einen Teilnehmenden „beisteuern". Für diese Teilnehmenden bestünde zudem ein größeres Re-Identifizierungsrisiko, weswegen ggf. besondere datenschutzrechtliche Vorkehrungen zu treffen wäre. Leichter ist demgegenüber die Nutzung von Abrechnungsdaten, die in den kassenärztlichen Vereinigungen abgebildet sind, wobei die o.g. Nachteile zu berücksichtigen sind (siehe Abschnitt 3.1.2.2).

Die dritte Herausforderung ergibt sich durch das im wesentlichen zweigliedrige Krankenversicherungssystem in Deutschland, in dem für GKV- und PKV-Daten unterschiedliche Datenschutzregeln gelten. Da die GKV-Daten zu den Sozialdaten gehören, ist für ihre Nutzung ein Antrag nach §75 SGB X zu stellen, wohingegen bei PKV-Daten eine Schweigepflichtentbindung nach §203 StGB einzuholen ist (Stallmann, 2018). Zudem weisen PKV-Daten in Teilen eine geringere Detailtiefe als GKV-Daten auf (Stallmann et al., 2015). Würde man wegen des möglichen höheren Aufwands bei der Datenharmonisierung und Einholung der Schweigepflichtentbindung PKVen bei CoVerlauf ausschließen, so würde sich der Stichprobenumfang weiter auf 1325 (69%) reduzieren.

Die vierte Herausforderung wird bei dem Vorgehen von SHIP deutlich, bei der ein exaktes Linkage verwendet wurde. Auf diese Weise wurden einzelne Versicherte ausgeschlossen, wenn keine exakte Übereinstimmung gefunden wurde, um sicherzustellen, dass die Anzahl falscher Zuordnungen reduziert wird. Dies wurde als wichtig erachtet, da sonst Abrechnungsdaten von Personen, die nicht an SHIP teilgenommen haben, fälschlicherweise verwendet werden könnten. Mit diesem Vorgehen konnten für über 90% der eingewilligten gesetzlich versicherten Teilnehmenden Abrechnungsdaten zugeordnet werden. Beim gewählten Verfahren in SHIP ist zu berücksichtigen, dass sich die niedrige Bevölkerungszahl in der Einzugsregion (im Zielaltersbereich ca. 200.000 Einwohner) positiv ausgewirkt hat, d. h. das Risiko von Verwechslungen ist niedriger.

Alles in allem zeigt sich, dass in der Praxis – trotz großer Anstrengungen bei der Einholung von Einwilligungserklärungen bei Kooperationsverhandlungen mit den Krankenkassen und bei der Einhaltung der Datenschutzregeln – ein großer Anteil an Daten nicht vollumfänglich genutzt werden kann. Die damit einhergehenden verringerten Fallzahlen und Selektionseffekte können bei statistischen Auswertungen zu unsichereren und verzerrten Ergebnissen führen. Zudem lassen sich zu speziellen Bevölkerungsgruppen (Privatversicherte oder schwer zu erreichenden Gruppen), häufig keine gesicherten Aussagen treffen, weil von ihnen keine Daten vorliegen.

### 3.1.4 Europäischer Vergleich

In einem systematischen Review wurden Zustimmungsraten für Record Linkage zwischen Primärdaten und administrativen Datenbanken bei unterschiedlichen internationalen Studien erfasst (da Silva et al., 2012). Ähnlich wie in Deutschland variierten die Zustimmungsquoten (39% – 97%), wobei häufig sehr hohe Quoten erreicht wurden (Median: 88%) (vgl. Tabelle 1). Allerdings sind beispielsweise in Dänemark zu CoVerlauf vergleichbare Studien auch ohne Primärdatenerhebung und Informed Consent möglich (Lund et al., 2021), da dort eine COVID-19-Registerkohorte aufgebaut wurde (Pottegård et al., 2020) und für Register-studien besondere Regeln gelten (Thygesen et al., 2011). Diese Registerkohorte umfasst neben demografischen Daten alle SARS-CoV-2-Testdaten aus Dänemark (Datum, Art des Tests und Testergebnis) und dazu weitreichend individuelle Informationen zu Verschreibungen und Behandlungen ähnlich den deutschen GKV-Daten sowie Laborergebnisse zu ver-schiedenen klinischen Tests. Diese Kohortendaten werden im Rahmen des „Danish Wealth and Wellbeing Survey" zudem mit Befragungsdaten auf Individualebene verknüpft (Rosendahl Jensen et al., 2021), was auch in Dänemark eines Informed Consent bedarf. Die Einholung kann in aller Regel über die eindeutige, permanente und persönliche ID durchgeführt werden (Bondonno et al., 2019). Dabei ist keine Genehmigung einer Ethikkommission nötig, eine Zustimmung der dänischen Datenschutzbehörde genügt (Bondonno et al., 2019; Rosendahl Jensen et al., 2021).

## 3.2 Use Case 2: Linkage von Krankenkassen- und Krebsregisterdaten (Projekt DFG-Linkage)

### 3.2.1 Hintergrund und Ziel

Neben den Krankenkassen (siehe Use Case 1) sind auch die epidemiologischen und klinischen Krebsregister (KR) der Länder (siehe Use Case 3) eine attraktive Datenquelle für die Gesund-heitsforschung in Deutschland (Stallmann et al., 2015). Sie sammeln die wesentlichen Infor-mationen zu Diagnose, Therapie und Verlauf von Krebserkrankungen in ihrem Einzugsgebiet. Dabei findet die Krebsregistrierung in jedem Bundesland getrennt statt. Die einzige Ausnahme davon stellt momentan die Krebsregistrierung der Länder Berlin und Brandenburg dar, die in dem klinisch-epidemiologischen Krebsregister Brandenburg-Berlin organisiert sind. Das Gemeinsame Krebsregister der Länder Berlin, Brandenburg, Mecklenburg-Vorpommern und Sachsen-Anhalt und der Freistaaten Sachsen und Thüringen (GKR) wurde zum 31. Dezember 2022 aufgelöst, um die Krebsregister in diesen Ländern neu zu strukturieren.

In dem DFG-Projekt „Evaluierung eines indirekten Linkage-Ansatzes anhand einer Beispielstudie zum Risiko einer Krebsneuerkrankung und der Krebsmortalität bei Patienten mit Typ-2-Diabetes unter Behandlung mit verschiedenen Antidiabetika" (DFG-Linkage) (Kollhorst et al., 2022; Pigeot et al., 2021a; Pigeot et al., 2021b) sollten die KR-Daten genutzt werden, um Krankenkassendaten der pharmakoepidemiologischen Forschungsdatenbank (GePaRD) (Pigeot & Ahrens, 2008) mit Informationen zu Tumorstadium und Todesursache anzureichern. Bei GePaRD handelt es sich um eine Datenbank, die vom Leibniz-Institut für Präventions-forschung und Epidemiologie – BIPS gepflegt wird und die Abrechnungsdaten von circa 25 Mio. Versicherten von vier großen Krankenkassen enthält. An diesem Projekt haben sich nur die AOK Bremen / Bremerhaven und die Handelskrankenkasse (hkk) beteiligt. Zudem wurden für dieses Projekt die Landeskrebsregister Bayern, Bremen, Hamburg, Niedersachsen und Schleswig-Holstein sowie das GKR angefragt, wobei letztlich das GKR und das Landeskrebsregister Schleswig-Holstein nicht an Projekt teilnahmen. Das Ziel des Projekts war es, einen probabilistischen Linkage-Ansatz basierend auf direkt personenidenti-fizierenden Daten mit einem datenschutzrechtlich weniger aufwendigen, deterministischen Ansatz basierend auf nur indirekt identifizierenden Merkmalen (indirekten Identifikatoren), die sowohl in den KR-Daten als auch in GePaRD enthalten sind, zu vergleichen. Daneben sollte das Projekt Hinweise liefern, wie die Abläufe bei beiden Ansätzen verbessert werden können, um die datenschutzrechtlichen Anforderungen kosten- und zeiteffizient umzusetzen. Zudem sollte die Validität von Studienergebnissen in Abhängigkeit beider Verfahren untersucht werden. Als Beispiel wurde dazu das Risiko einer Krebsneuerkrankung bei Einnahme unterschiedlicher Typ-2-Diabetes-Medikamente analysiert. Als Krebsarten wurden eine eher häufige (Darmkrebs) und eine eher seltene (Schilddrüsenkrebs) Entität betrachtet.

### 3.2.2 Technische Umsetzung

### 3.2.2.1 Anträge und Genehmigungen

Zur Nutzung der Krankenkassendaten wurde zunächst ein Antrag nach § 75 SGB X beim Bundesversicherungsamt (mittlerweile Bundesamt für Soziale Sicherung) für die hkk-Daten und bei der Senatorin für Wissenschaft, Gesundheit und Verbraucherschutz in Bremen für die AOK-Daten gestellt. Nach Erteilung der Genehmigungen wurde in GePaRD eine Kohorte bestehend aus Personen mit Typ-2-Diabetes unter Therapie mit oralen Antidiabetika sowie Insulin gebildet. Zur Nutzung der Krebsregisterdaten zu Darm- und Schilddrüsen-karzinomen mussten ebenfalls Genehmigungen bei den zuständigen Aufsichtsbehörden (Beirat des Bayerischen Krebsregisters; Senatorin für Wissenschaft, Gesundheit und Verbraucherschutz des Landes Bremen; Hamburger Behörde für Wissenschaft, Forschung, Gleichstellung und Bezirke; Niedersächsisches Ministerium für Soziales, Gesundheit und Gleichstellung; Ministerium für Soziales, Gesundheit, Jugend, Familie und Senioren des Landes Schleswig-Holstein; Verwaltungsausschuss des GKRs) eingeholt werden. Diese wurden für die KR Bayern, Bremen, Hamburg und Niedersachsen erteilt, wobei wegen datenschutzrechtlicher Bedenken in Niedersachsen und Hamburg modifizierte Verfahren umgesetzt werden mussten. Für die KR Schleswig-Holstein bzw. das GKR konnten keine Genehmigungen eingeholt werden.

### 3.2.2.2 Probabilistisches Record Linkage auf Basis von Kontrollnummern

Das probabilistische Record Linkage mit Kontrollnummern ist unter den Krebsregistern weitgehend standardisiert und seit vielen Jahren im Einsatz (Hentschel & Katalinic, 2008) (siehe auch Use Case 3). Zur Bildung der sogenannten Kontrollnummern werden die personenidentifizierenden Merkmale, wie beispielsweise Vorname, Nachname und Tagesangabe des Geburtstages, in den KR standardisiert, pseudonymisiert und verschlüsselt, um insgesamt 22 Kontrollnummern abzuleiten, die aus alphanumerischen Zeichenketten bestehen (Hentschel & Katalinic, 2008). Diese werden dann zuzüglich weniger Klartextdaten (Monat und Jahr des Geburtsdatums, Geschlecht und Gemeindekennziffer) anstelle der identifizierenden Daten für das Linkage verwendet.

Da jedoch weder diese Kontrollnummern noch die für deren Bildung notwendigen personen-identifizierenden Merkmale in GePaRD enthalten sind, musste diese Information von den Versicherungen abgefragt werden. Dazu wurden über die Vertrauensstelle von GePaRD die Krankenversichertennummern (KVNR) der Mitglieder der Kohorte mit einer zusätzlichen projektspezifischen ID an die jeweilige Krankenkasse übermittelt. Die Krankenkassen generierten daraufhin die Kontrollnummern der entsprechenden Versicherten aus den personenidentifizierten Merkmalen und sendeten diese mit der projektspezifischen ID und Klartextangaben zu Geschlecht, Gemeindekennziffer sowie Geburtsjahr und -monat zurück an die Vertrauensstelle. Diese Informationen wurden dann abhängig vom Wohnort an das jeweilige Krebsregister weitergeleitet, in dem dann das probabilistische Linkage (Fellegi & Sunter, 1969) durchgeführt wurde. Dazu wurden Übereinstimmungsgewichte für Datenpaare beider Datenbanken, bei denen die Blockingvariablen phonetisch standardisierter Nachname und Geburtsjahr exakt übereinstimmten, berechnet. Anhand einer oberen und unteren Schranke für diese Gewichte konnten Datenpaare in „übereinstimmend", „möglicherweise übereinstimmend" zur manuellen Prüfung und „nicht übereinstimmend" eingeteilt werden. Für die manuelle Prüfung wurden weitere Klartextangaben herangezogen. Schließlich wurde im letzten Schritt eine Datei mit projektspezifischer ID, Gewichten und medizinischen Variablen der Krebsregister (u. a. zu Diagnose, Datum der Diagnose und Tumorstadium) an das BIPS für Analysen weitergegeben. Alle Daten wurden stets verschlüsselt übermittelt. Der entsprechende Datenfluss ist in Abbildung 3 schematisch dargestellt.

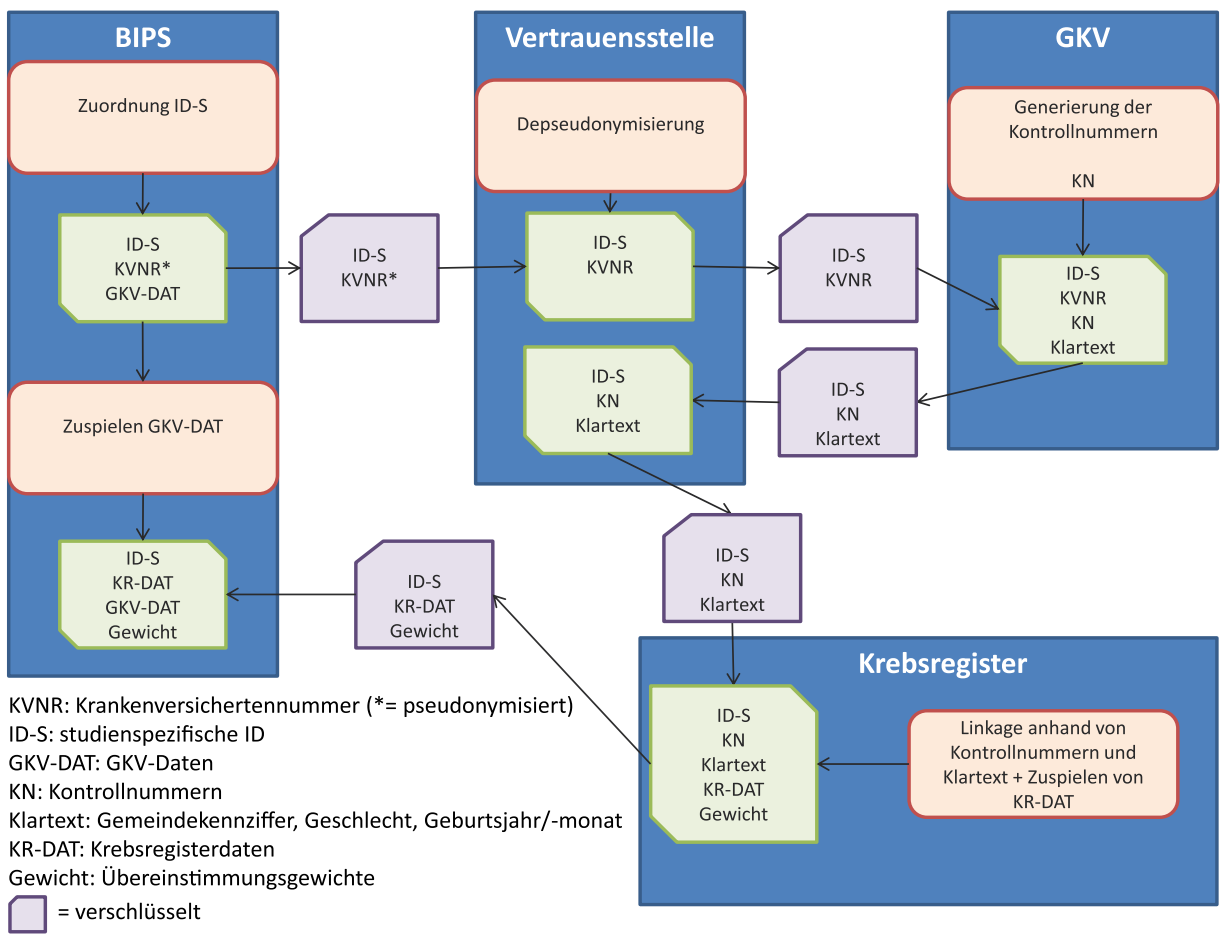

Abbildung 3: Vereinfachter schematischer Datenfluss für das probabilistische Record Linkage auf Basis von Kontrollnummern beim DFG-Linkage-Projekt (Pigeot et al., 2021b)

### 3.2.2.3    Deterministisches Record Linkage auf Basis von Geburtsjahr, Geschlecht und Gemeindekennziffer sowie medizinischen Angaben

Als erstes übermittelten die Krebsregister die Daten aller inzidenten Fälle von Darm- und Schilddrüsenkrebs mit den gleichen medizinischen Variablen wie beim obigen Record Linkage-Verfahren an das BIPS. Zusätzlich enthielten diese Daten personenbezogene Variablen zu Geburtsjahr, Geschlecht und Gemeindekennziffer, die nur für das Linkage verwendet wurden. Am BIPS wurde dann ein deterministisches Linkage durchgeführt. Für eine Verknüpfung wurden eine exakte Übereinstimmung von Geburtsjahr, Geschlecht, Gemeindekennziffer und Krebsart sowie eine Differenz bei dem Diagnosedatum von höchstens 90 Tagen in den Krebsregister- und den Krankenkassendaten vorausgesetzt. Wenn dies noch nicht zu einer eindeutigen Zuordnung führte, wurden folgende Kriterien schrittweise bis zur eindeutigen Zuordnung angewendet: 1. Übereinstimmung bei vier Stellen der ICD-Kodierung bei der Diagnose, 2. Übereinstimmung bei drei Stellen der ICD-Kodierung, 3. Berücksichtigung der stationären Diagnose aus GePaRD (unter Nicht-Beachtung der weniger validen ambulanten Diagnose) 4. geringere Differenz beim Diagnosedatum. Falls dann immer noch mehrere Zuordnungen möglich waren, wurde eine zufällig ausgewählt.

Aufgrund datenschutzrechtlicher Bedenken kam es zu folgenden Modifizierungen bei den beschriebenen Verfahren in Hamburg und Niedersachsen. Da das Hamburger Krebsregister keinen Kohortenabgleich durchführen durfte, konnten keine Daten von nicht-krebskranken Personen entgegengenommen werden. Aus diesem Grund wurde in Hamburg ein vom Krebs-register räumlich getrenntes Studienzentrum eingerichtet, in dem ein BIPS-Mitarbeiter die Linkage-Prozeduren durchführen konnte. Bei den niedersächsischen Krebsdaten hingegen wurde das Re-Identifizierungsrisiko durch die Information von Alter, Geschlecht, amtlicher Gemeindeschlüssel und Diagnose als zu hoch eingeschätzt. Aus diesem Grund wurde statt einer Gesamtstichprobe lediglich eine 50-prozentige Zufallsstichprobe bereitgestellt.

### 3.2.3  Möglichkeiten und Herausforderungen

Im Rahmen des DFG-Linkage-Projekts wurde das Kontrollnummern-Linkage als Goldstandard betrachtet und die Sensitivität sowie Spezifität des deterministischen Linkage berechnet (Kollhorst et al., 2022). Für Darm- und Schilddrüsenkrebs lag die Sensitivität bei 71,8% und 66,6%, wobei Modifizierungen des Linkage-Algorithmus (zum Beispiel bzgl. der erlaubten Diagnosedifferenz) keine oder keine wesentliche Erhöhung der Sensitivität bewirkte. Die Spezifität lag dabei stets über 99%. Der Hauptgrund für falsche Zuordnungen waren in GePaRD kodierte Krebsfälle, die, obwohl sie nicht im Krebsregister vorhanden waren, dortigen Krebsfällen zugeordnet wurden.

Darüber hinaus wurde untersucht, wie sich Ergebnisse von Ereigniszeitanalyen in Abhängigkeit der verschiedenen verknüpften Datensätze bedingt durch die angewendete Linkage-Methode unterscheiden können. Dazu wurde das Risiko für Darmkrebs in Abhängigkeit von Typ-2-Diabetes-Medikamenten (Sulfonylharnstoffe vs. Metformin) ermittelt. Es zeigte sich, dass beim deterministischen Linkage das Risiko um circa 10 Prozentpunkte gegenüber dem Kontrollnummer-Linkage verringert war (Hazard Ratio: 0,83 gegenüber 0,71).

Insgesamt zeigt sich, dass eine Umsetzung von bundesländerübergreifenden wissenschaftlichen Projekten zur gemeinsamen Analyse von Krankenkassen- und Krebsregisterdaten ohne Einholung von Einverständnissen oder Einwilligungen möglich ist – allerdings mit großen Einschränkungen. Dies gilt insbesondere bei dem deterministischen Ansatz, da die niedrige Sensitivität zu unsicheren und verzerrten Ergebnissen führen kann. Ein solches Verfahren ist daher im Allgemeinen nicht zu empfehlen.

Der probabilistische Ansatz basierend auf Kontrollnummern wurde in diesem Projekt zwar als Goldstandard verwendet. Allerdings stellt sich die Frage, ob dies immer gerechtfertigt ist, da Kontrollnummern für einen Abgleich unter den Krebsregistern entwickelt wurden und nicht für ein Linkage mit Krankenkassendaten, bei denen Merkmale für das Kontrollnummernverfahren möglicherweise nicht vollständig oder in einer anderen Form enthalten sein können. Die Erhebung der KVNR und die Einführung der dazugehörigen zusätzlichen Kontrollnummer in den Krebsregistern (Arbeitsgemeinschaft Deutscher Tumorzentren (ADT) & Gesellschaft der epidemiologischen Krebsregister in Deutschland (GEKID), 2014; Stegmaier et al., 2019) kann hier für zusätzliche Sicherheit beim Linkage sorgen, da die KVNR mittlerweile einmalig vergeben wird und lebenslang besteht.

In diesem Projekt wurde darüber hinaus deutlich, dass Studien dieser Art zeit- und kostenintensiv sind. Das hat mehrere Gründe. Erstens müssen die Krankenkassen einem solchen Projekt zustimmen, damit Anträge nach §75 SGB X gestellt werden können. Zweitens sind dafür je nachdem, ob Kassen in einem Bundesland oder deutschlandweit tätig sind, unterschiedliche Aufsichtsbehörden zuständig. Drittens muss für jedes Krebsregister ein separater Antrag nach einem uneinheitlichen Verfahren an die jeweilige Aufsichtsbehörde gestellt werden, wobei jeweils unterschiedliche Unterlagen notwendig sind. Viertens können unterschiedliche datenschutzrechtliche Bedenken vorgebracht werden, die länderspezifische Modifizierungen des ursprünglichen Verfahrens notwendig machen. Überdies können die verschiedenen Verfahren in den Bundesländern zu unterschiedlichen, teilweise ablehnenden Entscheidungen führen. Unter diesen Voraussetzungen erscheinen flächendeckende bundesweite Analysen nicht realisierbar. Aufgrund von regionalen Unterschieden und unterschiedlichen Krankenkassenpopulationen sind diese jedoch nötig zur Vermeidung von Stichprobenverzerrungen, die zur eingeschränkten Verallgemeinerbarkeit von Studienergebnissen führen können.

Das Zentrum für Krebsregisterdaten (ZfKD) sammelt zwar bereits Krebsregisterdaten auf Bundesebene. Diese können allerdings momentan aus rechtlichen Gründen und wegen fehlender geeigneter Identifikatoren nicht zum Zwecke des Record Linkage bereitgestellt werden (siehe Abschnitt 5.1.3 für eine rechtliche Betrachtung). Aus diesem Grund stellt das neue Gesetz zur Zusammenführung von Krebsregisterdaten (Gurung-Schönfeld & Kraywinkel, 2021) eine Möglichkeit zur Verbesserung dar. Das ZfKD ist zusammen mit anderen Organisationen (Arbeitsgemeinschaft Deutscher Tumorzentren, Deutsche Krebsgesellschaft, Deutsche Krebshilfe, Krebsregister und Patientenorganisationen) aufgefordert, bis Ende 2024 ein „Konzept zur Schaffung einer Plattform, die eine bundesweite anlassbezogene Datenzusammenführung und Analyse der Krebsregisterdaten aus den Ländern sowie eine Verknüpfung von Krebsregisterdaten mit anderen Daten ermöglicht", zu erarbeiten (§ 10 BKRG). Dieser Auftrag bietet somit die Möglichkeit, dass attraktive Lösungen für die Durchführung von bundesweiten Record Linkage-Projekten mit Krebsregisterdaten entwickelt werden, was die Wissenschaft und insbesondere Projekte wie die NFDI4Health (Fluck et al., 2021) im Rahmen des Aufbaus einer Nationalen Forschungsdateninfrastruktur als Aufruf begreifen sollten, geeignete Vorschläge zu erarbeiten.

### 3.2.4 Europäischer Vergleich

Wie in Deutschland findet in vielen europäischen Ländern eine populationsbasierte Krebsregistrierung statt (Forsea, 2016). Allerdings sind die Möglichkeiten der Gesundheitsforschung für ein Record Linkage zwischen Krebsregister- und administrativen Gesundheitsdaten in manchen Ländern ungleich größer. Das gilt insbesondere für die nordischen Länder Dänemark, Finnland, Island, Norwegen und Schweden, die eine lange Tradition an Gesundheitsregistern und der personenbezogenen Verknüpfungen zwischen diesen Registern vorweisen können (Pukkala et al., 2018; van Herk-Sukel et al., 2012). In diesen Ländern wird Record Linkage durch eine zentrale Antragstellung und eine eindeutige, permanente und persönliche ID vereinfacht (Pukkala et al., 2018; van Herk-Sukel et al., 2012). Daneben bieten aber mittlerweile auch weitere Länder sehr gute Möglichkeiten für eine Verknüpfung von Krebsregister- und administrativen Gesundheitsdaten. Beispielsweise weicht das Vorgehen in Estland nicht wesentlich von den Grundsätzen in den nordischen Ländern ab (Rahu et al., 2020; Rahu & Rahu, 2018). Zudem wurden in den Niederlanden (Kuiper et al., 2020; van Herk-Sukel et al., 2012), Schottland (van Herk-Sukel et al., 2012), England (Wood et al., 2021) und Wales (Jones et al., 2019) erfolgreich forschungsfreundliche Plattformen für Record Linkage zwischen einer Vielzahl an Registern aufgebaut.

## 3.3 Use Case 3: Linkage der Landeskrebsregister am Beispiel von Nordrhein-Westfalen und Niedersachsen

### 3.3.1 Hintergrund und Ziele

Die Krebsregistrierung ist in Deutschland föderal organisiert: Durch das Krebsfrüherkennungs- und -registergesetz (KFRG) ist jedes Bundesland verpflichtet, epidemiologische und klinische Krebsregistrierung sicherzustellen (Arndt et al., 2020). Dabei sind je nach Bundesland die teils seit Jahrzehnten bestehenden epidemiologischen Krebsregister zu klinischen Registern ausgebaut oder zusätzlich zum bestehenden epidemiologischen Register ein eng kooperierendes klinisches Register etabliert worden.

Die Landeskrebsregister sollen unter anderem das Auftreten und die Trendentwicklung von Tumorerkrankungen beobachten, statistisch auswerten und Grundlagen für die Gesund-heitsplanung bereitstellen. Außerdem sollen sie zu einer Bewertung präventiver und kurativer Maßnahmen sowie zur Qualitätssicherung der onkologischen Versorgung beitragen und Daten für die wissenschaftliche Forschung einschließlich der Ursachenforschung zur Verfügung stellen (siehe beispielsweise § 1 Absatz 2 Gesetz über das Epidemiologische Krebs-register Niedersachsen (GEKN), § 1 Absatz 3 Landeskrebsregistergesetz NRW (LKRG NRW)).

Die Landeskrebsregister pflegen umfangreiche Datenbanken, die durch Meldungen von Krankenhäusern, niedergelassenen Ärzt:innen, Instituten für Pathologie, Screeningeinheiten, Meldeämtern, Gesundheitsämtern und Statistischen Landesämtern gespeist werden. So werden die wesentlichen Informationen zu Diagnosen, Therapien und Verläufen sowie dem Vitalstatus im Zuge einer Krebsbehandlung standardisiert und strukturiert erfasst (Gurung-Schönfeld & Kraywinkel, 2021). Eine wesentliche Aufgabe ist dabei die Zusammenführung der Daten aus den verschiedenen Quellen und die Integration dieser personenbezogenen Informationen in eine konsolidierte Krankengeschichte (ein sogenanntes „Best-of"). Die Erhebung, Verarbeitung und Nutzung der Daten erfolgen dabei jeweils auf Basis der landesgesetzlichen Regelungen.

Zu den von den Registern erhobenen personenidentifizierenden Merkmalen (in den Gesetzen oft als „Identitätsdaten" bezeichnet) gehört seit dem Inkrafttreten des KFRG auch die Krankenversichertennummer (KVNR). Sie wird daher immer häufiger auch für das Record Linkage in den Registern verwendet. Da die KVNR jedoch für privatversicherte Personen und einen großen Teil älterer registrierter Erkrankungsfälle nicht vorliegt, ergänzt sie aktuell lediglich die bestehenden Verknüpfungsverfahren. Für einen Abgleich der Krebsregisterdaten mit Daten aus organisierten Früherkennungsprogrammen wird inzwischen explizit vorgeschrieben, dass ein aus dem unveränderbaren Teil der KVNR des Versicherten abgeleitetes Pseudonym zu verwenden ist (§ 25a Abs. 4 Satz 6 SGB V).

### 3.3.2 Technische Umsetzung

Ein von vielen Registern langjährig eingesetztes Verfahren für die Zusammenführung ist das probabilistische Record Linkage nach Fellegi & Sunter (1969), das häufig in einer Privacy-Preserving-Variante mithilfe sogenannter Kontrollnummern umgesetzt wird (siehe Use Case 2). Die wesentlichen Aspekte des Verfahrens und seiner Anwendung in den Krebsregistern wurde im Manual der epidemiologischen Krebsregistrierung publiziert (Hentschel & Katalinic, 2008).

Für das probabilistische Record Linkage werden aus den Identitätsdaten der Patient:innen eingehender Meldungen durch Normierung, Zerlegung und Bildung phonetischer Codes bis zu 30 Einzelmerkmale erzeugt. Anhand dieser können unter Berücksichtigung von ermittelten Merkmalsfehlerraten und relativen Häufigkeiten der konkreten Ausprägungen Übereinstimmungsgewichte mit bestehenden Meldungen berechnet werden. Die KVNR kann hier als ein Merkmal in die Gewichtsberechnung eingehen oder aber auch für eine Vorauswahl von Bestandsmeldungen genutzt werden, die zuerst mit der neu eingehenden Meldung abgeglichen werden sollen.

In der Kontrollnummern-Variante werden vor allem die direkt identifizierenden Merkmale wie Namen und deren phonetische Codes und ggf. Adressangaben vor der Nutzung ver-schlüsselt. Dies geschieht meist in einer dem Register vorgelagerten Vertrauensstelle. Das Verschlüsselungsverfahren hat dabei zwei Stufen und wird entsprechend einer Empfehlung in allen Landeskrebsregistern gleichartig durchgeführt (Appelrath et al., 1996): eine in allen Registern einheitliche erste Stufe (mittels Hashfunktion) und eine landesspezifische und reversible zweite Stufe (mittels symmetrischer Verschlüsselung). Da die Verschlüsselung deterministisch ist, können die für die Berechnung des Übereinstimmungsgewichts notwendigen Vergleiche auch auf den verschlüsselten Daten durchgeführt werden.

Bei Rücknahme nur der zweiten registerspezifischen symmetrischen Verschlüsselung und einer erneuten Überverschlüsselung mit einem z. B. projektspezifischen Schlüssel lassen sich auch die Daten verschiedener Landeskrebsregister auf Ebene der Kontrollnummern miteinander abgleichen, um z. B. doppelt vorhandene Erfassungen zu erkennen.

### 2.3.2.1    Das Routine-Linkage

Für die Verknüpfung einer neuen Routinemeldung wird nach bereits vorhandenen Einträgen im Bestand des Registers gesucht. Je nach Übereinstimmungsgewicht mit Bestandsmeldungen wird die neue Meldung einem bestehenden Patientendatensatz hinzugefügt (oberhalb der oberen Schranke), ein neuer Datensatz angelegt (unterhalb der unteren Schranke) oder die Meldung in die manuelle Nachbearbeitung geleitet (Graubereich zwischen den beiden Schranken). Dies wird durch Dokumentationskräfte des Registers vorgenommen, die dann zur Klärung auch noch weitere Informationen hinzuziehen, wie etwa medizinische Angaben. Ggf. erfolgen Rückfragen beim Melder oder beim Einwohnermeldeamt.

Bei einigen Krebsregistern wird durch Verschiebung der oberen und unteren Schranke Sensitivität und Spezifität speziell für verschiedene Meldungsarten gesteuert. Während für die medizinischen Daten gilt, dass im Zweifel eher die Zusammengehörigkeit von Datensätzen erkannt werden soll, gilt z. B. für vom Meldeamt übermittelte Namensänderungs- und Umzugsmeldungen, dass nur bei sehr hoher Übereinstimmung eine Zuordnung der Information zum Registerdatensatz erfolgen soll. In den folgenden Boxen wird das Routine-Linkage am Beispiel eines Bundeslandes mit integriertem epidemiologischen und klinischen Krebsregister (Box 3) und am Beispiel eines Bundeslandes mit getrennten epidemiologischen und klinischen Krebsregister (Box 4) veranschaulicht.

## Box 3: Das Routine-Linkage im Landeskrebsregister Nordrhein-Westfalen

Die Rechtsgrundlage für die Arbeit des Landeskrebsregisters NRW (LKR-NRW) ist das Gesetz über die klinische und epidemiologische Krebsregistrierung im Land Nordrhein-Westfalen. Im LKR-NRW lagen Ende 2021 ca. 12 Millionen Meldungen von ca. 2,7 Millionen Personen vor, die mit Informationen von 92 Instituten für Pathologie, 269 Krankenhäusern, mehr als 500 Praxen und medizinischen Versorgungszentren, 21 Screeningeinheiten sowie 396 Meldeämtern gespeist werden. Alle Meldungen werden mithilfe des probabilistischen Record Linkage auf Basis von Kontrollnummern mit den Bestandsdaten verknüpft. Dabei liegen die Klartextinformationen in der Vertrauensstelle nur zeitlich begrenzt vor – die Kontrollnummern hingegen für alle Meldungen in der nachgelagerten Registerstelle dauerhaft. Zusätzlich werden die Identitätsdaten als Block zu einem sogenannten Identitäts-Chiffrat verschlüsselt. Die Identitäts-Chiffrate dürfen jedoch nur zu bestimmten Zwecken entschlüsselt werden – unter anderem zur Aufklärung unklarer Zuordnungen beim Record Linkage. Von den ca. 150.000 eingehenden Meldungen pro Monat müssen 2,5 bis 5% manuell nachbearbeitet werden (Stand 2020).

Bei einer Evaluation des Routine-Linkage des LKR-NRW wurde für eine Stichprobe von 150.000 Meldungen eine Homonymfehlerrate von 0,015% und eine Synonymfehlerrate von 0,2% festgestellt (Schmidtmann et al., 2016). Die Fehlerraten können bei größeren Datenmengen jedoch deutlich steigen. Daher wurden vom LKR-NRW seitdem zahlreiche Optimierungsmaßnahmen am Verfahren umgesetzt (Berücksichtigung von Kreuzvergleichen zwischen Namensbestandteilen sowie Berücksichtigung der KVNR).

## Box 4: Das Routine-Linkage im Landeskrebsregister Niedersachsen

In Niedersachsen gibt es seit 2018 neben dem Epidemiologischen Krebsregister Niedersachsen (EKN) auch das Klinische Krebsregister Niedersachsen (KKN). Beide Register arbeiten eng zusammen, haben aber getrennte Datenbestände, sodass auf Basis derselben Meldungen zwei voneinander unabhängige RL-Verfahren im klinischen sowie im epidemiologischen Krebsregister zum Einsatz kommen. Während das RL-Verfahren des KKN vor allem auf eine schnelle Zuordnung von Meldungen zum aktuellen Erkrankungsfall unter Nutzung der KVNR ausgerichtet ist, legt das EKN einen Schwerpunkt auf die Zuordnung von Meldungen ohne Angabe der KVNR (wie z. B. bei Todesbescheinigungen, Umzugsmeldungen oder Namensänderungen). Dafür nutzt das EKN auch medizinische Angaben.

Im KKN wird eine Kombination aus einer exakten Zuordnung anhand der KVNR und einem probabilistischen Verfahren angewendet. Zum Auflösen von unsicheren Matches wird auf Identitätsdaten im Klartext zugegriffen. Im EKN kommt hingegen das etablierte probabilistische RL-Verfahren auf Basis der Kontrollnummern zum Einsatz. Das KKN und das EKN sammeln und analysieren Abweichungen zwischen beiden RL-Verfahren. Bei unauflösbaren Abweichungen werden Anfragen bei den zuständigen Meldebehörden gestellt, um anhand der Wohnort-Daten die Zuordnung zu kontrollieren.

Somit können durch die zwei unabhängigen RL-Verfahren falsche Zuordnungen schnell gefunden und korrigiert werden, womit es sich als ein zuverlässiges Element der internen Qualitätssicherung erwiesen hat.

### 3.3.2.2 Kohortenabgleich: Record Linkage mit externen Datenquellen

Analog zum Routine-Linkage erfolgt das Record Linkage mit externen Datenquellen zu Forschungszwecken, z. B. für Fall-Kontroll- oder Kohortenstudien (siehe auch Use Case 2). In den Landeskrebsregistergesetzen wird auch geregelt, unter welchen Bedingungen ein Record Linkage zwischen externen Daten und Krebsregisterdaten möglich ist. In der Regel ist ein Abgleich innerhalb des Krebsregisters eher zulässig als ein Record Linkage, bei dem die Einzelfalldaten des Registers herausgegeben werden sollen.

Im Fall einer Datenverknüpfung im Register müssen beim Register entsprechende Forschungsanträge gestellt und nach Freigabe die für das Linkage notwendigen Informationen dem Register zur Verfügung gestellt werden. Hier gibt es je nach Landeskrebsregister leicht abweichende Verfahren, unter anderem abhängig davon, ob die Daten im Klartext oder bereits pseudonymisiert zur Verfügung gestellt werden können. Bei diesem Vorgehen wird geprüft, für welche der übermittelten Datensätze Informationen im Krebsregister vorliegen und dann werden je nach Fragestellung und Datenschutzkonzept Einzeldaten oder aggregierte Daten zurückgegeben. Die eingehenden Studiendaten werden anschließend gelöscht und nicht in die Register übernommen.

Wenn in solchen Fällen größere Datenmengen zu verknüpfen sind, kann die dabei anfallende manuelle Nachbearbeitung nicht in jedem Fall geleistet werden. Es muss daher vor Projektbeginn unter anderem geklärt werden, ob nur sichere Treffer berücksichtigt werden sollen, oder ob ggf. vom Routine-Linkage abweichende obere und untere Schranken verwendet werden können. In den folgenden Boxen wird das Record Linkage zwischen Krebsregistern und externen Datenquellen mit pseudonymisierten Daten (Box 5) und mit Klartext-Daten (Box 6) an unterschiedlichen Beispielen aus den Krebsregistern Nordrhein-Westfalen und Niedersachsen veranschaulicht.

## Box 5: Record Linkage zwischen Krebsregistern und externen Datenquellen mit pseudonymisierten Daten im Krebsregister Nordrheinwestfalen

Das Landeskrebsregister NRW (LKR-NRW) nutzt die Möglichkeit der Kohortenabgleiche regelmäßig zum Abgleich mit anderen Datenquellen. Dazu gehören beispielhaft alle zwei Jahre erfolgende Abgleich mit Teilnehmendendaten des Mammographie-Screening-Programms zur Bestimmung von Intervallkarzinomen (Heidinger et al., 2012), die Anreicherung der NAKO-Gesundheitsstudie mit Daten des Krebsregisters, die Validierung von Todesursachen die algorithmisch aus Krankenkassendaten ermittelt wurden (Langner et al., 2019b) oder die Untersuchung der Krebsinzidenz bei Patient:innen mit Diabetes mellitus Typ 2 (Kajüter et al., 2014).

Dabei werden die Identitätsdaten der Personen bereits bei der sendenden Einrichtung aufbereitet, normiert, phonetische Codes gebildet und die erste Verschlüsselungsstufe für die Kontrollnummern durchgeführt. Diese werden dann Ende-zu-Ende-verschlüsselt und an die Kontrollnummernstelle des LKR-NRW bei der Kassenärztlichen Vereinigung Westfalen Lippe in Dortmund übermittelt. Dort wird dann die zweite Verschlüsselungsstufe angewendet und damit die NRW-spezifischen Kontrollnummern erzeugt, die dann dem LKR weitergeleitet werden.

Einige nicht direkt identifizierende Personenmerkmale wie Geschlecht, Monat und Jahr der Geburt, sowie PLZ und Wohnort werden zusammen mit einer Kommunikations-ID und ggf. projektspezifischen Nutzdaten an das LKR übermittelt. Dort werden die Daten dann wieder mit den Kontrollnummern verknüpft und für den Abgleich verwendet. Zur Unterstützung des Prozesses hat das LKR-NRW 2020 die Software Coho-Client entwickelt, die den Prozess der Bildung der Kontrollnummern und die Ende-zu-Ende-Verschlüsselung implementiert. Die Software wurde seitdem erfolgreich sowohl für den Intervallkarzinomabgleich des Mammographie-Screening-Programms als auch für das oben genannte Projekt zur Evaluation des Todesursachenalgorithmus eingesetzt.

## Box 6: Record Linkage zwischen Krebsregister und externen Datenquellen mit Klartext-Daten im Krebsregister Niedersachsen

Im § 11 GEKN wird auch der Abgleich mit externen Daten im Rahmen von Forschungsvorhaben geregelt. Nach Absatz 5 ist es zulässig, dass das EKN auch Daten von nicht an Krebs erkrankten Personen speichert, um prospektiv die Häufigkeit einer Tumorerkrankung in einer Gruppe von Personen mit einem gemeinsamen Merkmal zu beobachten. Dies kann z. B. eine Personengruppe sein, die einer besonderen Risiko-Exposition ausgesetzt war. Ein Beispiel für eine bevölkerungsbezogene Langzeit-Kohortenstudie ist die Untersuchung nach einem Gefahrstoffunfall mit Epichlorhydrin der Deutschen Bahn in Südniedersachsen im Jahr 2002. Für diese Kohortenstudie übermittelten die Meldebehörden der Vertrauensstelle aufgrund eines entsprechenden Antrags die personenidentifizierenden Angaben (Name, Vorname, Geburtsdatum, Wohnort als Gemeindekennziffer und Geschlecht) aller Einwohner:innen der Gemeinde zum Stichtag. Für spätere Vergleiche erhält die Vertrauensstelle zusätzlich in regelmäßigen Abständen die Daten zu Wegzügen aus der Gemeinde. Die Angaben werden in der Vertrauensstelle entsprechend dem Routineverfahren des Krebsregisters verschlüsselt und mit einem Projektschlüssel versehen beim Landesgesundheitsamt hinterlegt. In der Vertrauensstelle werden anschließend alle Daten gelöscht. Klatextangaben von Nicht-Betroffenen liegen in diesem Fall also nur für kurze Zeit zum Zwecke der Verschlüsselung vor.

Die Vertrauensstelle fordert die verschlüsselten Daten im fünfjährigen Abstand vom Landesgesundheitsamt an. Die Registerstelle gleicht diese Daten in pseudonymisierter Form mit dem Krebsregisterdatenbestand ab. Anschließend findet ein Vergleich der Krebshäufigkeit mit ausgewählten Vergleichsregionen statt. Auf diese Weise soll untersucht werden, ob die zum Zeitpunkt des Gefahrstoffunfalls in der Gemeinde lebenden Personen später häufiger an Krebs erkranken als die in der Vergleichsregion. Auch für Personen, die ihren Wohnsitz ändern, kann so das spätere Auftreten einer Krebserkrankung gegebenenfalls mit der früheren Gefahrstoff-Exposition in Verbindung gebracht werden.

### 3.3.3 Herausforderungen

Eine wesentliche Herausforderung des Record Linkage speziell in Krebsregistern von Flächenländern ist die große Anzahl an Meldungen, die sich über die Jahre im Register ansammeln und mit neuen Meldungen abgeglichen werden müssen. Hier stellt die gemäß KFRG flächendeckend zu erhebende KVNR eine wichtige Basis für die weitere Entwicklung dar. Dabei ist einschränkend zu sagen, dass die KVNR grundsätzlich nicht für die Verknüpfung der Daten von privatversicherten Personen sowie der Meldungen von Meldeämtern genutzt werden kann. Für eine Verknüpfung mit alten Erkrankungsfällen kann sie erst genutzt werden, nachdem eine Meldung mit der KVNR erstmals mithilfe des probabilistischen Linkage dem Fall hinzugefügt werden konnte.

Kritisch zu hinterfragen ist außerdem, ob die für die Abbildung von Ähnlichkeiten verwendete Kölner Phonetik noch zeitgemäß ist. Die Umstellung auf ein anderes Verfahren könnte jedoch

Register, die nicht dauerhaft über die Klartextdaten aller Patient:innen verfügen, vor große Herausforderungen stellen, da dann für viele Jahre zwei Systeme parallel zu führen wären.

Eine weitere Herausforderung ist die große Menge manueller Nachbearbeitungen, die im Rahmen von Kohortenabgleichen anfallen. Der Rückfall auf nur sichere Treffer oder eine Adjustierung der Grenzen kann hier zwar den Arbeitsaufwand reduzieren, aber auch die Ergebnisse verzerren, sodass dies von Projekt zu Projekt zu bewerten ist. Allerdings werden aktuell auch neue Ansätze des maschinellen Lernens verfolgt, um die manuelle Nachbearbeitung bei großen Abgleichprojekten wesentlich zu reduzieren. In NRW konnte dafür ein Ensemble von Machine-Learning Verfahren entwickelt und mithilfe von 73.000 manuellen Entscheidungen trainiert werden. Es erreicht unter den 80% am sichersten richtig zugeordnet Fällen eine korrekte Zuordnung von über 99% dieser Fälle (Siegert et al., 2016). Das heißt, damit würde sich eine wesentliche Reduktion der manuellen Nachbearbeitung auf 20% der Fälle realisieren lassen. Aktuell wird das System zur Entscheidungsunterstützung im Routine-Linkage implementiert.

### 3.3.4 Fazit

Die Krebsregister der Länder setzen seit vielen Jahren erfolgreich auf das probabilistische Record Linkage-Verfahren sowohl für die routinemäßige Zusammenführung der Daten aus den zahlreichen meldenden Einrichtungen als auch für anlassbezogene Abgleiche für wissenschaftliche Projekte und zu Zwecken der Qualitätssicherung von Screeningprogrammen. Dabei werden je nach Bundesland täglich mehrere Tausend Meldungen an Bestandsdaten mit Millionen von Meldungen vorbeigeführt. Die Verfahren sind bewährt und evaluiert und werden ständig verbessert. Mit der KVNR steht den Registern nun auch eine weitere wesentliche Information für die sichere Verknüpfung zur Verfügung, auch wenn diese nicht in allen Fällen, z. B. bei Privatversicherten, Meldeamtsdaten oder Altfällen angewendet werden kann. Zudem ermöglicht die Verwendung von Kontrollnummern in einigen Ländern Kohortenabgleiche, ohne dass vollständige Klartextinformationen zur Identifikation der Studienteilnehmenden an die Register übermittelt werden müssen.

## 3.4 Use Case 4: Linkage am Beispiel der molekularen SARS-CoV-2 Surveillance (DESH)

### 3.4.1 Hintergrund und Ziel

Ziel der seit Beginn der COVID-19 Pandemie im Frühjahr 2020 implementierten gesundheitspolitischen Maßnahmen ist es, durch eine nachhaltige Reduktion von Neuinfektionen die Anzahl an COVID-19-bedingten Krankenhauseinweisungen zu senken. Hierdurch soll eine Überlastung des Gesundheitswesens vermieden und darüber hinaus eine Reduktion bzw. Vermeidung von COVID-19-bedingter Übersterblichkeit erreicht werden (Flaxman et al., 2020).

Grundlage für politische Maßnahmen zur Pandemiesteuerung ist vor allem die detaillierte Erfassung der Inzidenz. Diese basiert auf den Fallzahlen von Sars-COV-2-PCR-Nachweisen, die Testlabore und behandelnde Ärzte gemäß Infektionsschutzgesetzes (IfsG) an die regional zuständigen Gesundheitsämter melden. Seit Juli 2020 erfolgt die automatisierte Weitergabe der Meldedaten von den Gesundheitsämtern an das Robert Koch-Institut (RKI) mittels des deutschen elektronischen Melde- und Informationssystems (DEMIS) (Diercke et al., 2021).

Mit dem Aufkommen neuer Virusmutationen, die sich gegenüber der Basisvariante durch eine z.T. deutlich erhöhte Infektiosität und Virulenz auszeichnen, sind die absoluten Fallzahlen für eine adäquate Pandemiesteuerung nicht mehr ausreichend. Die Erfassung spezifischer Virusmutationen lässt sich im Rahmen der bestehenden DEMIS-Architektur aber nicht umsetzen. Aus diesem Grund wurde mit dem deutschen elektronischen Sequenz-datenhub (DESH) ein Softwaresystem etabliert, das es im Einklang mit der Datenschutz-Grundverordnung (DSGVO) gestattet, DEMIS-Meldedaten mit Sars-CoV-2-Sequenzdaten auf Personenebene zu verknüpfen. Die räumlich und zeitlich exakte Nachverfolgung bekannter sowie das Screening auf neue Sars-CoV-2-Varianten werden hierdurch ermöglicht.

### 3.4.2 Technische Umsetzung

Die Architektur des entsprechend der Verordnung des Bundesgesundheitsministeriums vom 18.1.2021[3] entwickelten Systems zur molekularen Sars-CoV-2-Surveillance dient der Verknüpfung zweier getrennter Datensätze. Dabei findet zum einen der DEMIS-Datensatz Verwendung, der durch die Weitermeldung positiver Sars-CoV2-PCR-Befunde von

---

[3]  https://www.bundesgesundheitsministerium.de/presse/pressemitteilungen/2021/1-quartal/coronavirus-surveillanceverordnung.html, Zugriff 05.05.2022

Seiten der Gesundheitsämter an das RKI generiert wird. Neben Namen, Geburtsdatum, Geschlecht und Adresse an COVID-19 erkrankter Personen wurde diesem im Zuge der DESH-Entwicklung ein neues Datenfeld hinzugefügt, das für eine fallbezogene Pseudonymisierungs-ID vorgesehen ist, mittels derer die Verknüpfung mit den Sequenzierungsdaten aus DESH erfolgt. Zum anderen werden verordnungsgemäß anhand des neu entwickelten DESH-Backends die Sequenzierungsergebnisse von 5% aller PCR-positiven Sars-CoV2-Proben automatisiert an das Robert Koch-Institut übertragen und tagesaktuell in einem eigenen Datensatz zusammengefasst. Auf Seite der sequenzierenden Labore sind zwei Dateien in das DESH-System hochzuladen (siehe Abbildung 4): Eine Datei im sogenannten „FASTA"-Format (textbasiertes Format zur Darstellung und Speicherung von Nukleinsäuresequenzen) enthält die eigentlichen Gensequenzen der untersuchten Proben, versehen mit einer fallspezifischen FASTA-ID. Ein dazu korrespondierendes Datenfeld weist eine zweite CSV-Datei auf, die darüber hinaus Informationen zu Datum, Art, Sequenzierungsgrund, Publikationsstatus und einsendendem Labor einer gegebenen Probe umfasst. Auch dieser Datensatz ist – analog zur DEMIS-Datei – mit einem Datenfeld versehen, in das eine fallbezogene Pseudonymisierungs-ID eingetragen werden kann.

Der Prozess der Übertragung von DESH-Daten erfolgt wie in Abbildung 4 dargestellt: Die von den Labor-Informations-Management-Systemen (LIMS) bereitgestellten FASTA-Dateien werden zusammen mit den Daten, die zusätzliche Fallinformationen enthalten, automatisiert über eine REST-API an das DESH-Backend übertragen, das beide Datensätze über eine temporäre Datenbank an das RKI weiterprozessiert.

Zudem dient das Backend der Erstellung einer Pseudonymisierungs-ID. Diese stellt einen Universally Unique Identifier der Version 4 (UUID4) dar. UUIDs sind 128-Bit-Zahlen, die im Rahmen der ISO-Standards ISO / IEC 11578:1996 und ISO / IEC 9834-8:2005 als Kennziffern zur Identifikation spezifischer Daten etabliert sind[4]. Obwohl die beschränkte Anzahl von Bits keine eindeutige Zuordnung garantiert, ist die Anzahl möglicher IDs mit 2122 = 5,3 * 1036 so hoch, dass die Wahrscheinlichkeit für einen Homonymfehler für die praktische Anwendung hinreichend niedrig ist.

Nach Erstellung wird die UUID einerseits fallbezogen automatisch dem Datensatz der CSV-Datei hinzugefügt, der von den sequenzierenden Labors begleitend zu den

---

FASTA-Dateien zur Verfügung gestellt wird. Andererseits werden generierte UUID vom DESH-Backend an die Labore zurückgespielt, die die erhaltenen UUID ebenfalls an die entsprechende Meldung der Sars-CoV2-Neuerkrankung anhängen, indem die UUID in das dafür geschaffene Feld des DEMIS-Datensatzes eingepflegt wird. Verantwortlich für den letzten Schritt sind einzelne oder mehrere Labore, je nachdem, ob unterschiedliche Labore an PCR-Testung und Sequenzierung beteiligt sind. Nach Empfang der DESH-Daten durch das RKI durchlaufen diese zunächst einen Qualitätscheck zum Ausschluss duplizierter Fälle. Fälle sind dabei als Dubletten definiert, wenn sowohl deren genetische Sars-CoV2-Sequenzen sowie FASTA-ID und Zusatzinformationen gemäß CSV-Datei identisch sind. Die eigentliche Verknüpfung zwischen DESH- und Sars-CoV2-Meldedaten wird anschließend als exaktes Record Linkage über die in beiden Datensätzen enthaltene fallbezogene UUID durchgeführt.

### 3.4.3 Möglichkeiten und Herausforderungen

Zum Stand November 2021 sind von 110 Laboren insgesamt 354.911 auswertbare Gensequenzen für die molekulare Sars-CoV2-Surveillance zur Verfügung gestellt worden. Das DESH-System leistet damit einen kontinuierlichen Beitrag zur Überwachung des Verlaufs der COVID-19 Pandemie. Allerdings konnten von 336.961 übermittelten Sequenzen nur 142.888 entsprechende Meldungen im DEMIS-System zugeordnet werden, was insgesamt einer Quote von 42,4% entspricht. Die Quote der Synonymfehler beträgt somit insgesamt 57,6%. Die genaue Quote der Homonymfehler kann aktuell nicht angegeben werden, weil – abgesehen von der bereits möglichen Identifikation doppelter Fälle – weitere Ursachen aktuell noch vom RKI untersucht werden.

Die Ursache für die hohe Quote von Synonymfehlern besteht im Wesentlichen in dem erheblichen zusätzlichen Bedarf an zeitlichen und personellen Ressourcen, der für Labore aufgrund der Teilnahme am DESH-System entsteht. Sowohl die Erfassung und Bereitstellung der Zusatzinformationen, die in der zu übertragenden CSV-Datei hinterlegt sind, als auch die Übermittlung der durch das DESH-Backend erstellten UUID an das DESH- und das DEMIS-System obliegt allein den Laborbetreibern. Aufgrund der – pandemiebedingt – kurzen Zeitspanne, die für den Aufbau des DESH-Systems zur Verfügung stand, können die benötigten zeitlichen und personellen Kapazitäten nicht immer zuverlässig vorgehalten werden. Aufgrund der aktuellen Rechtslage ist es derzeit jedoch nicht möglich, ein System einzurichten, das eines geringeren Aufwands technischer und personeller Ressourcen bedarf. Ermöglicht würde dies zum Beispiel durch die Schaffung von gesetzlichen Rahmenbedingungen, die die Einrichtung einer gemeinsamen Treuhandstelle für Sequenz- und Meldedaten erlauben würde.

Sinnvoll wäre eine solche Treuhandstelle insbesondere vor dem Hintergrund, dass es für die Zukunft vorgesehen ist, das Record Linkage von Sequenz- und Meldedaten vollumfänglich in ein einziges System zu integrieren. Dabei sollen die im Rahmen von DESH erlangten Erfahrungen genutzt werden, um die DESH-Plattform für die Aufnahme und Verknüpfung von weiteren Sequenzdaten (z. B. von einem anderen Virus) anzupassen.

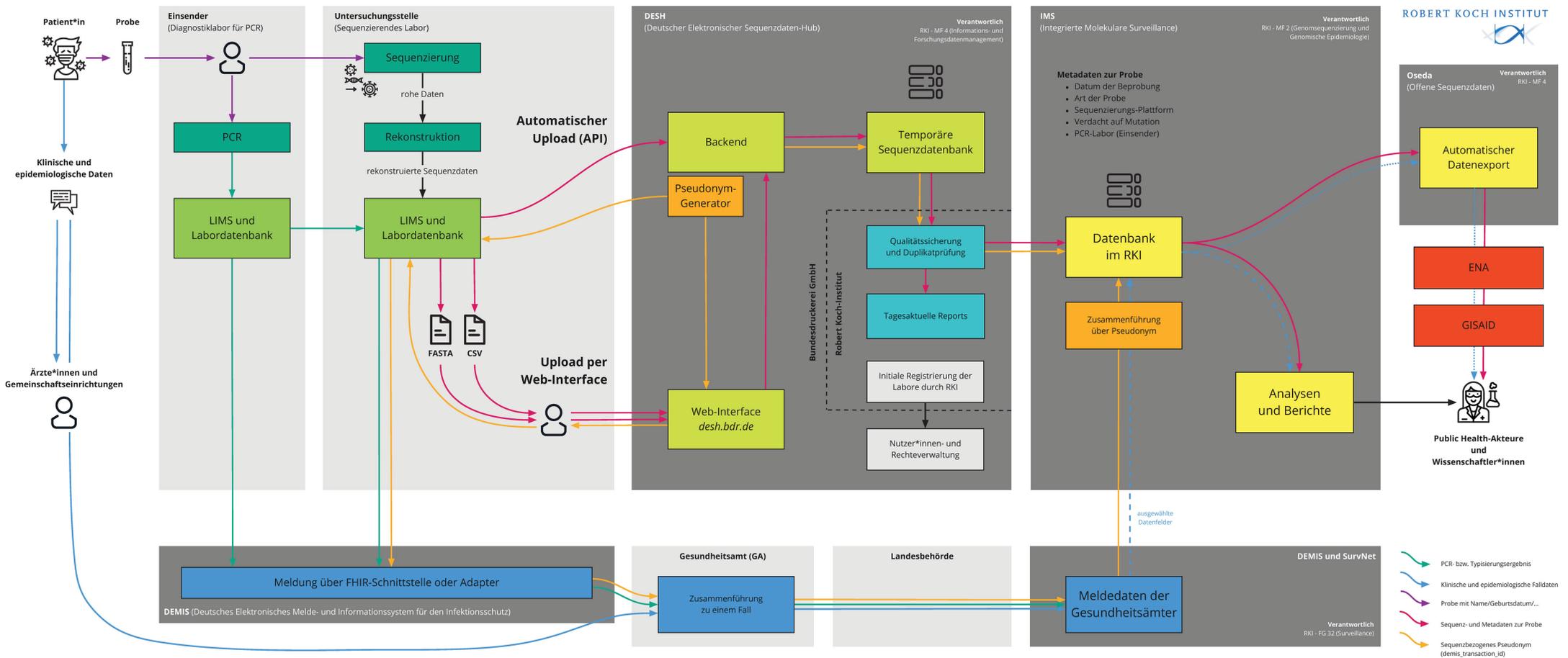

Abbildung 4: Vereinfachte schematische Darstellung der DESH-Systemarchitektur [5]

5   https://www.rki.de/DE/Content/InfAZ/N/Neuartiges_Coronavirus/DESH/DESH.html, Zugriff: 28.06.2023

## 3.5 Use Case 5: Linkage von Daten aus Routinedaten und -proben und klinischen Erhebungen innerhalb eines rechtlichen Trägers

### 3.5.1 Hintergrund und Ziele

Die Zusammenführung und Integration dezentral erhobener Daten in der patientennahen Krankenversorgung ist nach wie vor eine große Herausforderung im Gesundheitswesen. Ende der 1990er Jahre wurden sehr gut akzeptierte Dokumentationslösungen zunächst aus dem Auftrags- und Befundmanagement, dann rund um weitere Funktionen am klinischen Arbeitsplatz etabliert. In der Krankenversorgung ist für das Record Linkage eine synonymarme und homonymfreie Vergabe von Patienten- und Fallnummern essenziell. In der Regel ist das Record Linkage durch administrative Vorgänge (Abrechnung mit den Kostenträgern) über entsprechende Kommunikationsserver im Sinne eines exakten Record Linkage gelöst. Anhand der Identitätsprüfung bei der Patientenaufnahme können die Krankenversichertennummer (KVNR) aus der elektronischen Gesundheitskarte und die Patientennummer bzw. Fallnummer aus der Krankenhausdokumentation in der Regel eindeutig zugeordnet werden.

Die Grundlage für die Erhebung, Verarbeitung und Verknüpfung dieser Daten sind zunächst die Allgemeinen Vertragsbedingungen des Gesundheitsträgers. Der eigentliche Behandlungsvertrag und weiter gehende Einwilligungen sind Grundlage für jede Behandlung (siehe Beispiel der Universitätsmedizin Göttingen[6]). Die Patientenaufklärung bzw. Einwilligung gem. Art. 12ff DSGVO soll zukünftig im Sinne der Verbesserung der Nutzbarkeit der Daten eine weitergehende Einwilligung (Broad Consent) vereinheitlicht und umgesetzt werden[7].

Eine Herausforderung in den Kliniken in den 00er-Jahren war das Record Linkage der Biomaterialproben aus der Krankenversorgung und der patientennahen Forschung. Für die verwechslungsfreie Zuordnung der Biomaterialien, der dazugehörigen verschiedenen Einzelanalysen und der patientenbezogenen Befundberichte z. B. aus der klinischen Chemie, der Pathologie oder der Sequenzierungslabors wird ein weiteres exaktes Record Linkage benötigt. Diese Herausforderung wurde in der Überarbeitung der generischen Datenschutzkonzepte der TMF adressiert und mit Lösungsvorschlägen publiziert (Dangl et al., 2010; Pommerening et al., 2006; Prokosch et al., 2018).

---

[6] https://www.umg.eu/patienten-besucher/ihr-aufenthalt/behandlungsvertrag/, Zugriff 10.12.2021

[7] https://www.aerzteblatt.de/nachrichten/112355/Bundesweite-Patienteneinwilligung-schafft-Rechtssicherheit, Zugriff 10.12.2021

Parallel hierzu haben sich – meist aus dem Medizincontrolling motiviert – so genannte Clinical Data Warehouses entwickelt. Die entsprechenden Daten aus der Krankenversorgung können für administrative Anfragen z. B. des medizinischen Dienstes der Krankenkassen oder auch von klinischen Forschende beantragt werden (Ganslandt et al., 2011).

### 3.5.2 Technische Umsetzung

Im Folgenden stellen wir passend zu den einleitend genannten Herausforderungen beispielhaft einige Implementierungen von Record Linkage im Krankenhausumfeld vor. Innerhalb eines rechtlichen Trägers sind die Lösungen vergleichsweise einfach, solange im Behandlungskontext die Klarnamen und die Versichertennummer als Identifikatoren verwendet werden können. Im Forschungskontext werden die identifizierenden Daten jedoch pseudonymisiert, wodurch eine Verbindung von den identifizierenden Daten in der Krankenversorgung und den in der Forschung genutzten pseudonymisierten Daten für die Dauer der intendierten Nutzung gewährleistet sein muss.

### 3.5.2.1 Record Linkage im Clinical Data Warehouse

Grundlage eines klinischen Data Warehouse sind meist Daten aus der Abrechnung von Patient:innen. Diese liegen aufwandsarm regelmäßig als nach § 301 SGB V formatierte Daten zu Patient:innen, Fällen, Diagnosen, Prozeduren und entgeltfähigen Medikationen vor. Weitere Daten aus zusätzlichen klinischen Bereichen wie Intensivmedizin oder Kardiologie sowie Daten aus hoch automatisierten Bereichen wie klinische Chemie, Radiologie und Pathologie lassen sich über Patientennummer und Fallnummer hinzufügen. Etwas komplexer ist die Verknüpfung mit einzelnen Biomaterialproben und deren Analyse, da das Datenmodell aufgrund der Relationen Patient-Fall-Biomaterial-Aliquot vielschichtiger ist. Dies betrifft beispielsweise Daten aus der Hochdurchsatzsequenzierung in der Onkologie.

Die Grundlage für das verwendete exakte Linkage bildet die Festlegung eines „Stammdatenmasters", bei dem Aufnahme, Änderung und Korrektur der technischen Identitäten, Verlegung und Entlassung von Patient:innen gespeichert wird. Die entsprechenden Informationen werden standardisiert und direkt nach einem Ereignis (Aufnahme, Verlegung, Entlassung) an alle angeschlossenen Systeme versendet.

### 3.5.2.2    Record Linkage in einem MII-Datenintegrationszentrum (DIZ)

Um die Routineversorgungsdaten aus den deutschen Krankenhäusern transparenter und besser zugänglich zu machen, waren im Rahmen der ersten Förderperiode der Medizininformatik-Initiative des BMBF in allen deutschen Universitätskliniken so genannte Medizinische Datenintegrationszentren (DIZ) organisatorisch und technisch aufzubauen (Semler et al., 2018). Ziel ist es, ausgewählte Daten aus den Universitätskliniken anhand des Kerndatensatzes der MII zur gemeinsamen Nutzung durch die jeweiligen Antragsteller entdeckbar zu machen. Der MII-Kerndatensatz wurde nach internationalen IT- und Terminologie-Standards festgelegt. Diese Datensätze sind die Mindestanforderung an alle DIZe, welche für die stationäre Patientendaten vorgehalten werden müssen (Ammon et al., 2019). In weiteren Ausbaustufen werden Erweiterungsmodule und nicht-universitäre Standorte aufgenommen. Die DIZ der MII sind nicht homogen aufgebaut. In den meisten Fällen werden die relevanten Datenquellen bereits im Versorgungskontext über die Patientennummer zusammengeführt (exaktes Linkage). Für die Datennutzung werden die Daten jeweils für den Antragsteller mit eigenen Pseudonymen versehen. Hierfür wird in der Regel ein Pseudonymisierungsdienst eingesetzt, um bei Ergänzungen oder Korrekturen des Datensatzes eines Patient:innen immer das gleiche Pseudonym zuordnen zu können.

Neben der Abfrage über die Use and Access Committees am einzelnen Standort sind die Daten der MII-DIZe übergreifend über das Forschungsdatenportal Gesundheit[8] entdeckbar und nutzbar. Für die Details zu den standortübergreifenden Abfragen im Rahmen der MII sei auf Abschnitt 3.9 verwiesen.

### 3.5.2.3    Record Linkage in klinischen Forschungsprojekten

In standortinternen Forschungsprojekten werden meist sehr einfache Record Linkage-Ansätze gewählt. Hier sind Projekte der Grundlagenforschung und klinische Forschungsgruppen bzw. Sonderforschungsbereiche zu unterscheiden, die translational am Übergang zwischen Grundlagenforschung und Klinik angesiedelt sind.

Meist werden für den Forschungsbereich sehr einfache listenbasierte Pseudonymisierungs-verfahren gewählt. Durch die Verknüpfung der Vergabe des jeweiligen Pseudonyms an die informierte Einwilligung des Patient:innen ist im Prozess eine Synonymarmheit und eine Homonymfreiheit in der Regel gegeben. In translationalen Projekten sind pro Standort neben den neu zu gewinnenden Forschungsdaten z. B. aus Fragebögen oder Biomarkeranalysen häufig auch Daten aus der Patientenakte oder aus Registern zu integrieren. Hierfür sind etwas aufwendigere Prozesse für die fehlerfreie Zuordnung der Pseudonyme und ebenso für die ggf. notwendige De-Pseudonymisierung zu etablieren.

---

8        https://forschen-fuer-gesundheit.de/, Zugriff: 01.12.2022

### 3.5.3 Möglichkeiten und Herausforderungen

Die standortinterne Verknüpfung von personenbezogenen Datensätzen lässt sich über die eindeutigen Patientennummern, die in Pseudonymisierungslisten erfasst sind, relativ einfach herstellen. Die Qualität des damit erfolgten Record Linkage ist als sehr gut zu bezeichnen.

Perspektivisch wäre im Rahmen der Einführung der elektronischen Patientenakte (ePA) nach § 363 SGB V zu betrachten, inwiefern die dann pro Patient:in lebenslang eindeutigen Identifier (eIDs) ein sehr viel einfacheres standortübergreifendes Record Linkage erlauben würde. Der große Vorteil der Nutzung der eIDs bestünde darin, dass der aufwendige und bisweilen fehleranfällige Prozess der Identitätsprüfung in wohldefinierten Prozessen rund um den Personalausweis durchgeführt würde[9] und ggf. weiter Datenbestände über das Forschungs-datenzentrum nach § 363 SGB V bzw. § 303c beantragbar wären.

## 3.6 Use Case 6: Linkage von klinischen Routine- und Studiendaten über mehrere rechtliche Träger hinweg am Beispiel des nationalen Netz-werks Genomische Medizin Lungenkrebs

### 3.6.1 Hintergrund und Ziel

In der personalisierten Medizin werden genomische und molekulare Daten (Biomarker) genutzt, um auf einzelne Patient:innen individuell zugeschnittene Therapien zu ermöglichen (Steffen & Steffen, 2013). Um in der Onkologie u. a. die genetischen Veränderungen in den Tumorzellen zu identifizieren, wird eine molekulare Diagnostik benötigt, die nur in hoch-spezialisierten Zentren durchgeführt werden kann. Um trotzdem vielen Patient:innen unabhängig vom Ort der behandelnden Einrichtung Zugang zu solchen innovativen Verfahren zu ermöglichen, sind standort- und sektorenübergreifende Netzwerke notwendig. Eines der größten Netzwerke im Bereich der molekularen Diagnostik in Europa stellt das nationale Netzwerk Genomische Medizin Lungenkrebs (nNGM) dar. Der von der

---

9 https://www.personalausweisportal.de/Webs/PA/DE/buergerinnen-und-buerger/eID-karte-der-EU-und-des-EWR/eid-karte-der-eu-und-des-ewr-node.html, Zugriff: 01.12.2022

Deutschen Krebshilfe geförderte Zusammenschluss von derzeit 20 Krebszentren (Büttner et al., 2019) umfasst alle hochspezialisierten Comprehensive Cancer Center (CCC), die Zentren des Deutschen Konsortiums für Translationale Krebsforschung (DKTK) (Joos et al., 2019) sowie das Deutsche Krebsforschungszentrum (DKFZ). Diese Zentren arbeiten mit mehr als 300 klinischen Netzwerkpartnern aus Kliniken und niedergelassenen Onkolog:innen zusammen. Das übergeordnete Ziel von nNGM ist die flächen-deckende Implementierung einer umfassenden, harmonisierten molekularen Diagnostik und die Etablierung des Zugangs zu personalisierten Therapien für alle Patient:innen mit fortgeschrittenem Lungenkrebs. Die zweite Aufgabe des nNGM ist die Bereitstellung der anfallenden Daten für die Krebsforschung. Durch die Forschung in standortübergreifenden Verbünden können Erkrankungen als Folge individueller genetischer Mutationen oder immer kleiner werdende molekulare Krankheitsunter-gruppen mit ausreichenden Fallzahlen erforscht und Therapien verbessert werden (Schillinger & Kron, 2019).

Die Verbindung einer so großen Anzahl von Partnerstandorten – von denen die meisten über sehr heterogene IT-Infrastrukturen verfügen – zu einem homogenen Daten- und Prozess-netzwerk erfordert neben einer flexiblen und skalierbaren IT-Infrastruktur (Abbildung 5) eine effiziente Record Linkage-Lösung zur Datenzusammenführung sowohl im Behandlungs-kontext als auch im Rahmen der Forschung über Institutsgrenzen hinweg. Dabei kommt die sogenannte „Mainzelliste" zum Einsatz, ein webbasierter Pseudonymisierungsdienst, der Personenidentifikatoren (Patientenpseudonyme) aus identifizierenden Daten (IDAT) wie Name und Vorname erzeugt und seine Funktionen über eine flexible Web-Schnittstelle (REST) bereitstellt (Lablans et al., 2015b).

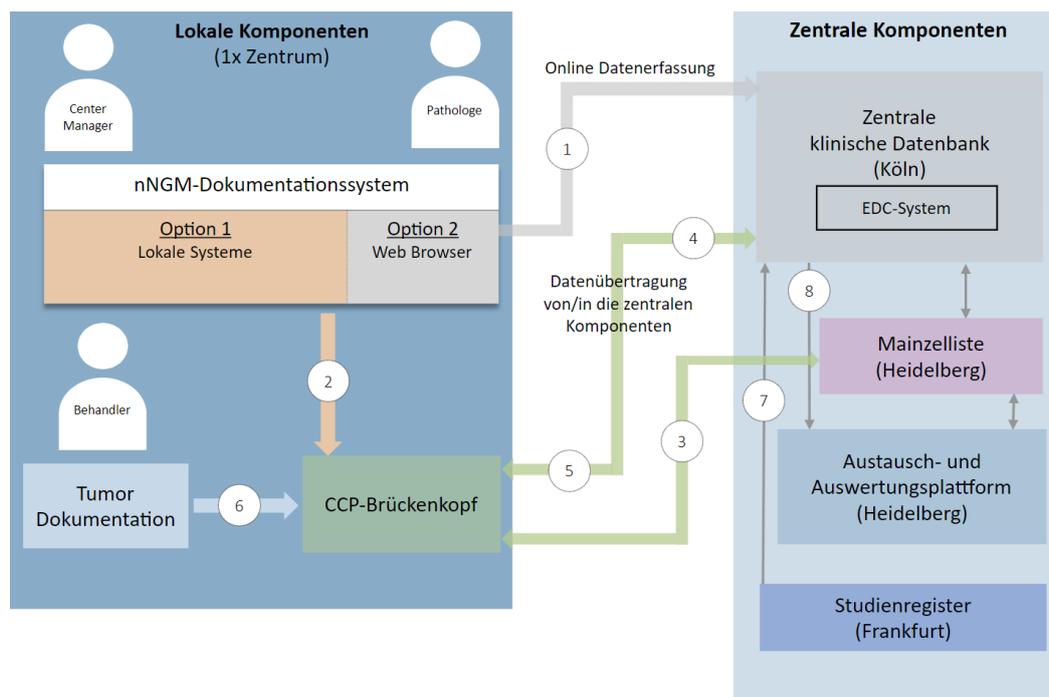

Abbildung 5: nNGM-Architektur aus zentralen und lokalen IT-Komponenten

Die nNGM-Zentren können zwei Dokumentationsoptionen nutzen, um Patientendaten an die zentrale klinische Datenbank in Köln zu übermitteln (vgl. Abbildung 5). Zum einen kann die Eingabe direkt über den Webbrowser erfolgen (1); hierfür stellt die klinische Datenbank eine Dokumentationsoberfläche (EDC-System) zur Verfügung. Dasselbe zentrale EDC-System nutzen auch Onkolog:innen aus Praxen und Kliniken, um Anfragen zur molekularen Diagnostik für ihre Patient:innen an ein Netzwerkzentrum ihrer Wahl zu senden, bereits gemeldete Daten einzusehen und die Nachdokumentation manuell zu erfassen. Zum anderen können die Zentren neben dieser manuellen Meldung einen lokalen Webservice (Brückenkopf) nutzen, um auch umfangreiche Daten automatisiert zu melden (2). Hierzu überträgt jedes nNGM-Zentrum zunächst relevante Daten im sogenannten HL7 FHIR-Format, einen etablierten Standard für den elektronischen Datenaustausch im Gesundheitswesen, an den Brückenkopf (2). Der Brückenkopf bereitet diesen Datensatz dann in mehreren Schritten für die Übermittlung an die zentrale klinische Datenbank vor: Pseudonymisierung und Record Linkage mit der Mainzelliste (3), Validierung gegen die FHIR-Spezifikation und schließlich die Datenübermittlung an das EDC (4). Andersherum lädt der Brückenkopf – regelmäßig alle Daten zu Patient:innen „seines" nNGM-Zentrums herunter (5) und kombiniert sie mit Daten aus anderen Quellen zur Nachnutzung (6) für Forschung und Wissenschaft. Nachdem die Molekularpathologie des nNGM-Zentrums die molekulare Diagnostik für interne oder externe Onkolog:innen durchgeführt hat, kann der sogenannte Case-Manager des Zentrums neben möglichen Therapieoptionen Informationen über geeignete klinische Studien aus dem

Studienregister (7) in den finalen Befund aufnehmen. Schließlich werden ausgewählte Daten aus der zentralen klinischen Datenbank zur weiteren Datenanalyse in eine Analyseplattform übertragen (8).

Bislang haben sich 16 Zentren für die manuelle Meldung über den Webbrowser und 6 Zentren für die automatisierte Meldung über den Brückenkopf entschieden. Bezogen auf die Pseudonymisierung wurden im Rahmen der Anforderungsanalyse folgende Hauptanforderungen von der zuständigen IT-Task-Force identifiziert:

- Die Empfehlungen des generischen Datenschutzleitfadens der TMF e.V. sind sowohl in Bezug auf das Datenschutzkonzept als auch auf die Trennung der Daten nach medizinischen (MDAT) und identifizierenden Daten (IDAT) zu berücksichtigen.

- Das zentrale EDC-System für das nNGM-Netzwerk besteht aus einer Dokumentationssoftware und einer zentralen klinischen Datenbank. Das System kann sowohl direkt für die Dateneingabe als auch für den Datenimport je nach Standortbedarf genutzt werden. Es erhält zentral nie IDATs, sondern nur Pseudonyme.

- Jede Instanz im Netzwerk (z. B. ein lokales Zentrum) kennt für die Patient:innen nur die eigenen Patientenpseudonyme, sodass unkontrollierte Datenzusammenführungen verhindert werden.

- Die berechtigten Nutzenden (z. B. behandelnde Onkolog:innen) sehen im Behandlungskontext über das zentrale EDC-System ihre Patient:innen im Klarnamen. Da im EDC-System keine IDAT gespeichert sind, werden IDAT aus dem Pseudonymisierungsdienst (Mainzelliste) in den Webbrowser der Nutzenden geladen.

- Die zentral gesammelten klinischen Daten müssen für die Nachnutzung sowohl dem Dateneigentümer am Standort als auch dem Netzwerk nach einem geregelten Verfahren zur Verfügung gestellt werden. Dafür müssen andere Patientenpseudonyme generiert werden als die, die im Behandlungskontext zum Einsatzkommen. Da die Dateneinsicht bei der Nachnutzung nicht im Behandlungskontext erfolgt, dürfen hier keine Klarnamen angezeigt werden.

## 3.6.2 Technische Umsetzung

**Anforderung: Berücksichtigung der Empfehlungen des generischen Datenschutzleitfadens des TMF e.V., sowohl in Bezug auf das Datenschutzkonzept als auch auf die Trennung der Daten nach medizinischen (MDAT) und identifizierenden Daten (IDAT).**

Das in Anlehnung an den TMF-Leitfaden entwickelte Datenschutzkonzept wurde mit allen Datenschutzbeauftragten der Standorte abgestimmt. Zudem wurde dafür ein positives Votum der TMF-AG Datenschutz eingeholt (Pommerening et al., 2014). Dieses Konzept klassifiziert die Variablen nach MDAT und IDAT.

Die IDAT umfassen Vor- und Nachname sowie Geburtsdatum. Diesen drei Variablen dienen als Identifikatoren beim Record Linkage. Zusätzlich wird die Versichertennummer (KVNR) erfasst und der Mainzelliste übermittelt. Sie wird nicht als Identifikator beim Record Linkage verwendet, sondern nur für die Plausibilitätsprüfung genutzt: Die Mainzelliste meldet einen Fehler, falls die neu übermittelten IDAT mit bereits gespeicherten Daten in Konflikt stehen:

- Die Patient:innen haben bereits eine andere KVNR in der Mainzelliste Datenbank.

- Die KVNR der Patient:innen sind bereits anderen Patient:innen zugeordnet.

Wenn für Patient:innen keine KVNR vorliegen, werden das Ersatzverfahren (Ersatznummern) der Krebsregister angewendet (Tabelle 2). Dies gilt z. B. für Selbstzahlende und Geflüchtete. Für diese speziellen Nummern werden die vorgesehenen Plausibilitätsprüfungen über­sprungen. U. a. für Abrechnungszwecke wurden weitere IDAT wie postalische Adresse in die Mainzelliste aufgenommen, aber nicht beim Record Linkage berücksichtigt.

Tabelle 2: Ersatzkodes für die Versichertennummer

| Versichertengruppe | Ersatzkode |
|---|---|
| Asylbewerber:innen | 970100001 |
| Keine Angabe zum Kostenträger | 970000099 |
| Kostenträger ohne IK-Nummer | 970001001 |
| Privatversichert, Kasse unbekannt | 970000022 |
| Selbstzahlende | 970000011 |

**Anforderung: Jede Instanz im Netzwerk kennt für die Patient:innen nur die eigenen Patienten­pseudonyme, sodass unkontrollierte Datenzusammenführungen verhindert werden.**

Die Mainzelliste erzeugt beim Anlegen der neuen Patient:innen nach dem Record Linkage jeweils einzigartige netzwerkweite IDs (ID$_{PL}$), die jedoch nicht nach außen kommuniziert werden. Stattdessen werden anfragenden Systeme und Nutzenden je nach ihren Berechtigungen andere ID-Typen übermittelt, etwa standort- oder systembezogene Pseudonyme. Hierzu findet in der Mainzelliste bei Anfrage nach einem bestimmten ID-Typen (z. B. ID$_{EDC}$) eine interne Umschlüsselung statt (ID$_{PL}$ → ID$_{EDC}$). So müssen diese neu erzeugte IDs (Patientenpseudonyme) nur temporär für die Anfrage erzeugt und nicht in der Mainzelliste-Datenbank persistent gespeichert werden.

**Anforderung: Das EDC-System für das Netzwerk erhält zentral nie IDATs (auch nicht temporär), sondern nur Pseudonyme.**

Um dies zu realisieren, setzen die Tumorzentren die sog. CCP-Brückenkopf-Software (Lablans et al., 2015b) ein. Der Brückenkopf ist lokal an jedem Zentrum installiert und kann direkt mit der zentralen Mainzelliste kommunizieren (Abbildung 6).

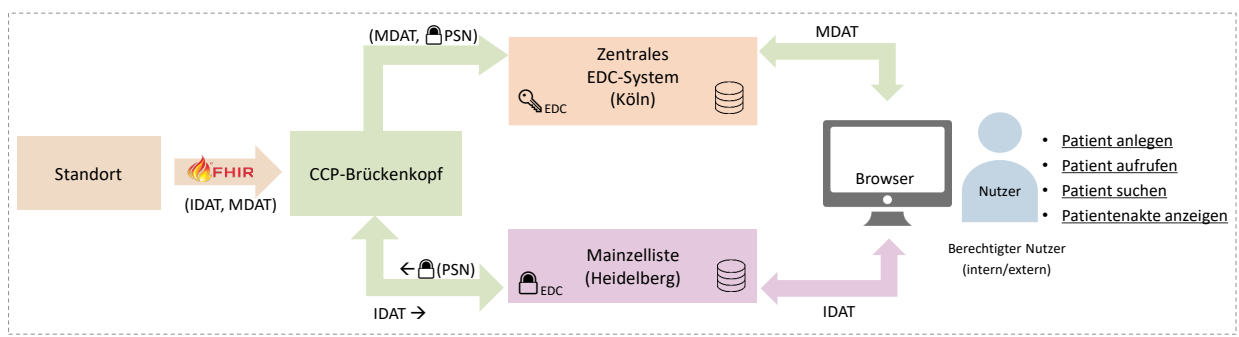

Abbildung 6: Schematischer Datenfluss zur Trennung und Zusammenführung von IDAT und MDAT (CCP: Clinical Communication Platform, PSN: Pseudonym, FHIR: Fast Healthcare Interoperability Resources, „Schlüssel-EDC:" privater EDC-Schlüssel, „Schloss-EDC:" öffentlicher EDC-Schlüssel)

Gemäß Abbildung 6 nimmt der Brückenkopf Daten im FHIR-Format entgegen, extrahiert die IDAT und sendet sie an die Mainzelliste (ID-Type: ID$_{EDC}$). Die Mainzelliste führt das Record Linkage aus, erzeugt die ID$_{EDC}$ und gibt sie zurück an den Brückenkopf. Der Brückenkopf ersetzt die IDAT durch das neue Pseudonym und leitet es via FHIR-Schnittstelle an das zentrale EDC-System in Köln. Um zu verhindern, dass der Standort das EDC-Pseudonym einsieht, verschlüsselt die Mainzelliste das Pseudonym zusätzlich mit einem öffentlichen Schlüssel des EDC-Systems. Das EDC-System muss dann schließlich das Pseudonym mit seinem privaten Schlüssel entschlüsseln und zusammen mit den MDAT abspeichern. Das zentrale EDC-System inkl. klinischer Datenbank wird vom Universitätsklinikum Köln betrieben, während die Mainzelliste bei der DKFZ-Abteilung Datenschutz angesiedelt ist.

**Anforderung: Die berechtigten Nutzenden (z. B. behandelnde Onkolog:innen) sehen im Behandlungskontext ihre Patient:innen im Klarnamen (transparente Re-Identifikation)**

Im nNGM werden Daten von Tumorpatient:innen, die an den nNGM-Zentren behandelt werden, erhoben und verarbeitet. Sie werden zum größten Teil aus vorhandenen Datenverarbeitungs-systemen (z. B. Software zur Tumordokumentation) in die zentralen IT-Komponenten des nNGM eingebracht (siehe Abbildung 5). Darüber hinaus erlaubt das zentrale EDC die manuelle Eingabe von Daten direkt in das System. Dies ist erforderlich z. B. zur Erfassung von Folgeuntersuchungen, die von Netzwerkpartnern wie Krankenhäusern und niedergelassene Onkolog:innen durchgeführt wurden. Dabei können die behandelnden Ärzt:innen neue Patient:innen anlegen, bereits gespeicherte Patient:innen aufrufen oder einfach nach Patient:innen suchen. Im EDC-System können die behandelnden Personen die Patientenakten (MDAT) aufrufen, die IDAT der dazugehörigen Patient:innen werden dafür von der Mainzelliste abgerufen.

**Anforderung: Die zentral gesammelten klinischen Daten müssen für die Nachnutzung sowohl dem Dateneigentümer am Standort als auch dem Netzwerk nach einem geregelten Verfahren zur Verfügung gestellt werden. Dabei müssen andere Patientenpseudonyme als im Behandlungskontext zum Einsatz kommen.**

Damit der Standort seine zentral gesammelten Daten lokal für nNGM oder andere Verbünde (z. B. DKTK, German Biobank Alliance, MII-Konsortien) kontrolliert für Forschungszwecke freigeben kann, kommt der Brückenkopf aus der Clinical Communication Platform (CCP) des DKTK zum Einsatz. Der Brückenkopf ist eine IT-Komponente, die die projektbezogenen Daten der jeweiligen Einrichtung vorhält. Über diese Brückenköpfe können Forschende einerseits pseudonymisierte Daten der Patient:innen an ihren jeweiligen Standorten für standortübergreifende Forschungsprojekte oder klinische Studien zur Verfügung stellen oder beziehen. Andererseits erlauben die Brückenköpfe es Forschenden, standortübergreifend mithilfe von Abfragemasken gezielt nach Patient:innen für klinische Studien oder nach z. B. Gewebeproben für transnationale Forschungsprojekte zu suchen (föderierte Fallzahlabfrage). Zu diesem Zweck lädt jeder Brückenkopf an einem nNGM-Standort in regel-mäßigen Abständen die medizinischen Daten zu „seinen" Patient:innen aus dem zentralen EDC-System herunter. Dabei werden die Pseudonyme wie oben beschrieben neu verschlüsselt.

### 3.6.2.1 Mainzelliste und Record Linkage

Die Mainzelliste ist ein webbasierter Pseudonymisierungsdienst, der Patientenpseudonyme aus identifizierenden Daten (IDAT) wie Name und Vorname erzeugt und seine Funktionen über eine flexible Web-Schnittstelle (REST) bereitstellt (Lablans et al., 2015a). Die Mainzelliste führt ein probabilistisches Record Linkage durch und verarbeitet und speichert im nNGM-Kontext die Klartext-IDAT, um sie mit bereits früher gespeicherten zu vergleichen. Je nach Ergebnis dieses Abgleichs erzeugt sie entweder ein neues Pseudonym IDPL oder gibt ein bestehendes zurück. Hierfür führt die Mainzelliste vor dem eigentlichen Record Linkage eine Plausibilitätsprüfung der IDAT sowie eine Zeichentransformation und -bereinigung durch. So werden z. B. Kleinbuchstaben in Großbuchstaben umgewandelt und Umlaute durch ihre ASCII-Äquivalente ersetzt. Diese Schritte zielen darauf ab, den Prozess des Record Linkage zu vereinfachen. Beim tatsächlichen Record Linkage werden dann die neu einzutragenden IDAT mit den bereits in der Mainzelliste-Datenbank gespeicherten IDAT abgeglichen. Hierzu werden sogenannte Ähnlichkeits-Scores gebildet. Je nach Übereinstimmungsgrad erzeugt die Mainzelliste entweder ein neues Pseudonym IDPL oder gibt ein bestehendes zurück. Der Ähnlichkeitsscore wird berechnet als die gewichtete Summe der Ähnlichkeitswerte aller IDAT-Variablen. Abhängig vom verwendeten Schwellenwert werden die Datenpaare in drei Kategorien eingestuft: sichere Übereinstimmung, sichere Nicht-Übereinstimmung und unsichere Übereinstimmung (Contiero et al., 2005). Die Mainzelliste führt zwar das Record Linkage auf Basis von IDAT aus, speichert aber zusätzlich für mehr Flexibilität (z. B. Ausdehnung des Projekts auf den ambulanten Sektor) die entsprechenden einwegverschlüsselten Bloomfilter (Rohde et al., 2021).

### 3.6.3 Möglichkeiten und Herausforderungen

Die Datenintegration im nNGM mithilfe der Mainzelliste wurde im Rahmen einer umfangreichen Pilotierung mit sechzehn Krebszentren geprüft. Die Prüfung der Anforderungen sowohl aus Nutzersicht als auch aus technischer Sicht verlief erfolgreich. Mit der Pilotierung konnte gezeigt werden, dass, trotz der Heterogenität der IT-Systeme und Organisationen sowie der getrennten Speicherung von IDAT und MDAT, die Dokumentationsabläufe im Hintergrund ohne zusätzlichen Aufwand für die Nutzenden durch die Pseudonymisierung laufen können. Der hier beschriebene Anwendungsfall ist als Video[10] verfügbar.

---

10      https://www.youtube.com/watch?v=JMCuR5yjfIs, Zugriff 12.12.2021

Im Vergleich zur Pilotierung birgt der Regelbetrieb weitere Herausforderungen:

- Verfügbarkeit der IT-Systeme: Bei der Vernetzung der nNGM-Mitglieder und -Partner muss eine prospektive, fallbezogene Ablaufdokumentation durch unterschiedliche Akteure und Organisationen erfolgen. Dabei ist die Anforderung an die Verfügbarkeit der Mainzelliste ausgesprochen hoch, da ein Ausfall zur Störung des Ablaufs führt.

- Keine einheitliche KVNR: Beim Einpflegen der KVNR können Verwechselungen und Fehler bei der Dokumentation mithilfe der Mainzelliste entdeckt werden. Da allerdings in der privaten Krankenversicherung anders als in der gesetzlichen Krankenversicherung keine lebenslange, eindeutige KVNR etabliert ist, kann bei Privat-Versicherten keine entsprechende Überprüfung stattfinden.

- Datenlöschung: Bei Widerruf von Patienteneinwilligungen oder Löschung von Datensätzen aus dem EDC-System sind automatisch auch Pseudonyme aus der Patientenliste zu entfernen.

- Datenzusammenführung/-trennung: Zur Vermeidung von falsch positiven Zusammenführungen wurde der Schwellenwert für die Übereinstimmung beim probabilistischen Record Linkage sehr hoch gesetzt. Menschliche Fehler können allerdings nicht ausgeschlossen werden, die z. B. zur Vergabe von mehr als einem Pseudonym pro Patient:in führen können. Hierfür muss neben einer Schnittstelle auch für den Betreiber der Patientenliste eine nutzerfreundliche Lösung bereitgestellt werden, um bei Bedarf Duplikate zusammenzuführen oder fälschlicherweise zusammengeführte Datensätze zu trennen.

- Die Verwendung vom Record-Linkage in einem klinischen Kontext birgt auch die Herausforderung, dass sich IDAT im Laufe der Behandlung ändern können, was zusätzlich zu Fehlern bei der Verknüpfung führen kann. Hier sind Prozesse der IDAT Qualitätssicherung zu etablieren.

Weitere Herausforderungen waren die hohen regulatorischen Hürden durch die erforderliche Abnahme des Datenschutzkonzeptes durch die einzelnen Netzwerkzentren. Hier führten – trotz Vorliegen eines einheitlichen Datenschutzkonzepts mit positivem Votum der AG Datenschutz des TMF e.V. – unterschiedliche Interpretationen durch die zuständigen

Datenschutzbeauftragten zu unterschiedlichen Bewertungen sowie teils inkompatiblen Anpassungswünschen und somit zu hohem zusätzlichem zeitlichem Aufwand zur Konsensfindung, Berücksichtigung der Bewertungen und Anpassungswünsche.

## 3.7 Use Case 7: Standortübergreifendes Record Linkage im Deutschen Konsortium für Translationale Krebsforschung (DKTK)

### 3.7.1 Hintergrund und Ziel

Das Deutsche Konsortium für Translationale Krebsforschung (DKTK) (Joos et al., 2019) wurde im Jahr 2012 als Deutsches Zentrum für Gesundheitsforschung im Bereich der Onkologie gegründet. Durch eine Förderung des BMBF und der beteiligten Bundesländer in Höhe von ca. 28 Millionen jährlich sind mehr als 300 Personen an zehn Standorten in diesem nationalen Krebsforschungsnetzwerk tätig, um Erkenntnisse aus der Grundlagenforschung in die klinische Praxis zu übertragen. Für die dabei anfallenden immensen Datenmengen hat das DKTK eine verteilte Struktur etabliert. Grundlage dieser föderierten Arbeitsweise ist die Clinical Communication Plattform (CCP) (Lablans & Schmidt, 2020), mit deren Hilfe neben Record Linkage und Pseudonymisierung auch Authentifizierung und Datenaustausch sowohl innerhalb eines Standorts als auch standortübergreifend erfolgt.

Die Besonderheit im DKTK besteht darin, dass bereits 2013 die Strategie einer föderierten Speicherung und Nutzung vorhandener (historischer) Daten verfolgt wurde. Während prospektiv im DKTK erhobene Datensätze auf Rechtsgrundlage einer informierten Patienten-einwilligung verarbeitet werden, liegt eine solche Einwilligung für historische oder in der Routine entstehenden Datensätze selten vor. In DKTK-Kontext unterscheidet man daher zwischen Daten, für die eine explizite Einwilligung vorliegt, und Daten ohne eine solche.

Diese besondere Ausgangslage macht das DKTK aus drei Gründen zu einem interessanten An-wendungsfall: Erstens führen die unterschiedlichen Einwilligungs-Situationen zu einer Kom-bination unterschiedlicher Record Linkage-Verfahren. Zweitens wird mit dem sogenannten „Brückenkopf" ein Modell der dezentralen Datenhaltung verfolgt. Die dezentrale Datenhaltung hat sich über viele Jahre auch über die Onkologie hinaus bei der wissenschaftlichen Nachnutzung von klinischen Routinedaten etabliert – in ähnlicher Weise etwa in den Datenintegrationszentren der Medizininformatik-Initiative. Drittens liegt hier nach unserer Kenntnis seit ihrer Inbetriebnahme im Jahr 2014 die erste produktiv genutzte Implementierung einer föderierten Treuhandstelle (federated Trusted Third Party, fTTP, siehe Glossar) vor.

## 3.7.2  Technische Umsetzung

### 3.7.2.1    Lokales vs. zentrales Record Linkage

Im DKTK werden zwei verschiedene Varianten des Record Linkage durchgeführt:

- Beim **lokalen Record Linkage** wird der IDAT-Vergleich innerhalb eines Standorts durchgeführt. Nur die Datensätze des jeweiligen Standorts werden untereinander abgeglichen. Dafür enthält jeder Brückenkopf eine lokale Patientenliste auf Basis der Mainzelliste.

- Beim **zentralen Record Linkage** erfolgt der Abgleich standortübergreifend mithilfe eines zentralen Identitätsmanagements an zwei Standorten.

Bei der Implementierung dieser Varianten sind folgende Rahmenbedingungen zu beachten:

1. IDAT zu Patient:innen, die nicht explizit eingewilligt haben, dürfen den Standort grundsätzlich nicht verlassen. Demzufolge kommt hier nur ein lokales Record Linkage zum Einsatz.

2. Klartext-IDAT zu Patient:innen, die explizit eingewilligt haben, sollten nicht außerhalb der behandelnden Einrichtung gespeichert werden. Vielmehr müssen verschlüsselte Identifikatoren verwendet werden, die eine Re-Identifizierung möglichst schwierig, aber einen Ähnlichkeitsvergleich möglichst einfach machen (Privacy Preserving Record Linkage).

3. Manche Prozesse, z. B. die Klärung unsicherer Übereinstimmungen oder die Zusammenführung von Patientendatensätzen, funktionieren wesentlich besser auf Basis von Klartext-IDAT. Für diesen Zweck werden Klartext-IDAT lokal innerhalb der Mainzelliste am jeweiligen Standort gespeichert.

### 3.7.2.2    Matching-Algorithmus

Für das Record Linkage in DKTK wird ein probabilistischer Algorithmus der EpiLink-Software (Contiero et al., 2005) verwendet, der ein fehlertolerantes Record Linkage ermöglicht. Das Ergebnis wird durch einen Ähnlichkeitsscore dargestellt, der mithilfe der vordefinierten Schwellenwerte die Kategorisierung der IDAT-Paare in folgende Gruppen ermöglicht:

- **Sichere Übereinstimmung**: Das Ergebnis ist höher als der Schwellenwert für sichere Übereinstimmungen – Patient:in wurde gefunden.

- **Sichere Nicht-Übereinstimmung**: Das Ergebnis ist niedriger als der Schwellenwert für sichere Nicht-Übereinstimmungen – Patient:in ist neu.

- **Unsichere Übereinstimmung**: Es ist nicht möglich sicher zu bestimmen, ob es sich um eine Übereinstimmung oder Nicht-Übereinstimmung handelt – zusätzliche Schritte sind notwendig, zum Beispiel die manuelle Überprüfung der IDAT.

Die folgenden Merkmale der Patienten-IDAT werden sowohl für das lokale auch als auch für das zentrale Record Linkage verwendet:

- Vorname

- Nachname

- Frühere Namen

- Geburtsdatum

- Staatsangehörigkeit

- Geschlecht.

Die IDAT-Merkmale werden paarweise verglichen, um die Ähnlichkeit der Patient:innen zu berechnen. Außerdem können für manche Merkmale so genannte Exchange Groups definiert werden, die einen Vergleich aller Felder-Kombinationen innerhalb einer Gruppe ermöglichen. Diese Option wird verwendet, um bspw. Vertauschungen von Vor- und Nachnamen zu erkennen.

### 3.7.2.3    Föderiertes Record Linkage

Das standortübergreifende (föderierte) Record Linkage basiert auf dem Prinzip des Privacy-Preserving Record Linkage (Vatsalan et al., 2013). Im Unterschied zum klassischen Record Linkage werden IDAT nicht im Klartext verglichen, sondern mittels sogenannter Bloomfilter (Schnell et al., 2009b) in nicht oder kaum umkehrbare, aber gut vergleichbare Bitfolgen umgewandelt. Diese einwegverschlüsselten Bitfolgen eignen sich trotz ihrer Transformation gut zum Vergleich im EpiLink-Verfahren und werden anstelle von IDAT im Record Linkage verwendet. Die beschriebene Transformation wurde ursprünglich von Kontrollnummern (vgl. Use Case 2 und 3) inspiriert, die in epidemiologischen Krebsregistern verwendet werden.

Die Transformation wird mithilfe eines zentralen Hashers durchgeführt. Diese Komponente übernimmt die IDAT-Kodierung mithilfe einer Hashfunktion und wird normalerweise an einem anderen DKTK-Standort als die Patientenliste betrieben.

In den mittlerweile fast acht Jahren ihres Bestehens wurde das Record Linkage im DKTK kontinuierlich fortentwickelt. Im Folgenden wird ein beispielhafter Überblick über die verschiedenen Record Linkage-Varianten im DKTK gegeben. Nach Übermittlung der Klartext-IDAT an den lokalen Brückenkopf prüft dieser stets im ersten Schritt den Einwilligungsstatus der betroffenen Patient:innen. In Abhängigkeit davon wird der Datensatz entweder nur lokal oder zusätzlich zentral pseudonymisiert (vgl. Abbildung 7):

- Fall A: Die zu behandelnde Person hat nicht explizit für die Pseudonymisierung ihrer Daten im DKTK eingewilligt (rote Pfeile). In diesem Fall werden die Klartext-IDAT stets bereits lokal kodiert. Dies erfolgt mithilfe eines lokalen Hashers unter Anwendung einer Hashfunktion und eines geheimen kryptographischen Schlüssels, der die Rückumwandlung zu IDAT erschwert (bspw. bei Wörterbuchattacken). Das Record Linkage wird auf den kodierten IDAT durchgeführt. Anschließend werden lokale Pseudonyme generiert.

- Fall B: Für die zu behandelnde Person liegt eine explizite Einwilligung für die Pseudonymisierung ihrer Daten im DKTK vor (grüne Pfeile). Die Klartext-IDAT können entweder im lokalen oder zentralen Identitätsmanagement mithilfe eines lokalen Hashers in kodierte IDAT umgewandelt werden. Damit diese kodierten IDAT standortübergreifend vergleichbar sind, muss derselbe kryptographische Schlüssel Einsatz kommen. Hierzu werden zwei Varianten unterstützt: Entweder die Transformation mit einem lokalen Hasher, die jedoch anders als in Fall A alle mit demselben Schlüssel initialisiert wurden, oder aber die Verwendung eines zentralen Hashers. In beiden Fällen wird das föderierte Record Linkage basierend auf den so kodierten IDAT durchgeführt und standortübergreifend verknüpfbare Pseudonyme generiert.

In beiden Fällen können die IDAT auch im Klartext in der lokalen Patientenliste abgespeichert werden, was Prozesse wie eine Korrektur von IDAT oder eine nachträgliche Zusammenführung der Patientendaten erleichtert.

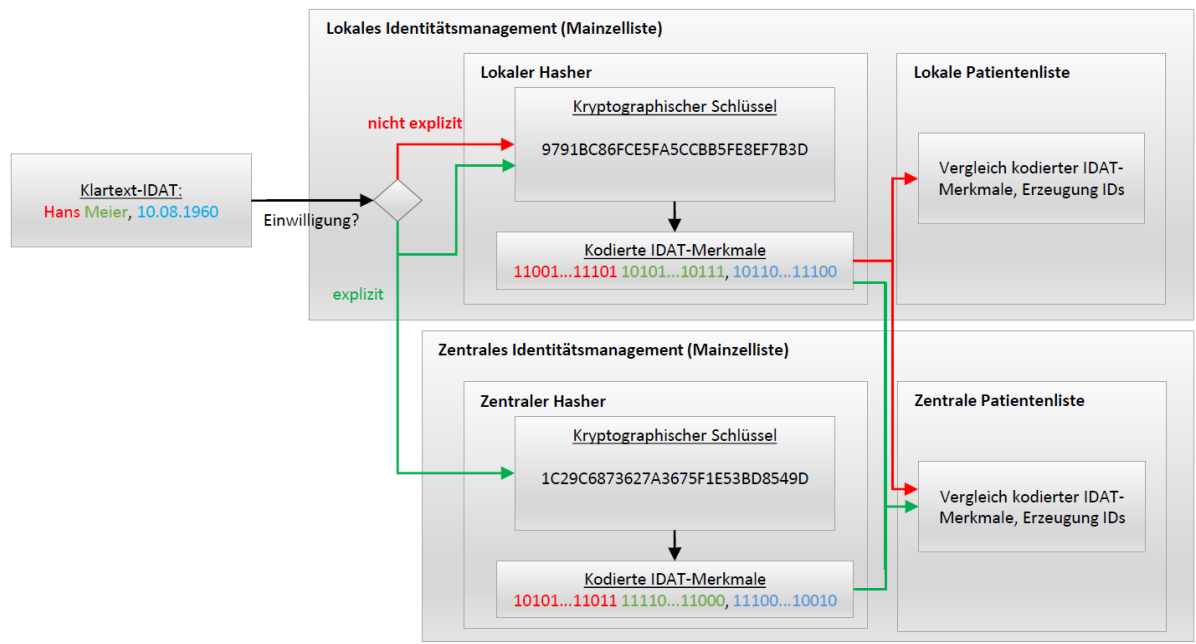

Abbildung 7: Schematische Darstellung des lokalen und zentralen Record Linkage im DKTK

### 3.7.2.4  Umgang mit unsicheren Übereinstimmungen

Die Ergebnisse des Record Linkage werden im DKTK wie folgt berücksichtigt:

- Bei einer **sicheren Übereinstimmung** wird das vorhandene Pseudonym als Antwort auf die Pseudonymisierungsanfrage zurückgegeben.

- Bei einer **sicheren Nicht-Übereinstimmung** werden die neuen IDAT (in ihrer gehashten Form) in die Mainzelliste eingetragen. Ein neu generiertes Pseudonym wird als Antwort zurückgegeben. Analog zum nNGM kommen verschiedene Pseudonyme zum Einsatz (Beschreibung siehe Use Case 6).

- Bei einer **unsicheren Übereinstimmung** ist es nicht möglich sicher festzustellen, ob für Patient:innen in der Patientenliste nicht schon Einträge existieren. Daher wird wie im Fall einer sicheren Nicht-Übereinstimmung verfahren, der Datensatz jedoch als *tentativ* markiert. Für eine spätere Daten-zusammenführung wird das Pseudonym des zum Matching-Zeitpunkt ähnlichsten Patienten ebenfalls gespeichert. Die Patientenpaare mit tentativen Übereinstimmungen können später von einer berechtigten Person (Administrator, Datenschützer) geprüft werden, um die Einträge der Patient:innen zusammenzuführen oder zu trennen. Dies führt zur Entfernung der *tentativ*-Markierung.

### 3.7.2.5 Verwendete Tools

Zur Anwendung kommt die etablierte open source Software Mainzelliste (Lablans et al., 2015a). Zu Beginn wurde der DKTK-Pseudonymisierungsprozess fest in Programmcode formuliert und war damit nicht einfach les- oder anpassbar. Später wurde dieser alte Ansatz durch eine neue, einfach gehaltene Pseudonymisierungssprache namens MAGICPL ersetzt (Tremper et al., 2021). Der Vorteil dieses Ansatzes ist, dass Operationen wie die Entscheidung für eine Pseudonymisierungsvariante oder die Transformation von IDAT nur einmalig implementiert werden müssen und in verschiedenen Projekten wiederverwendet werden können. Diese Komponenten laufen sowohl zentral (DKTK-Standorte Frankfurt und Mainz) als auch an jedem einzelnen DKTK-Standort. Die Komponenten wurden an jedem Standort als Bestandteil des DKTK-Brückenkopfes ausgerollt, sodass eine einheitliche und auch mit beschränktem Personalstamm leicht zu pflegende lokale Vertrauensstelle an jedem Standort vorliegt.

### 3.7.3 Herausforderungen

Im DKTK liegt die erste Implementierung eines föderierten Record Linkage vor. Sie befindet sich seit vielen Jahren im produktiven Einsatz mit echten Patientendaten. Dadurch konnten erste Erfahrungen für den Betrieb föderierter Record Linkage-Infrastrukturen gewonnen werden.

Die größten Herausforderungen liegen in der unterschiedlichen technischen und personellen Ausstattung der Standorte und der Heterogenität der Daten in den Quellsystemen. Zur Bewältigung der technischen Probleme ist ein gutes Projektmanagement erforderlich, das für die Standorte minimale technische Anforderungen formuliert und ihre Erfüllung nachverfolgt. Dieser Prozess erforderte im DKTK mehrere Jahre. Hinzu kommen organisatorische Hürden, die sich von Standort zu Standort unterscheiden und nicht immer zentral gelöst werden können. Dies können unterschiedliche Landesgesetzgebungen sein (Schneider, 2015) oder auch eine unterschiedliche Auslegung durch die zuständigen Datenschutzbeauftragten im Rahmen ihrer (zuweilen sehr unterschiedlich wahrgenommenen) Ermessensspielräume.

Ein drittes Problem sind Patienteneinwilligungen, die zwar seitens der CCP im DKTK gut vorbereitet werden konnten, aber deren Rollout inkl. ärztlichem Aufklärungsgespräch in der klinischen Praxis schlicht nicht flächendeckend funktioniert. Es kann also auch für prospektiv entstehende (Routine-) Daten nicht von einer flächendeckenden informierten Patienteneinwilligung ausgegangen werden. Die Lösung hierfür kann in einer breiten Umsetzung des Broad Consents der Medizininformatik-Initiative liegen, die aber naturgemäß nur prospektiv gelingt und naturgemäß keine Spezifika von DKTK-initiierten Studien abdecken.

### 3.7.4 Ausblick

Die Pseudonymisierung im DKTK wird in enger Kooperation mit weiteren acht Einrichtungen, die zur Software Mainzelliste beitragen, aktiv weiterentwickelt. Aktuell wird im DKTK zusätzlich eine projektbezogene Pseudonymisierung (Projektpseudonymisierung) in Betrieb genommen, die eine Pseudonymisierung auch für DKTK angegliederte Projekte und Studien erlaubt. Ein weiteres Feature ist eine Nutzeroberfläche, entwickelt in einem parallelen Projekt, die Operationen wie die Zusammenführung von Patientendatensätzen über die Web-Schnittstelle (REST) der Mainzelliste abbildet. Die Einführung der Nutzeroberfläche ermöglicht die manuelle Nachbesserung der Record Linkage im Fall der unsicheren Übereinstimmung, was zu höherer Qualität der Record Linkage führt.

Eine vielversprechende methodische Weiterentwicklung stellt eine neue Generation besonders sicherer PPRL-Verfahren auf Basis von Secure Multi Party Computation (Mainzelliste Secure EpiLinker, MainSEL (Stammler et al., 2022)) dar. Sie bieten höhere, mathematisch beweisbare Sicherheitsgarantien gegenüber den 2009 entwickelten Bloomfiltern und erfordern insbesondere keine Speicherung von IDAT (auch nicht in kodierter Form) außerhalb des Standorts. Ein solches Verfahren konnte auf Basis der Mainzelliste an neun Universitätskliniken erprobt werden (auf Testdaten, aber im realen Setting) (Kussel et al., 2022). Erste datenschutzrechtliche Untersuchungen deuten darauf hin, dass dieses Verfahren keinen Datentransfer, insbesondere Transfer von identifizierenden Daten, im Sinne der DSGVO darstellen und damit neue, bislang rechtlich verschlossene Einsatzmöglichkeiten bieten könnte. Im DKTK sind solche neuen Verfahren aufgrund der einheitlichen Umgebung an allen beteiligten Universitätskliniken mit den DKTK-Brückenköpfen verhältnismäßig einfach umzusetzen; Ergebnisse werden darüber hinaus open-source als Erweiterungen der Mainzelliste zur Verfügung gestellt.

## 3.8 Use Case 8: Spezialfall Treuhandstelle – die Unabhängige Treuhandstelle der Universitätsmedizin Greifswald

*Anmerkung der Autoren: Der folgende Absatz, sowie die Abschnitte 3.8.2.1 und 3.8.3 wurden in Teilen bereits vorab online veröffentlicht [11].*

### 3.8.1 Hintergrund und Ziele

Mit Blick auf die umfassenden Anforderungen des TMF-Datenschutzleitfadens wurde 2014 die Unabhängige Treuhandstelle der Universitätsmedizin Greifswald (THS) per Beschluss des Vorstandes der Universitätsmedizin Greifswald (UMG) als zentrale Einrichtung der Universitätsmedizin errichtet und ihr die Unabhängigkeit und Weisungsfreiheit gegenüber der UMG zugesichert. Das Datenschutz- und IT-Sicherheitskonzept der THS wurde bereits 2014 vom Landesdatenschutzbeauftragten Mecklenburg-Vorpommern positiv votiert. Das Konzept der Treuhandstelle wurde in den vergangenen Jahren erfolgreich etabliert und an die Datenschutz-Grundverordnung (DSGVO) angepasst.

Das Konzept der Treuhandstelle unterstützt die informationelle Gewaltenteilung nach TMF-Leitfaden (siehe Abbildung 8). Dies beinhaltet die separate Speicherung und das einheitliche fehlertolerante Record Linkage von personenidentifizierenden Daten, die Verwaltung und Bereitstellung patientenspezifischer (und ggf. anwendungsspezifischer) Pseudonyme für Studien, Forschungsdatenbanken und Bioproben sowie die Verwaltung von Einwilligungen und Widerrufen.

Die organisatorisch selbständige Treuhandstelle der Universitätsmedizin Greifswald entwickelt und betreibt zum Schutz der besonders schützenswerten personenidentifizierenden Daten spezialisierte Software und bietet damit datenschutzkonforme Lösungen für die medizinische Forschung[12]. Die Treuhandstelle der Universitätsmedizin Greifswald ist mit ihren Lösungen Partner in zahlreichen deutschlandweiten und einzelnen internationalen Projekten. Dazu zählen im Rahmen des Netzwerks Universitätsmedizin (NUM) die NUM Klinische Epidemiologie- und Studienplattform (NUKLEUS) und die NUM-Routinedatenplattform (NUM-RDP) (Heyder et al., 2023) (siehe auch Kap. 3.9). Weitere Beispiele sind das

---

[11] https://www.ths-greifswald.de/ueber-uns, Zugriff 20.12.2022

[12] https://www.ths-greifswald.de/forscher/, Zugriff 25.02.2022

Deutsche Zentrum für Herz-Kreislauf-Forschung (DZHK[13]), die NAKO-Gesundheitsstudie[14], das INTERREG-Vorhaben „Baltic Fracture Competence Center" (BFCC[15]) sowie das MIRACUM-Konsortium[16] der Förderinitiative Medizininformatik.

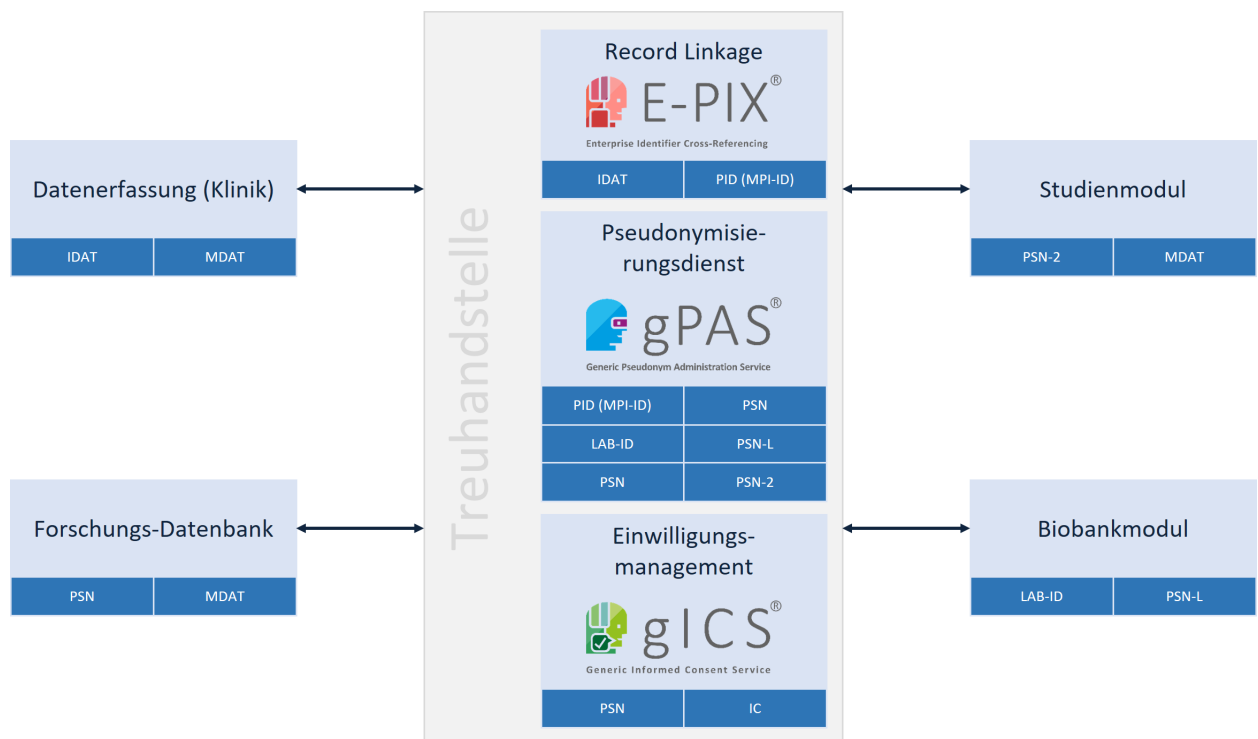

Abbildung 8: Informationelle Gewaltenteilung in Anlehnung an den TMF-Leitfaden mit den Lösungen der THS Greifswald (IDAT: identifizierende Daten, MDAT: medizinische Daten, PSN: Pseudonym, IC: Informend Consent, MPI-ID: Master-Patienten-Index-ID, PID: Projekt-ID)

### 3.8.2 Technische Umsetzung

Basierend auf den Erfahrungen beim standortübergreifenden Record Linkage in den Projekten der THS Greifswald lassen sich folgende Aspekte und Randbedingungen im praktischen Betrieb von Record Linkage Lösungen ableiten:

**Vergleichsalgorithmen, Namensräume und Datenquellen:** Grundsätzlich können beim Record Linkage einfache String-Ketten-Vergleiche zum Einsatz kommen. Die Verwendung von Phonetik-Algorithmen ist nur begrenzt geeignet, da dies eine Festlegung auf den zugrunde liegenden Sprachraum erfordert (z. B. Kölner Phonetik). Bei Namensvergleichen mit gemischten Namensräumen (z. B. bei Projekten mit Beteiligung unterschiedlicher Nationen oder bei Projekten mit Teilnehmenden aus Ballungszentren) hat sich in der Praxis die Verwendung der Levenshtein-Distanz (Levenshtein, 1966) durchgesetzt. Bei der Verwendung von mehreren zusammenzuführenden Datenquellen sollten vor Beginn der Datenerhebung nach Möglichkeit alle für die Datenzusammenführung relevanten Datenquellen berücksichtigt werden. Ein nachträgliches Hinzufügen von Datenquellen kann je nach verwendeter Record Linkage-Lösung zu Problemen führen. Zudem sollte mit Blick auf die zu erwartende Qualität der Übermittlungen der einzelnen Datenquellen eine „vertrauenswürdige Datenquelle" definiert werden, um bei komplexen Linkage-Szenarien automatisiert eine geeignete Entscheidung fällen zu können.

---

13      https://dzhk.de/, Zugriff 23.02.2022

14      http://www.nako.de/, Zugriff 23.02.2022

15      https://www.bfcc-project.eu/registry.html, Zugriff 23.02.2022

16      https://www.miracorg/, Zugriff 23.02.2022

**Konfiguration:** Die Bestimmung der Identifikatoren, Schwellenwerte und Matching-Parameter erfolgt in der THS der Universitätsmedizin auf Basis von Erfahrungswerten, die u. a. in den vorstehenden Projekten gesammelt wurden. Laut Fellegi & Sunter (1969) ließen sich optimale Parameter errechnen, wenn ein „Goldstandard" für Datensätze mit personenidentifizierenden Daten mit bekanntem Outcome (Anzahl enthaltener Synonyme, Anzahl fehlerbehafteter Datensätze) zur Verfügung stünde, um Systeme, Algorithmen und existierende Record Linkage-Lösungen in Bezug miteinander zu vergleichen. Dies würde z. B. eine Bewertung der Qualität ermöglichen.

**Performance:** Die Geschwindigkeit des Record Linkage-Prozesses auf Basis von personenidentifizierenden Daten ist grundsätzlich abhängig von der verfügbaren Infrastruktur, den Hardware-Ressourcen, dem verwendeten Tool und der Anzahl der durchzuführenden Abgleiche. Insofern wirkt sich die Größe des Datenbestands auf die Dauer des Matching-Prozesses aus. Aus den Projektanforderungen kann sich die Notwendigkeit eines Matching „in Echtzeit" ergeben. Hampf et al. (2020) haben die Performance des E-PIX® im Setting eines Forschungsvorhabens an der Charité – Universitätsmedizin Berlin mit mehreren Millionen Patientendaten umfassend untersucht und dessen Eignung für sehr große Vorhaben erfolgreich belegt[17].

---

[17] https://www.ths-greifswald.de/forscher/e-pix/, Zugriff am 23.02.2022, Details dazu finden Sie in Tabelle „Übersicht Tools Record Linkage" im Anhang.

**Doppler-Auflösung:** Wie in den Beispielprojekten gezeigt kann ein Großteil der personen-identifizierenden Daten automatisiert verknüpft werden. Nur in Einzelfällen sind manuelle Eingriffe erforderlich. Im Rahmen des Record Linkage-Prozesses erkannte potenzielle Synonymfehler sollten stets aufgelöst werden, um die Qualität der Datenzusammen-führung verknüpfter medizinischer Datensätze zu verbessern. Dieser Prozess ist personal-intensiv und verursacht Kommunikationsaufwand (Sicherstellen der Korrektheit der personenidentifizierenden Daten) und Dokumentationsaufwand (Nachvollziehbarkeit der Zusammenführung/Trennung von Datensätzen bei falscher Zusammenführung). Record Linkage-Lösungen vereinfachen allerdings dabei den Umgang mit unvollständigen personenidentifizierenden Daten und potenziellen Synonymen in der täglichen Praxis. Homonymfehler sind weiterhin möglich.

### 3.8.2.1 Personeller und technischer Aufwand

Der operative Betrieb einer Treuhandstelle geht weit über das Betreiben und/oder Hosting von Software-Lösungen hinaus. Die Administration und Weiterentwicklung der Infrastrukturen und Softwarekomponenten sowie Datenpflege und Kommunikation erfordern personelle Ressourcen vielfältiger Fachrichtungen. Dieser personelle Aufwand ist bei der Konzeption einer Treuhandstellenlösung zu berücksichtigen. Form und Umfang sind dabei stets vom individuellen Projektsetting abhängig.

Das Team der Treuhandstelle Greifswald besteht aus einer Vielzahl von Mitarbeitenden. Informatiker:innen, Java- und Web-Spezialist:innen, medizinische Dokumentar:innen, jeweils eine Juristin und ein Physiker sowie studentische Hilfskräfte bilden die Brücke zwischen Studienteilnehmenden, Forschenden und Projektpartnern, um datenschutzkonforme Gesundheitsforschung zu ermöglichen.

Die Treuhandstelle Greifswald betreibt für jeden Mandanten (Forschungsprojekte, Forschungseinrichtungen, Studien) separate IT-Infrastrukturen. Dies macht ein hohes Maß an Organisation erforderlich. Zum Regelbetrieb zählen auch die Pflege und Qualitätsprüfung der innerhalb der Treuhandstelle ge-speicherten Informationen. Der Umfang der erforderlichen personellen Ressourcen ist abhängig von der Größe des Forschungsprojektes (Anzahl Studienzentren, Anzahl Datenquellen, Anzahl geplante Rekrutierung) und den individuellen Projekt-prozessen (mit/ohne Qualitätssicherung von Einwilligungen, Qualitätssicherung der übermittelten Daten, Auflösung möglicher Synonymfehler, Beratungsbedarfe, schnittstellen-seitige Anpassungsbedarfe).

### 3.8.3  Record Linkage auf Basis personenidentifizierender Daten in der Praxis

Neben der Verwaltung der Einwilligungen und Widerrufe von betroffenen Personen sowie der Bereitstellung und Administration kontextspezifischer Pseudonyme stellen der standort-übergreifende **Abgleich und die zentrale Verwaltung personenidentifizierender Daten ein wesentliches Aufgabenfeld des regulären Treuhandstellenbetriebs** dar. Dies umfasst die automatisierte Prüfung, Zusammenführung und Speicherung von personenidentifizierenden Daten (IDAT) betroffener Personen sowie das Auflösen potenzieller Synonymfehler. Dieser Record Linkage-Prozess innerhalb der Treuhandstelle stellt die Grundlage für die eindeutige und standortübergreifende Identifikation von Patient:innen und Studienteilnehmenden im Rahmen eines Forschungsvorhabens dar und ermöglicht somit die korrekte Zusammen-führung zugehöriger medizinischer Daten.

Der nachfolgende Abschnitt illustriert an ausgewählten Projekten der THS Greifswald diesen Record Linkage-Prozess. Die Universitätsmedizin Greifswald führt regelmäßige Kennzahlenerhebungen bei den Anwendern der Treuhandstellenlösungen durch. Nachfolgende Ausführungen basieren auf dem Erhebungsstand vom Juni 2021.

Das **Klinische Krebsregister Mecklenburg-Vorpommern** (KKR-MV) erhebt medizinische Daten auf gesetzlicher Grundlage gemäß DSGVO Artikel 6 Absatz 1 lit. e. Die Erfassung und das Record Linkage der Krebspatientendaten über Registerstellen in Mecklenburg-Vorpommern erfolgen im Krebsregistrierungssystem des Gießener Tumordokumentations-systems und über einen teilautomatisierten Melderegisterabgleich innerhalb der THS. Patient:innen haben ein Widerspruchsrecht gegen die Verarbeitung ihrer personenbezogenen Daten gemäß DSGVO Art 21. Absatz 1.

Um die medizinischen Daten der Krebspatient:innen von den einzelnen Registerstellen und der Leichenschauscheininformationen korrekt mittels E-PIX® zusammenführen zu können, werden die personenidentifizierenden Daten der Patient:innen (Vornamen, Nachname, Geschlecht, Geburtsdatum, PLZ des Hauptwohnsitzes) als Identifikatoren verwendet.

Unterstützt wird diese Datenzusammenführung durch Daten der elektronischen Meldeauskunft MV[18].

Die Schwellwerte für das Record Linkage sind so gewählt, dass eine automatische Zusammenführung nur bei exakter Übereinstimmung der Identifikatoren stattfindet. Es wird jedoch in Fällen mit moderater Ähnlichkeit noch eine manuelle Zusammenführung ermöglicht. In Summe wurden auf diese Weise mittels E-PIX® in der THS des KKR-MV personenidentifizierende Daten von ca. 275.000 Personen verwaltet. In den Daten wurden knapp 10.800 (3,9%) potenzielle Synonymfehler erkannt, die durch THS-Personal manuell überprüft und aufgelöst wurden.

**In der NAKO-Gesundheitsstudie** erfolgen die Datenerhebung und Untersuchung nach Einholung einer informierten Einwilligung der Teilnehmenden (gemäß DSGVO Artikel 6 Absatz 1 lit. a). Potenzielle Studienteilnehmende werden per Zufallsauswahl der Melderegister ermittelt und bereits vor deren Einwilligung mit ihrer Adresse in einer Datenbank zur Kontaktaufnahme und zum Teilnehmermanagement erfasst (MODYS) (Reineke et al., 2019). Dieses System steuert die Einladung zur Teilnahme an der NAKO-Gesundheitsstudie. Teilnehmende werden in der Treuhandstelle anhand personenidentifizierender Informationen mittels E-PIX® abgeglichen, identifiziert und ggf. zusammengeführt. Dadurch werden Mehrfach-Einladungen vermieden. Etwaige Totalverweigerer werden in der THS in einer separaten sogenannten Robinson-Liste geführt.

Für die eindeutige Identifikation der Teilnehmenden werden in der NAKO Vorname, Name, ausgewählte Adress-Bestandteile (Straße, Hausnummer, Postleitzahl) und das Geburtsjahr berücksichtigt. Der Vorname der Teilnehmenden ist im Vergleich zu den anderen Identifikatoren hoch gewichtet. Da oft kein vollständiges Geburtsdatum vorhanden ist, wird das Geburtsjahr separat erhoben. Die von den Meldeämtern übermittelten Stichproben können auch Doppler enthalten. Dies und das teilweise fehlende Geburtsdatum führen zur Erhöhung des manuellen Auflösungsaufwands. In Summe wurden in der THS der NAKO im Juni 2021 ca. 1.373.000 (potenzielle) Studienteilnehmende verwaltet. In 7,2% der Fälle wurden mögliche Synonymfehler erkannt. Diese mehr als 98.500 Fälle wurden in der THS manuell überprüft und aufgelöst.

---

[18]  https://www.mv-serviceportal.de/leistung/?leistungId=105601006, Zugriff: 21.06.2023

In den mehr als 20 Studien und Registern des **Deutsches Zentrum für Herz-Kreislauf-Forschung e.V.** (DZHK) erfolgt die medizinische Datenerhebung auf Basis der Einwilligung der Teilnehmenden (gemäß DSGVO Artikel 6 Absatz 1 lit. a). Die Anlage der Teilnehmenden innerhalb der THS und der Abgleich mit bereits vorhandenen Teilnehmenden mittels E-PIX® und auf Basis der personenidentifizierenden Daten im Rahmen des Record Linkage-Prozesses erfolgen erst nach Vorliegen der Einwilligung.

Dieser Abgleichsprozess umfasst Standard-Identifikatoren ergänzt um den (besonders hoch gewichteten) Geburtsort. Insbesondere in anderen EU-Staaten wird der Parameter „Geburtsort" häufig nicht oder falsch ausgefüllt (z. B. Nennung des Landes, der Region oder der Ortsangabe in der Landessprache). Dies führt zu manuellem Auflösungsaufwand. In Summe werden in der THS des DZHK zum Erhebungszeitpunkt ca. 12.000 Personen verwaltet. In 1,3% der Fälle wurden potenzielle Synonymfehler erkannt. Diese Fälle wurden manuell überprüft und aufgelöst, um die korrekte Zusammenführung medizinischer Daten von den mehr als 200 DZHK-Standorten zu ermöglichen.

## 3.9 Use Case 9: Zukünftig – standortübergreifendes Record Linkage in der Medizininformatik-Initiative

### 3.9.1 Hintergrund und Ziel

Die Medizininformatik-Initiative (MII) (Semler et al., 2018) ist ein vom Bundesministerium für Forschung und Bildung gefördertes Forschungsprojekt. Mithilfe der MII soll eine Infrastruktur aufgebaut werden, die der medizinischen Forschung die Nutzung klinischer Daten aus der Versorgung von Patient:innen ermöglicht. Die MII hat zudem das Ziel, mit IT-Lösungen sowohl die Forschungsmöglichkeiten als auch die Versorgung von Patient:innen zu verbessern. Hierzu soll die Nutzung von medizinischen Daten über Standortgrenzen hinweg ermöglicht werden. Daten, die zu einer Person an unterschiedlichen Standorten gesammelt wurden, sollen verknüpft werden. Dies ermöglicht, umfangreiche und qualitativ hochwertige Datensätze für die medizinische Forschung bereitzustellen. Dabei sollen die medizinischen Daten, die im Versorgungskontext erhoben worden sind, auch für Forschungszwecke zugänglich gemacht werden. Die rechtliche Grundlage besteht hierbei aus einer informierten Einwilligung, mit der eine Person das Einverständnis über diese Nutzung gewährt. In der MII wurde hierfür der Broad Consent entwickelt und wird von den Standorten erhoben und verwaltet. Die standortübergreifende Zusammenführung von medizinischen Daten stellt dabei eine

Herausforderung dar, die durch ein standortübergreifendes Record Linkage gelöst wird. Hierbei werden Personendatensätze auf Basis von Identifikatoren abgeglichen, sodass die gleiche Person über mehrere Standorte hinweg erkannt wird und die Datensätze miteinander verknüpft werden können.

An der MII sind alle Universitätskliniken Deutschlands beteiligt, die wiederum in vier Konsortien aufgeteilt sind. Die vier Konsortien DIFUTURE (Prasser et al., 2018), HiGHmed (Haarbrandt et al., 2018), MIRACUM (Prokosch et al., 2018) und SMITH (Winter et al., 2018) verfolgen bei der Errichtung einer forschungskonformen und versorgungsunterstützenden IT-Infrastruktur jeweils unterschiedliche Lösungsansätze. Es gibt jedoch Themen, die nur konsortiumsübergreifend bearbeitet werden können. Zu diesen gehört das standortübergreifende Record Linkage, dessen Konzeption von der konsortialübergreifenden Taskforce Datenschutz der MII zusammengestellt wurde. Damit wird sichergestellt, dass in der MII einheitliche Verfahren angewendet werden und Daten von Patient:innen auch über die Grenzen eines Konsortiums hinweg zusammengeführt werden können.

Die MII sieht die Etablierung eines Datenintegrationszentrums (DIZ) an jedem Standort vor (Semler et al., 2018). Darüber hinaus muss jeder Standort eine lokale Treuhandstelle im DIZ ansiedeln oder als separate Organisationseinheit aufbauen. Diese Treuhandstellen dienen zur Verwaltung von Identitätsdaten, Pseudonymen und Einwilligungen im Forschungskontext. Die MII sieht vor, dass die Standorte als dezentrale Strukturen die Verantwortlichkeit für die lokal erfassten Daten behalten. So dient der Aufbau zentraler bzw. föderierter Strukturen nur dazu, die dezentralen Strukturen zu verbinden. Standortübergreifende Personenabgleiche, die beispielsweise durch ein föderiertes Record Linkage erreicht werden, ersetzten dabei nicht das an den Standorten angesiedelte lokale Record Linkage. Darüber hinaus werden Daten weitestgehend dezentral an den jeweiligen Standorten verwaltet und nicht in einer gemeinsamen Datenbasis zusammengeführt.

### 3.9.2 Technische Umsetzung

### 3.9.2.1 Record Linkage-Systematisierung

Die Taskforce Datenschutz der MII hat mehrere grundlegende Konzepte für ein standortübergreifendes Record Linkage erarbeitet und systematisiert (siehe Abbildung 9). Grundsätzlich sollen mehrere Konzepte von der MII verfolgt werden. Eine technische Spezifikation konkreter Konzepte wurde bis Ende 2021 noch nicht durch die MII er-arbeitet. Jedoch dienen diese Konzepte als Grundlage für andere Projekte und wurden teilweise bereits anderweitig (z. B. NUM-CODEX bzw. dem Folgeprojekt NUM-RDP[19], siehe Use Case 3.9) technisch spezifiziert und umgesetzt, sodass diese bereits eine praktische Erprobung durchlaufen.

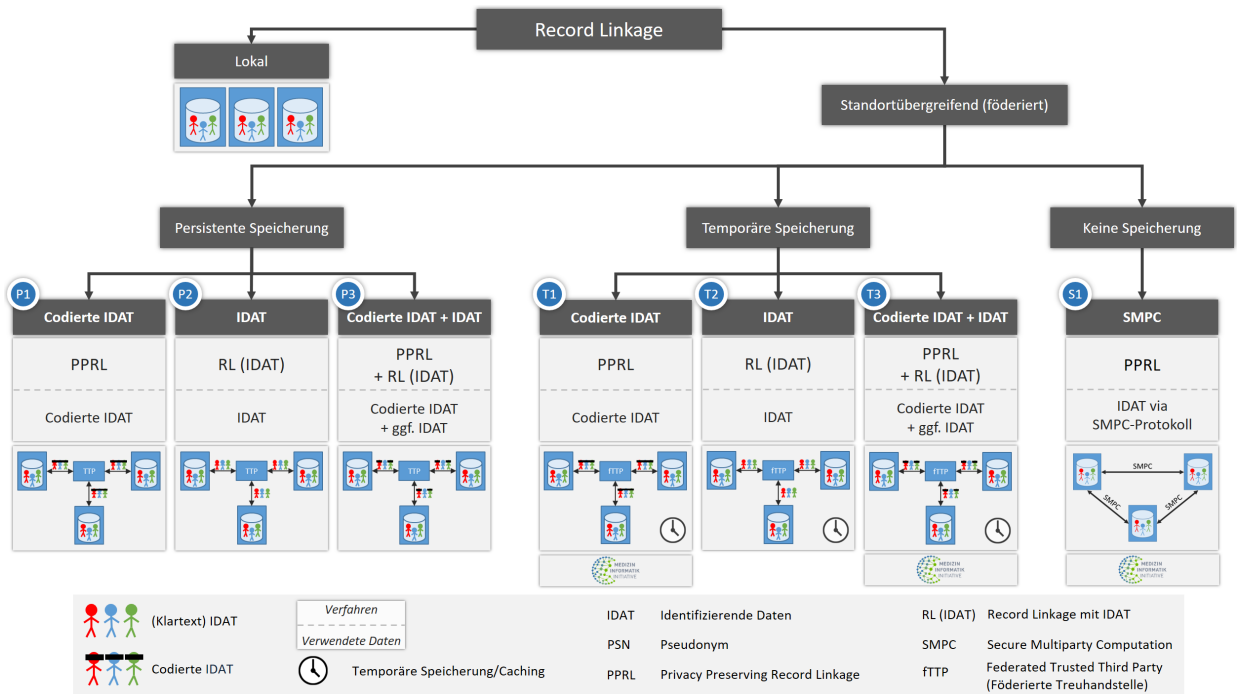

Abbildung 9: Systematisierung von Record Linkage Verfahren in der MII (Hampf et al., 2019)

Werden personenidentifizierende Daten (IDAT) (z. B. Vorname, Nachname, Geburtsdatum, Geschlecht usw.) mehrfach erfasst (z. B. zu unterschiedlichen Zeitpunkten in verschiedenen Abteilungen oder an unterschiedlichen Standorten), können sich diese unterscheiden. Gründe hierfür können Namensänderungen (z. B. bei Heirat) oder Fehler bei der manuellen Erfassung sein (z. B. Tippfehler, Zahlendreher oder verschiedene Schreibweisen wie Meier statt Meyer). Diese zwar zur selben Person zugehörigen, jedoch sich unterscheidenden Daten, werden als Personenidentitäten bezeichnet. Aufgabe eines jeden Standorts ist es, etwaig vorhandene Personenidentitäten einer tatsächlichen Person korrekt zuzuordnen. Dies wird durch ein lokales Record Linkage realisiert und ist Voraussetzung für ein standortübergreifendes Record Linkage. Hierbei kommen die dem Standort ohnehin bekannten IDAT zum Einsatz. Die korrekte Personenidentität, also bei der z. B. durch Rückfragen festgestellt wurde, dass es keine Fehler enthält, wird als Hauptidentität bezeichnet. Die anderen Identitäten werden meistens als Nebenidentitäten bezeichnet und werden der Hauptidentität angefügt. So kann später auch bei abweichenden Schreibweisen die korrekte Person referenziert werden.

Für die Erstellung von Forschungsdatensätzen aus den Daten mehrerer Standorte bedarf es eines standortübergreifenden Record Linkages mit dem Ziel, alle medizinischen Daten (MDAT) einer Person so zu verknüpfen, dass deren zugehörige Teildatensätze unter einem einzigen, jedoch projektspezifischen Pseudonym im Forschungsdatensatz vorliegen. In diesem Sinne ist das Record Linkage der Erzeugung und Zuordnung von Pseudonymen vorangestellt. Die MII verfolgt die in Abbildung 9 dargestellten Konzepte T1, T3 und S1 zum standortübergreifenden / föderierten Record Linkage. Im Folgenden werden die Konzepte T1 und T3 detaillierter erläutert, da die grundlegenden Methoden bereits in mehreren Projekten praktisch eingesetzt werden. S1 ist ein wissenschaftlich interessanter Ansatz, dessen Praxistauglichkeit zunächst in der MII erforscht und evaluiert werden muss (Hampf et al., 2019).

### 3.9.2.1.1 T1: Föderiertes Record Linkage basierend auf Wahrscheinlichkeiten (PPRL)

Um bei standortübergreifenden Forschungsvorhaben datensparsam und datenschutzfreundlich zu arbeiten, sollen die technischen Personenidentitäten mittels eines fehlertoleranten Record Linkage verglichen werden, ohne dass dabei Rückschlüsse auf die Identität einer tatsächlichen Person gezogen werden können (Hampf et al., 2019). Für diese Form des Privacy-Preserving Record Linkage (PPRL) werden die IDAT vor dem Verlassen des Standorts mithilfe von Bloomfiltern (siehe Tabelle Anhang 2) kodiert und von einer übergeordneten föderierten Treuhandstelle (fTTP) wahrscheinlichkeitsbasiert verglichen. Es wird somit auf Basis von Wahrscheinlichkeiten ermittelt, ob die Datensätze zur selben Person zugeordnet werden können. Dabei wird anhand von definierten Schwellenwerten festgelegt, wie sich zwei Personen-Kandidaten hinreichend ähneln müssen, um einer tatsächlichen Person zugeordnet werden zu können oder ob es sich um zwei verschiedene Personen handelt. Voraussetzung ist hierbei, dass angebundene Standorte mit einheitlichen Verfahren die kodierten IDAT auf Basis der am Standort vorliegenden Klartext-IDAT erzeugen. Dies kann bei unterschiedlichen Schreibweisen (z. B. durch Tippfehler) der standortspezifischen Klartext-IDAT zu unterschiedlichen kodierten IDAT führen, die dann in der fTTP (fTTP-Wahrscheinlichkeit) einer tatsächlichen Person zugeordnet werden können.

### 3.9.2.1.2 T3: Föderiertes Record Linkage basierend auf Eindeutigkeit

Wenn beim Abgleich von kodierten IDAT keine eindeutige Personenzuordnung ermittelt werden kann, kann für diese Datensätze zusätzlich ein Abgleich auf Klartext-IDAT durchgeführt werden. Hierzu wird die fTTP um einen Clearing-Prozess erweitert, welcher in einer getrennten Komponente umgesetzt wird. Dieser ermöglicht die nachgelagerte Übermittlung von Klartext-IDAT, um Eindeutigkeit herzustellen (diese Komponente wird deshalb als fTTP-Clearing oder fTTP-Eindeutigkeit bezeichnet). Nur im Fall eines vorhergehenden wahrscheinlichkeitsbasierten Abgleichs mit mehrdeutigem Ergebnis, werden die IDAT an die fTTP-Eindeutigkeit übermittelt bzw. von der fTTP angefordert. Für den Zweck des Record Linkage werden die Klartext-IDAT nur temporär gespeichert und nach Auflösung möglicher Dubletten restlos und unwiederbringlich bei der fTTP gelöscht. Somit ist es möglich, einen manuellen Abgleich durchzuführen, ähnlich wie es in konventionellen Treuhandstellen der Fall ist. Somit können z. B. erforderliche Rücksprachen mit den Standorten gehalten werden, um Dubletten eindeutig auflösen zu können oder auf mögliche Eingabefehler hingewiesen zu werden. Der Clearingprozess ermöglicht damit Forschungsvorhaben, bei denen die geringe Anzahl an Teilnehmenden ein hohes Maß an Qualität erfordert, um belastbare Forschungsergebnisse zu erzielen. Insbesondere werden in diesem Prozess auch Synonymfehler verhindert, die zur Entstehung von Dubletten und zur Mehrfachverwaltung von Personen im Bestand führen (Hampf et al., 2019).

### 3.9.3 Ausblick

Im Rahmen der MII wurde bislang noch kein standortübergreifendes Record Linkage implementiert. Die Taskforce Datenschutz der MII erarbeitet schrittweise die Anforderungen für einzelne Anwendungsfälle. Im Fokus stehen dabei zunächst Machbarkeitsanfragen, verteilte Analysen und die Datenbereitstellung für Dritte. Da derzeit noch keine praktische Umsetzung eines MII-weiten Record Linkage existiert, können hierzu noch keine Bewertungen getroffen werden.

## 3.10 Use Case 10: Standortübergreifendes Record Linkage im Netzwerk Universitätsmedizin (NUM-RDP)

### 3.10.1 Hintergrund und Ziel

Zur Unterstützung der Bekämpfung der weltweiten Corona-Pandemie wurde das Netzwerk Universitätsmedizin (NUM) gegründet, welches durch das Bundesministerium für Forschung und Bildung gefördert ist. Das Ziel von NUM war und ist die COVID-19-Forschung in den deutschen Universitätskliniken zu harmonisieren. Hierzu sollten Strukturen und Prozesse geschaffen werden, die die Kompetenzen aller Universitätskliniken in Deutschland bündeln[20]. Das NUM bestand initial aus 13 Teil-Projekten[21], die jeweils unterschiedliche Themen in klinikübergreifenden Forschungsvorhaben bearbeiten. Dabei ist das Ziel, belastbare Erkenntnisse zur Beantwortung dringender Forschungsfragen zu gewinnen. Das Teil-Projekt CODEX (COVID-19 Data Exchange Platform) (Prokosch et al., 2022) hatte die Realisierung einer zentralen Plattform zum Ziel, welche medizinische Daten in einer strukturierten Form an Forschende bereitstellt. Das Projekt wurde Ende 2021 abgeschlossen und wird im Folgeprojekt NUM-RDP (Routine Data Platform) (Heyder et al., 2023) basierend auf den bereits geschaffenen Strukturen erweitert. NUM-RDP ist strukturell in dezentrale Komponenten (NUM-Knoten) und zentrale Komponenten organisiert, die die Datenbereitstellung und -nutzung unterstützen. Mithilfe der NUM-Knoten werden an den jeweiligen Universitätskliniken Infrastrukturen und zugehörige Applikationen für die lokale Datenbereitstellung zur Verfügung gestellt. Die NUM-Knoten bauen hierbei auf bereits existierenden Infrastrukturen der Medizininformatik-Initiative (MII) auf. In der MII wurden hierzu an fast allen deutschen Universitätskliniken lokale Treuhandstellen und Datenintegrationszentren (DIZ) eingerichtet. Die zentrale RDP-Plattform stellt Forschenden umfangreiche Forschungsdaten in Form von standardisierten GECCO-Datensätzen (Sass et al., 2020) zu COVID-19 bereit. GECCO (German Corona Consensus Dataset) ist ein standardisierter Datensatz zur Erfassung von COVID-19 relevanten Forschungsdaten[22]. Die zur Verfügung gestellten Daten werden in NUM-RDP um Routinedaten erweitert, sodass künftig nicht nur Daten für die Corona-Forschung bereitstehen. Für die Übertragung der Forschungsdatensätze von den NUM-Knoten zur zentralen Plattform wird eine Komponente für den Datentransfer (NUM Transfer Hub) dazwischengeschaltet. Eine standortübergreifende und einheitliche Pseudonymisierung wird von der federated Trusted Third Party (fTTP, föderierte Treuhandstelle) implementiert. Diese ermöglicht auch eine zusätzliche Pseudonymisierungsstufe beim Datentransfer und bei Datenexporten an Forschende. Um die medizinischen Daten (MDAT), die an mehreren

---

20 https://www.netzwerk-universitaetsmedizin.de/aufgaben-und-ziele, Zugriff 24.11.2021

21 https://www.netzwerk-universitaetsmedizin.de/projekte, Zugriff 15.08.2021

22 https://www.bihealth.org/de/forschung/wissenschaftliche-infrastruktur/core-facilities/interoperabilitaet/home/gecco, Zugriff 13.03.2022

Standorten zu einer Person gesammelt worden sind, zusammenführen zu können, implementiert die fTTP ein Privacy-Preserving Record Linkage. Die rechtliche Grundlage für eine Datenübermittlung für Forschungszwecke ist der Informed Consent der zu behandelnden Person. Diese Einwilligung wird an dem jeweilig Standort eingeholt und verwaltet. Die Basis der Einwilligung stellt der MII Broad Consent dar [23]. Im Fall eines späteren Widerrufs, werden die in der Plattform vorliegenden Daten, sowie in der fTTP vorliegenden kodierten IDAT gelöscht. Entsprechende Prozesse werden im Rahmen von NUM-RDP bereits eingesetzt.

### 3.10.2 Technische Umsetzung des föderierten Record Linkages

Um Patient:innen an den teilnehmenden Standorten standortübergreifend eindeutig identifizieren und die entsprechenden MDAT miteinander verknüpfen zu können, wurden Anforderungen der MII für das föderierte Record Linkage in NUM-CODEX und -RDP technisch implementiert und spezifiziert. Dabei spielt eine föderierte Treuhandstelle (fTTP) eine wesentliche Rolle. Die fTTP besteht aus den zwei Komponenten fTTP-Wahrscheinlichkeit und fTTP-Eindeutigkeit / Clearing, deren grundsätzliche Arbeitsweisen im Abschnitt 3.9.2.1.1 bereits beschrieben sind. Das Ziel ist es, die MDAT einer Person über Standorte hinweg eindeutig zuzuordnen, um anschließend eine Datenzusammenführung zu ermöglichen. Die Speicherung des Verbundforschungsdatensatz im GECCO-Format bzw. der strukturierten Routinedaten erfolgt in der zentralen Plattform.

Zur Unterstützung der informationellen Gewaltenteilung werden für die Datenbereitstellung in der NUM-Infrastruktur IDAT und MDAT stets voneinander getrennt gehalten. Dafür werden an jedem Standort die IDAT in einer lokalen Treuhandstelle (THS) und die MDAT im lokalen DIZ verwaltet. Wie in der MII sind auch diese organisatorischen Aufteilungen eine Voraussetzung in NUM-CODEX und NUM-RDP. Mit dem Ziel, die MDAT zu verknüpfen bzw. zu pseudonymisieren, müssen diese, mit den entsprechenden DIZ-Pseudonymen versehen, zunächst von den Standorten an den NUM Transfer Hub (NTH) übermittelt werden (vgl. Abbildung 10A, gekennzeichnet als GECCO Transfer Hub). Der NTH führt mithilfe der fTTP eine Um-Pseudonymisierung durch, indem das mitgelieferte standortspezifische Pseudonym (DIZ-Pseudonym) durch das standortübergreifende Pseudonym ersetzt wird. Dadurch werden die MDAT einer Person von verschiedenen Standorten mit einem einheitlichen standort-

---

übergreifenden Pseudonym versehen und anschließend durch den NTH an die zentrale Plattform übermittelt. Der Standortbezug der einzelnen MDAT-Datensätze kann allein anhand der übergreifenden Pseudonyme nicht ermittelt werden. Wie in Abbildung 10A gezeigt, übermittelt ein DIZ-Standort die kodierten IDAT in Form eines Bloomfilters an die fTTP. Diese führt daraufhin ein probabilistisches Privacy-Preserving Record Linkage durch und liefert dem Standort ein entsprechendes DIZ-Pseudonym zurück. Das angewendete Verfahren zur Generierung der Bloomfilter muss bei allen Standorten (lokale THS) gleich sein, damit später die fTTP ein projektbezogenes PPRL auf Basis von allen zuvor übermittelten Bloomfiltern (auch anderer Standorte) durchführen kann. Um die Standorte zu unterstützen, werden von der fTTP entsprechende Tools (z. B. das durch die Unabhängige Treuhandstelle Greifswald bereitgestellte Identitätsmanagement E-PIX®) zur Verfügung gestellt. Die kodierten IDAT werden dabei projektbezogen in der fTTP gespeichert, um künftige Abgleiche durchführen zu können. Sowohl das Record Linkage als auch das Speichern des Bloomfilters erfolgt mit dem Tool E-PIX® in der fTTP, das um die Funktionalität des Bloomfilter-Vergleichs erweitert wurde (siehe Tabelle Anhang 2). Ein Abgleich kann dabei in Abhängigkeit der verwendeten Schwellwerte folgende Ergebnisse liefern (Hampf et al., 2020):

- **Perfekte Übereinstimmung:** Zwei kodierte IDAT sind identisch und haben wahrscheinlich dieselben IDAT zur Grundlage. Da die kodierten IDAT bereits vorliegen, werden diese nicht erneut gespeichert.

- **Automatische Zusammenführung:** Zwei kodierte IDAT weisen geringfügige Abweichungen auf (beispielsweise wegen Tippfehlern). Die verglichenen kodierten IDAT werden derselben Person als Personenidentitäten zugeordnet.

- **Mögliche Übereinstimmung:** Zwei kodierte IDAT weisen Ähnlichkeiten auf, jedoch ist die Abweichung zu groß, um eine zweifelsfreie Zuordnung zur selben Person zu erlauben. Hier kann ein nachgelagerter Abgleich von Klartext-IDAT erforderlich sein, wozu die Übermittlung der Klartext-IDAT an die fTTP für das Clearing erfolgen muss.

- **Nicht-Übereinstimmung:** Zwei kodierte IDAT weisen keine oder nur eine geringfügige Ähnlichkeit auf, sodass die IDAT zu zwei unterschiedlichen Personen gehören.

Können den kodierten IDAT im Vergleich keine in der fTTP hinterlegten kodierten IDAT zugeordnet werden, wird ein neuer Eintrag angelegt, ein neues standortübergreifendes Pseudonym und ein neues standortbezogenes DIZ-Pseudonym generiert. Letzteres wird dem Standort übermittelt und dort hinterlegt. Das standortübergreifende Pseudonym wird lediglich in der zentralen Plattform verwendet und niemals an einen Standort übermittelt. Diese Pseudonym-Hierarchie verhindert (1), dass die zentrale Plattform anhand des Pseudonyms Rückschlüsse auf die besuchten Standorte einer Person schließen kann und (2), dass die Standorte unautorisiert die personenbezogenen Daten untereinander abgleichen können. Wird hingegen eine Determinierung der zuzuordnenden Person auf Basis der verschiedenen kodierten IDAT aus verschiedenen Standorten festgestellt (perfekte Übereinstimmung, automatische Zusammenführung), werden standortspezifische DIZ-Pseudonyme und ein einheitliches standortübergreifendes Pseudonym generiert. Während die DIZ-Pseudonyme den Standorten übermittelt werden, wird das standortübergreifende Pseudonym für anschließende Um-Pseudonymisierung der entsprechenden Datensätze in der fTTP hinterlegt. Dieser Prozess ist beispielhaft für den Abgleich der von zwei Standorten übermittelten kodierten IDAT mit anschließender Pseudonymvergabe in Abbildung 10B dargestellt.

Darüber hinaus kann in bestimmten Fällen der Vergleich eine mögliche Übereinstimmung der Personenidentitäten ergeben. Infolgedessen werden die kodierten IDAT zunächst so gespeichert, als wären sie einer separaten Person zugeordnet. Dafür könnten grundsätzlich voneinander abweichende Bloomfilter-Kodierungen durch unterschiedliche Ausprägungen der IDAT (z. B. wegen Tippfehler, geänderter Name, verschiedene Schreibweisen) zu Grunde liegen. In diesen Fällen kann ein exaktes Record Linkage durch die fTTP-Clearing erfolgen (siehe Abbildung 10A). Dies ist beispielsweise für die Forschung zu seltenen Erkrankungen besonders wichtig, da diese Patient:innen häufig mehrere Standorte aufsuchen bis sie eine Diagnose erhalten. Aufgrund der geringen Anzahl an Patient:innen kann es zudem entscheidend sein, dass für einzelne Patient:innen ein eindeutiges Record Linkage erfolgt. Die fTTP-Clearing ermöglicht zudem, Rücksprache mit den Standorten zu halten und ggf. fehlerhafte Daten zu korrigieren. Nur in diesen Einzelfällen senden die Standorte, die zuvor die nicht eindeutig zuzuordnenden kodierten IDAT übermittelt haben, die dazugehörigen Klartext-IDAT (klassischerweise bestehend aus Vornamen, Nachname, Geburtsdatum, Geschlecht und ggf. weiteren Daten wie Geburtsname oder Adressdaten) an die fTTP-Clearing.

Das Ergebnis dieses Abgleichs wird von der fTTP-Clearing an die fTTP-Wahrscheinlichkeit übermittelt und entsprechend den kodierten IDAT zugeordnet. Die an die fTTP-Clearing übermittelten Klartext-IDAT werden unwiederbringlich nach dem Vorgang des Record Linkage gelöscht.

**A**

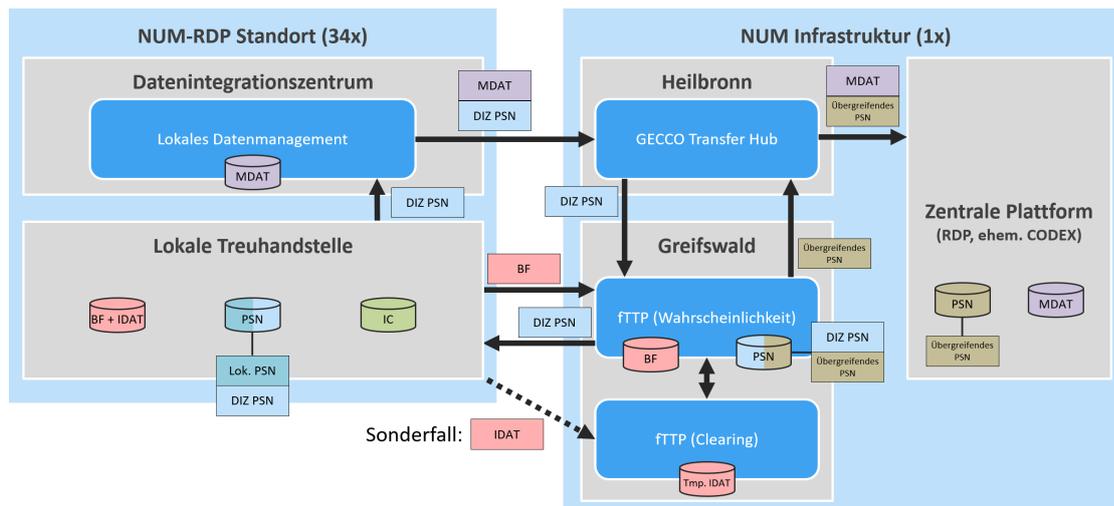

**B**

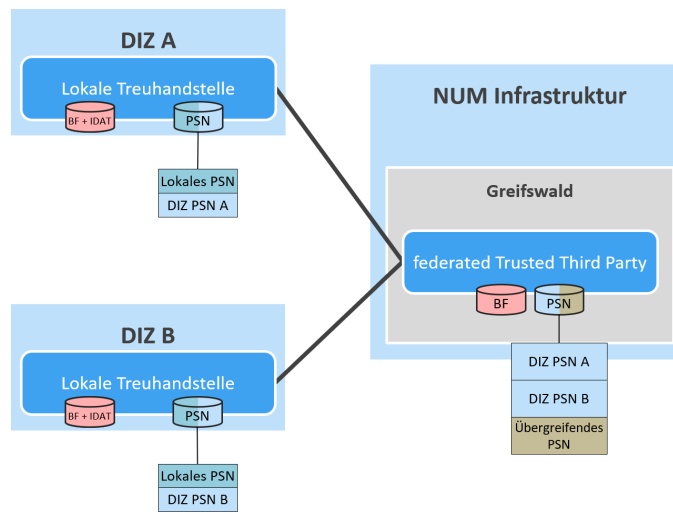

Abbildung 10: Infrastrukturkomponenten NUM-RDP
**A** Vereinfachter schematischer Aufbau eines NUM-RDP-Standorts und der NUM-Infrastruktur (DIZ: Datenintegrations-zentrum, MDAT: medizinische Daten, BF: Bloomfilter, PSN: Pseudonym, IC: Informed Consent, fTTP: federated Trusted Third Party); **B** Vergabe von Pseudonymen durch die fTTP. DIZ A und DIZ B haben für dieselbe Person jeweils standort-spezifische DIZ-Pseudonyme erhalten. Die Zusammenführung über das standortübergreifende Pseudonym ist nur der fTTP bekannt (DIZ: Datenintegrationszentrum, BF: Bloomfilter, IDAT: identifizierende Daten, PSN: Pseudonym, fTTP: federated Trusted Third Party)

### 3.10.3 Ausblick und Herausforderungen

Von Ende März 2021 bis Ende 2022 wurden sukzessiv alle NUM-Standorte an die NUM Infrastruktur angebunden und die Übermittlung von Testdaten in der Praxis erprobt[24]. Der Produktivbetrieb ist im Mai 2022 gestartet. Bis Juli 2023 wurden nur wenige hundert Patient:innen durch die Standorte registriert. Durch diese geringe Anzahl an Personen-datensätzen, sind noch keine nicht-eindeutigen Übereinstimmungen aufgetreten, sodass ein Clearing bisher nicht erforderlich war. Die aufgebaute Infrastruktur wird perspektivisch als Grundlage für andere MII-Teilprojekte dienen. Erprobte Prozesse und Erfahrungen können so wieder zurück in die MII gespielt werden.

---

24  https://www.ths-greifswald.de/erfolgreiche-live-demo-der-foederierten-treuhandstelle-fttp-in-num-codex/, Zugriff 23.11.2021

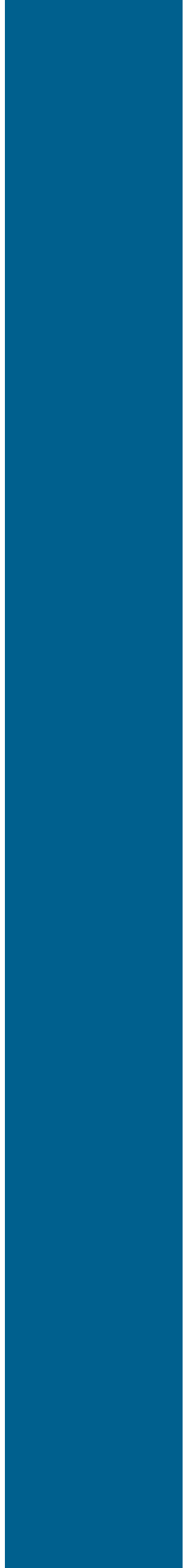

# 4 Allgemeine Vorgehensweise beim Record Linkage

## 4.1 Prinzipien des Record Linkage am Beispiel der Use Cases

In den Use Cases wurde gezeigt, auf welch unterschiedliche Weise und mit welchem Aufwand Forschende Record Linkage betreiben und mit welchen Herausforderungen und Problemen sie dabei konfrontiert werden, um dringende und komplexe Fragen der Gesundheitsforschung bearbeiten zu können. Eine Übersicht der diskutierten Use Cases bezüglich der organisatorischen Einordnung und der darin verwendeten Record Linkage-Verfahren ist in Tabelle 3 dargestellt.

In diesen geschilderten Use Cases werden mehrheitlich direkte Identifikatoren für die Umsetzung des Record Linkage eingesetzt. Diese Identitätsdaten (IDAT wie z. B. Name, Vorname, Geburtsdatum, Geschlecht und Adresse) werden entweder kodiert oder im Klartext verglichen. Der Abgleich von direkten, unverschlüsselten Identifikatoren und der damit verbundenen Verknüpfung von Daten einer Person auf Basis der exakten Übereinstimmung („exaktes Linkage") ist in lokalen klinischen Systemen gängige Praxis, wie in Use Case 5 geschildert. Aber auch außerhalb des lokal-klinischen Kontextes werden Daten auf diese Weise verlinkt. So werden in den beiden in Use Case 1 beschriebenen Projekten (CoVerlauf und SHIP) ebenfalls auf den exakten Vergleich von direkten Identifikatoren beruhende Verfahren verfolgt. Bei diesem Linkage-Verfahren erfolgt nur dann eine Verknüpfung, wenn die zu vergleichenden Identifikatoren exakt übereinstimmen. Fehler und unterschiedliche Schreibweisen der Identifikatoren werden nicht berücksichtigt. Im Gegensatz zu CoVerlauf, wo nur ein Identifikator Verwendung findet (KVNR), wurde in SHIP determiniert, welche Teilsätze der IDAT übereinstimmen müssen (u. a. phonetisch standardisierte Namen).

Die meisten der hier vorgestellten Use Cases verwenden probabilistische Record Linkage-Ansätze. Aufgrund der Fehlertoleranz dieser Methode muss jedoch die Notwendigkeit manueller Überprüfung der Zuordnungen in Betracht gezogen werden. In diesem Zusammenhang kann abhängig von den Qualitätsansprüchen und den möglichen negativen Auswirkungen auf die Analyse der Forschungsfrage ein aufwendiger händischer Vergleich der IDAT vonnöten sein (Hampf et al., 2019). Der probabilistische Ansatz kann entweder auf Klartext-IDAT (Use Case 6 „nNGM") oder auf kodierten IDAT (siehe Use Cases 2, 3, 4, 8, 9) organisiert werden. Die kodierten IDAT besitzen Eigenschaften, die die Bestimmung von Ähnlichkeiten bzw. Distanzen zwischen verschiedenen Ausgangsdatensätzen (IDAT) zulassen.

Die entsprechenden Linkage-Verfahren basieren auf dem Abgleich von Kontroll-nummern (siehe Use Cases 2 und 3) oder von in Bloomfilter tran formierten IDAT (siehe Use Case 7 DKTK, Use Case der MII sowie der künftige Use Case der MII). Damit verlassen keine IDAT im Klartext den Ort der Datenquelle (Forschungsdatengeber), das Verfahren entspricht insofern den Vorgaben des Privacy-Preserving Record Linkage (PPRL).

Im Allgemeinen wird in den Use Cases bei der Umsetzung des Record Linkage die informationelle Gewaltenteilung berücksichtigt. In diesem Sinne werden z. B. die nötigen Forschungsinfrastrukturen in unabhängige Teilstrukturen bzw. Module aufgeteilt und / oder es wird ein Identitätsmanagement eingesetzt, das weisungsfrei von den einzelnen Institutionen des Forschungsvorhabens agieren kann. Das Identitätsmanagement wurde in den meisten Use Cases über die Einbindung einer zentralen, unabhängigen und vertrauenswürdigen Stelle (Treuhandstelle oder Trusted Third Party) realisiert. Eine zentrale Einordnung dieser legitimierten Institution innerhalb des Forschungsverbunds ermöglicht die Gewährleistung eines hohen Sicherheitsstandards (Pommerening et al., 2014). Im Datenschutzleitfaden der TMF wird der Treuhänder als Einrichtung oder Person definiert, die zwischen Forschungs-datengebern und der Nachfrageseite von Forschungsdaten als eine unabhängige und neutrale Vertrauensinstanz sicher und gesetzeskonform operiert (Pommerening et al., 2014). Die informationelle Gewaltenteilung wird über geregelte Autorisierungen auf verschiedenen Ebenen der Datenbestandsverwaltung garantiert. Dabei werden die für die Forschung relevanten Gesundheitsdaten von den direkt personenidentifizierenden Daten separat und unabhängig voneinander administriert. Hierbei wird auf die elektronisch geführte und beim Treuhänder verortete Patienten-/Probandenliste, in der der Bezug der identifizierenden Daten (IDAT) zu Pseudonymen hinterlegt ist, großen Wert gelegt. Demzufolge werden die Verantwortlichkeiten einer Treuhandstelle üblicherweise u. a. in der Führung der Patienten-/Probandenliste und der Umsetzung der Pseudonymisierungsstrategien gesehen. Der Zugang und die Administration dieser Aufgabenbereiche sollten unabhängig, verfahrenstransparent und vertrauenswürdig umgesetzt werden (Kaulke et al., 2020).

Auch die in den Use Cases beschriebenen Verfahren basieren auf derartigen zentralen Linkage-Einheiten (Treuhandstellen). Dabei können sie entweder im Kontext des PPRL dem Abgleich von normierten, vergleichbaren und menschenunlesbaren IDAT (kodierten IDAT) oder aber dem Abgleich von Klartext-IDAT dienen (unkodierte IDAT), die vom Datengeber zur

Verfügung gestellt werden. Potenzielle Doppler werden auf dieser Basis erkannt und aufgelöst. Die Verknüpfung der Identitäten ist dabei mit der Pseudonymzuweisung eng verzahnt.

Im Rahmen klinischer Forschungsinfrastrukturen, bei denen eine Verknüpfung klinischer Daten aus mehreren Standorten erfolgt (vgl. Use Case 7, Use Case NUM-RDP und der künftige Use Case der MII), werden klinische Daten zunächst auf lokaler Ebene zusammengeführt. Bei dem sogenannten föderierten Record Linkage unter Verwendung von federated Trusted Third Parties (fTTP) erfolgt keine dauerhafte einrichtungsübergreifende Speicherung personen-identifizierender Informationen an zentraler Stelle. Auf Basis von beispielsweise Bloom-filtern, die lokal an den Standorten berechnet werden, kann ohne Kenntnis der ursprünglichen identifizierenden Daten ein einrichtungs- und datensatzübergreifendes Record Linkage vorgenommen werden. Somit kann über Einrichtungs- und Datensatz-Grenzen hinweg erkannt werden, ob Daten aus verschiedenen Quellen zu ein und derselben Person gehören. Beim föderierten Record Linkage werden identifizierende Daten (IDAT) nur in den lokalen Treuhandstellen verwaltet und verlassen nicht den Standort, an dem sie gewonnen wurden.

## 4.2    Allgemeine Herausforderungen beim Record Linkage in den Use Cases

Allen Anwendungsfällen ist gemeinsam, dass die Fehlerraten für Homonyme und insbesondere für Synonyme nicht unbeträchtlich sind. Der Use Case 4 „DESH" sticht hierbei mit einer Synonymfehlerquote von 57,6% heraus – allerdings können auch in anderen Use Cases Forschungsergebnisse durch erhöhte Fehlerquoten verzerrt werden. Die Folgen der Fehler beim Record Linkage werden in Use Case 2 „DFG-Linkage" demonstriert, bei dem gezeigt wurde, dass das Krebsrisiko aufgrund dieser Fehler deutlich unterschätzt wird. Daher ist je nach Szenario eine Risikoabschätzung bezüglich der Fehlerart notwendig. Diese Abschätzung resultiert beispielsweise bei krankheitsbezogenen Registern in erhöhten Personalbedarfen (Brenner & Schmidtmann, 1998). Dies gilt insbesondere für die Krebsregister, die einen hohen Aufwand betreiben, um durch manuelle Nachbearbeitung das Record Linkage-Ergebnis zu verbessern (siehe Use Case 3 „Krebsregister"). Die manuelle Nachbearbeitung ist aber nicht immer durchführbar (Use Case 6 „nNGM"). Eine Möglichkeit zur Verbesserung kann die Verwendung der Krankenversichertennummer (KVNR) als direkten Identifikator darstellen. Einerseits kann die KVNR der gesetzlichen Krankenversicherung (GKV) aufgrund ihrer Eindeutigkeit und lebenslangen Gültigkeit Identitätsverwechselungen im Linkage-Prozess vorbeugen, andererseits ist diese jedoch bei den privaten Krankenversicherungen (PKV) nicht etabliert. Überdies ist die KVNR in manchen Datenquellen gar nicht oder nicht immer vorhanden oder darf dort nicht gespeichert werden. Zudem gibt es weitere Personengruppen ohne KVNR (z. B. Asylbewerber:innen und Selbstzahlende). Neben dem Aufwand für die manuelle Nachbearbeitung stellt auch der Aufwand für die Einholung von Genehmigungen eine Herausforderung bei Record Linkage-Projekten dar. Dies gilt insbesondere, wenn Datenquellen über mehrere Dateneigner verteilt vorliegen und unter unterschiedliche rechtliche Regelungen fallen – wie dies bei Landeskrebsregisterdaten oder Abrechnungsdaten aus regionalen oder überregionalen sowie privaten und gesetzlichen Krankenkassen der Fall ist. Erschwerend kommt hinzu, dass bei Anwendungsfällen, die als Rechtsgrundlage für das Record Linkage informierte Einwilligungen von Patient:innen oder Proband:innen benötigen, oft eine reduzierte Einwilligungsquote bei der nachträglichen Einholung derselbigen zu verzeichnen ist. Für einen Überblick über die gesetzlichen Regelungen sei auf Kapitel 5 verwiesen.

Insgesamt ist festzuhalten, dass das Record Linkage für die Gesundheitsforschung in Deutschland ein aufwendiges Unterfangen darstellt, bei dem den Beteiligten je nach Einsatzbereich, rechtlicher Grundlage, Datenschutzmodell und verfügbaren Identifikatoren aus verschiedenen Verfahrensoptionen auswählen müssen. So müssen für jeden Anwendungsfall und jedes Record Linkage-Projekt einzelfallspezifische Lösungen entwickelt, geprüft, ggf. modifiziert und – falls positiv beschieden – umgesetzt werden. Dabei ist die sorgsame Wahl des Verfahrens mit direkten oder indirekten Identifikatoren eine Voraussetzung für erfolgreiches Record Linkage. In der Praxis ist diese Wahl zahlreichen rechtlichen, organisatorischen, personellen und zeitlichen Restriktionen unterworfen, sodass erheblicher Verbesserungsbedarf beim Record Linkage für die Gesundheitsforschung besteht.

Tabelle 3: Einordnung der in Kapitel 3 diskutierten Use Cases hinsichtlich Rechtsgrundlage, Datenschutz, Identifikatoren, Linkage-Verfahren und Herausforderungen

| Use Case | Rechtliche Grundlage | Datenschutz Modell | Umsetzung des Datenschutz | Identifikatoren | Linkage – Verfahren | Herausforderungen und Probleme |
|---|---|---|---|---|---|---|
| **Standortübergreifende Forschungsinfrastrukturen mit Primär- und Sekundärdaten** | | | | | | |
| 1a (CoVerlauf) | Einwilligung | informationelle Gewaltenteilung | zentrale, rechtlich, räumlich und personell von der Forschung getrennte Treuhandstelle | kodierte IDAT | exaktes Linkage | Höhe der Einwilligungsquote; Kooperationsaufwand durch hohe Anzahl an GKVen und PKVen; rechtlicher und inhaltlicher Unterschied zwischen GKV- und PKV-Daten |
| 1b (SHIP) | Einwilligung | informationelle Gewaltenteilung | räumlich und personell getrennte Datengeber und Datenempfänger | unkodierte IDAT | deterministisches bzw. exaktes Linkage | Höhe der Einwilligungsquote, insbesondere bei nachträglicher Einholung; Ausschluss einzelner Versicherter zur absoluten Vermeidung von falschen Verknüpfungen |
| **Standortübergreifendes Record Linkage mit Registerdaten** | | | | | | |
| 2a (DFG-Linkage) | § 75 SGB X und länderspezifische Krebsregistergesetze | informationelle Gewaltenteilung | zentrale, rechtlich, räumlich und personell von der Forschung getrennte Treuhandstelle | kodierte IDAT | probabilistisches Linkage | Linkage mit Kontrollnummern für Daten der Krankenversicherung nur teilweise passend; KVNR nicht bzw. mittlerweile nur teilweise verfügbar; zusätzliche Verwendung von KVNR zur Plausibilitätsprüfung nötig; rechtliche und organisatorische Herausforderungen bei der Einbeziehung von mehreren GKVen und Krebsregistern aus mehreren Bundesländern |
| 2b (DFG-Linkage) | § 75 SGB X und länderspezifische Krebsregistergesetze | informationelle Gewaltenteilung | Linkage auf indirekten Identifikatoren ohne den Einsatz einer Treuhandstelle | keine IDAT, indirekte Identifikatoren | deterministisches Linkage | weniger sensitives Linkage im Vergleich zu 2a und Verzerrung der Ergebnisse; rechtliche und organisatorische Herausforderungen bei der Einbeziehung von mehreren GKVen und Krebsregistern aus mehreren Bundesländern |

Tabelle 3: Einordnung der in Kapitel 3 diskutierten Use Cases hinsichtlich Rechtsgrundlage, Datenschutz, Identifikatoren, Linkage-Verfahren und Herausforderungen

| | | | | | | |
|---|---|---|---|---|---|---|
| 3 (Krebsregister) | Landeskrebs-registergesetze und Einwilligung | informationelle Gewaltenteilung | zentrale, rechtlich, räumlich und personell von der Forschung getrennte Treuhandstelle | kodierte IDAT | probabilistisches Linkage | Fehlerrate beim Linkage bei größeren Datenmengen ansteigend; große Mengen manueller Nachbearbeitung im Rahmen von Kohortenabgleichen; eingeschränkte Nutzung der KVNR; verwendete Phonetik nicht für Variation der Namen in Deutschland passend |
| 4 (DESH) | Verordnung des BMG | informationelle Gewaltenteilung | zentrale, rechtlich, räumlich und personell von der Forschung getrennte Treuhandstelle | kodierte IDAT | probabilistisches Linkage | hohe Fehlerrate; Homonymfehlerrate nicht ermittelbar; zeitliche und personelle Ressourcenknappheit in Laboren (insbesondere während der Pandemie); rechtliche Rahmenbedingungen für Aufbau einer Treuhandstelle |
| **Standortübergreifende Forschungsinfrastrukturen mit klinischen Daten** | | | | | | |
| 6 (nNGM) | Einwilligung | informationelle Gewaltenteilung | zentrale, rechtlich, räumlich und personell von der Forschung getrennte Treuhandstelle | unkodierte IDAT und zur Plausibili-tätsprüfung Ver-sichertennummer (KVNR) | probabilistisches Linkage | Verfügbarkeit der IT-Systeme; keine einheitliche Versichertennummer; unterschiedliche Landesgesetzgebungen und ihre Auslegung; manuelle Fehler und manueller Aufwand bei nachträglicher Datenverknüpfung/ Trennung |
| **Standortinterne Verknüpfung klinischer Daten** | | | | | | |
| 5 (Universitäts-kliniken) | Landeskranken-hausgesetze | eingeschränkte informationelle Gewaltenteilung | zentrale, räumlich und personell von der Forschung getrennte Treuhandstelle | unkodierte IDAT | exaktes Linkage | – |

Tabelle 3: Einordnung der in Kapitel 3 diskutierten Use Cases hinsichtlich Rechtsgrundlage, Datenschutz, Identifikatoren, Linkage-Verfahren und Herausforderungen

| Verteilte Erschließung von klinischen Sekundärdaten | | | | | | |
|---|---|---|---|---|---|---|
| 7 (DKTK) | Einwilligung | informationelle Gewaltenteilung (verteilte TTP-Services mit Brückenköpfen) | zentrale, rechtlich, räumlich und personell von der Forschung getrennte Treuhandstelle | lokal: unkodierte IDAT; zentral: kodierte IDAT | probabilistisches Linkage | technische und personelle Ausstattung der Standorte; Heterogenität der Daten bei den Quellsystemen; unterschiedliche Landesgesetzgebungen und ihre Auslegung; muss auch Bestandsdaten ohne Patienteneinwilligungen unterstützen |
| **Verteilte Erschließung von klinischen Sekundärdaten in der MII und NUM-RDP** | | | | | | |
| MII | Einwilligung und Landeskrankenhausgesetze | informationelle Gewaltenteilung (verteilte TTP-Services) | zentrale rechtlich, räumlich und personell von der Forschung getrennte Treuhandstelle | kodierte IDAT | probabilistisches Linkage | Use Case noch nicht in der Praxis |
| NUM-RDP | Einwilligung und Landeskrankenhausgesetze | informationelle Gewaltenteilung (verteilte TTP-Services) | zentrale, rechtlich, räumlich und personell von der Forschung getrennte Treuhandstelle | fTTP- Wahrscheinlichkeit: kodierte IDAT; fTTP-Eindeutigkeit: unkodierte IDAT | probabilistisches Linkage | Use Case in Testung |

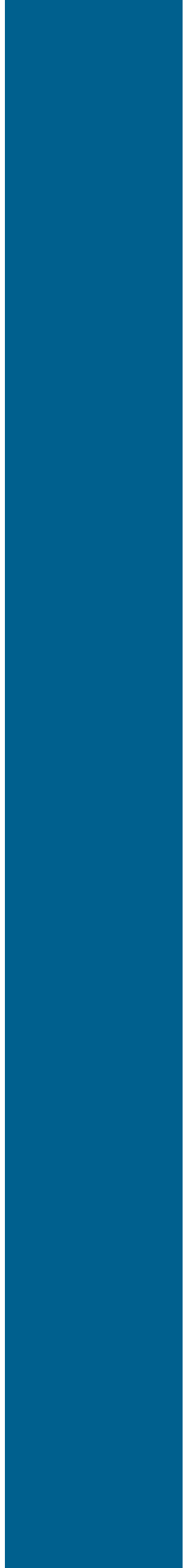

# 5 Rechtliche Möglichkeiten und Anforderungen zur Verknüpfung personenbezogener Daten zu Forschungszwecken

## 5.1 Rechtliche Rahmenbedingungen für die Durchführung von Record Linkage

Im Rahmen der wissenschaftlichen Forschung sind für die Verwendung von Sozialdaten bzw. Gesundheitsdaten umfassende datenschutzrechtliche Vorgaben zu beachten. Die Nutzung einzelner sowie insbesondere die Verknüpfung unterschiedlicher Datenquellen (insb. Primär- und Sekundärdaten) unterliegt einem komplexen datenschutzrechtlichen Regelungsregime, das überdies Besonderheiten im Hinblick auf die wissenschaftliche Forschung enthält, in deren Rahmen sich deshalb auch besondere organisatorische Herausforderungen stellen (March et al., 2015).

### 5.1.1 Grundlagen, Datenkategorien und Use Cases

Zur Legitimation der Datenverarbeitung einschließlich des Record Linkage innerhalb der Use Cases können entweder die datenschutzrechtliche Einwilligung der betroffenen Person oder ein gesetzlicher Erlaubnistatbestand herangezogen werden. Für die einzelnen Use Cases sind dabei je nach Verarbeitungssituation und Datenkategorie unterschiedliche Rechtsgrundlagen relevant, die im Folgenden anhand der datenschutzrechtlichen Legitimationsinstrumente dargestellt werden.

### 5.1.2 Datenschutzrechtliche Einwilligung

Nach wie vor zentrales Instrument zur Legitimation und Durchführung des Record Linkage von personenbezogenen Daten ist die Einwilligung der betroffenen Person, weil gesetzliche Erlaubnistatbestände rar und nicht selten mit hohen Abstimmungsaufwänden verbunden sind, sobald landes- oder sektorübergreifende Datenquellen verlinkt werden sollen. So kann die Einwilligung zur Erhebung von Primärdaten, zur Erhebung der Krankenversicherungsnummer (KVNR) und der Krankenversicherung, der Übermittlung von Versichertendaten/Sekundärdaten und zur Verknüpfung von Daten in Abhängigkeit der Use Cases verwendet werden. Dementsprechend kann die Einwilligung im Ergebnis auch bei allen Use Cases als Erlaubnisgrundlage für die Datenverarbeitung herangezogen werden:

- Linkage von Primärdaten und Krankenkassendaten (Use Case 1)

- Linkage von Krankenkassen- und Landeskrebsregisterdaten (Use Case 2)

- Linkage von Primär- und Krebsregisterdaten (Use Case 3)

- Linkage von Melde- /Labordaten nach Infektionsschutzgesetz (Use Case 4)

- Linkage von Daten aus Routinedaten und -proben und klinischen Erhebungen innerhalb eines rechtlichen Trägers (Use Case 5)

- Linkage von Daten aus Routinedaten und -proben und klinischen Erhebungen über mehrere rechtliche Träger in nNGM (Use Case 6)

- Standortübergreifendes (föderiertes) Record Linkage im DKTK (Use Case 7)

Wird die Verarbeitung zu Forschungszwecken auf die Einwilligung gestützt, muss neben der Freiwilligkeit der Einwilligung sichergestellt werden, dass die betroffene Person diese jederzeit widerrufen kann. Daneben muss die Einwilligung nach dem Grundsatz der Transparenz informiert erfolgen. Problematisch kann der Zeitpunkt der Einholung der Einwilligung sein, denn insbesondere bei der Sekundärnutzung von Daten ist zum Zeitpunkt der Erhebung häufig noch nicht bekannt, dass diese Daten im Rahmen von Forschungsprojekten mit Daten aus anderen Quellen verlinkt werden sollen. Um wiederholte Einwilligungsprozesse zu vermeiden, die nicht nur zu erheblichen Aufwänden bei Forschenden, sondern nicht selten auch zu einer sogenannten Einwilligungsfatigue bei Patient:innen führen, wird zunehmend ein sogenannter Broad Consent verwendet, der gewisse Verlinkungen bereits bei der erstmaligen Einwilligung anlässlich der Datenerhebung bzw. Datenumwidmung vorsieht (ein solcher Broad Consent wird bereits bei der MII und der NAKO eingesetzt). Dieses Vorgehen ist jedoch nach wie vor ebenfalls mühsam, weil sich der Standard der MII noch nicht allgemein bzw. überall durchgesetzt hat. Eine oftmals geforderte Alternative zur Einwilligung im Sinne einer Widerspruchsmöglichkeit einer Datenverarbeitung, namentlich Opt-out, ist im Forschungs-kontext nach jetziger Rechtslage nicht umsetzbar. Die DSGVO fordert die aktive und eindeutige informierte Zustimmung einer betroffenen Person ein, insbesondere bei der Verarbeitung be-sonderer Kategorien personenbezogener Daten, und nicht die aktive Ablehnung einer solchen Datenverarbeitung.

Für die Verarbeitung von Sozialdaten, die ebenfalls personenbezogene Daten sind (§ 67 Abs. 2 S. 1 SGB X), gelten gem. § 67b SGB X bereichsspezifische Anforderungen. Dies folgt aus der Vorschrift in § 35 Abs. 2 S. 1 SGB I, wonach die sozialrechtlichen Datenschutzvorschriften grundsätzlich abschließend gelten. Auch hier ist die Möglichkeit des Broad Consent vorgesehen, indem in § 67b Abs. 3 SGB X festgestellt wird, dass die Einwilligung zur Verarbeitung personenbezogener Daten zu Forschungszwecken für ein bestimmtes Vorhaben oder für bestimmte Bereiche der wissenschaftlichen Forschung erteilt werden kann. Eine besondere Vorschrift im Sozialrecht zur Übermittlung von Sozialdaten für die Forschung und Planung enthält mit Blick auf die Einwilligung der § 75 SGB X. Die Vorschrift, auf die im Folgenden im Rahmen der gesetzlichen Erlaubnistatbestände vertieft eingegangen wird, ist insoweit hervorzuheben, als sie gesetzlich festgelegte tatbestandliche Erfordernisse mit der Einwilligung kombiniert. Dabei gilt, dass eine Übermittlung ohne Einwilligung der betroffenen Person nicht zulässig ist, soweit es zumutbar ist, ihre Einwilligung einzuholen.

### 5.1.3 Gesetzliche Erlaubnistatbestände

### 5.1.3.1 Sozialdaten

Bereits im Rahmen des Einwilligungstatbestandes wurde die Vorschrift des § 75 SGB X als besondere Regelung aus dem Sozialdatenschutz skizziert, da sie gesetzliche Anforderungen mit der Einwilligung verknüpft. Demgemäß ist die Übermittlung von Sozialdaten zulässig, soweit sie u. a. erforderlich ist für ein bestimmtes Vorhaben der wissenschaftlichen Forschung im Sozialleistungsbereich und schutzwürdige Interessen der betroffenen Person nicht beeinträchtigt werden oder das öffentliche Interesse an der Forschung das Geheimhaltungsinteresse der betroffenen Person erheblich überwiegt. Dabei gilt aber, dass eine Übermittlung ohne Einwilligung der betroffenen Person nicht zulässig ist, soweit es zumutbar ist, ihre Einwilligung einzuholen. Lediglich Angaben über den Namen und Vornamen, die Anschrift, die Telefonnummer sowie die für die Einleitung eines Forschungsvorhabens zwingend erforderlichen Strukturmerkmale der betroffenen Person können für Befragungen auch ohne Einwilligung übermittelt werden. Die Ausnahmen vom Einwilligungserfordernis werden für diese Forschungsklausel teilweise eng verstanden. Argumentiert wird, dass als Grenze der Zumutbarkeit lediglich ein unverhältnismäßiger Aufwand im Einzelfall gelten kann und der Verzicht auf die Einwilligung der betroffenen Person deshalb lediglich auf absolute Ausnahmefälle beschränkt ist (Herbst, 2021). Dem kann jedoch nicht ohne Weiteres gefolgt werden, denn vielfach dürfte die Durchführung von Großforschungsvorhaben mit

Krankenkassendaten mit einer zusätzlichen Einwilligung nicht zumutbar sein. Gerade für den Bereich der Forschungsdatenverarbeitung lässt sich die DSGVO durchaus auch so verstehen, dass kein Primat der Einwilligung gilt. Für die Datenverarbeitung nach § 75 SGB X ist weiterhin zu berücksichtigen, dass eine entsprechende Übermittlung von Sozialdaten nur zulässig ist, soweit grundsätzlich die vorherige Genehmigung durch die oberste Bundes- oder Landesbehörde, die für den Bereich, aus dem die Daten herrühren, zuständig ist, vorliegt. § 75 SGB X besitzt eine Use Case-übergreifende Relevanz (Use Cases 1 und 2), soweit Sozialdaten verarbeitet werden.

### 5.1.3.2 Krebsregisterdaten

Die Erhebung von Daten aus den Krebsregistern der Länder ist aufgrund der unterschiedlichen datenschutzrechtlichen Rahmenbedingungen als Folge des Föderalismus mit Herausforderungen verbunden. Die Ermächtigungsgrundlagen in den jeweiligen Landesgesetzen gestatten den Krebsregisterbehörden, Daten zu Forschungszwecken an Dritte zu übermitteln.[25]

Für die Verwendung von Krebsregisterdaten (Use Cases 2 und 3) ergeben sich Änderungen durch das Gesetz zur Zusammenführung von Krebsregisterdaten vom 18. August 2021 (Gurung-Schönfeld & Kraywinkel, 2021). Das Zentrum für Krebsregisterdaten hat nach § 2 BKRG die Aufgaben der Zusammenführung und der Prüfung der von den Krebsregistern übermittelten Daten gem. § 6 Abs. 2 Nr. 1 BKRG sowie der Erstellung eines Datensatzes nach Maßgabe des § 6 Abs. 2 Nr. 2 BKRG und der Durchführung von Studien und Analysen nach Maßgabe des § 6 Abs. 2 Nr. 3 BKRG. Die Förderung der wissenschaftlichen Nutzung der beim Zentrum für Krebsregisterdaten vorliegenden Daten nach Maßgabe des § 8 BKRG sowie die Einrichtung einer zentralen Antrags- und Registerstelle nach Maßgabe des § 10 BKRG sind ebenfalls Aufgaben des Zentrums. Die Berechtigung zum Empfang der Daten aus den Landeskrebsregistern ergibt sich mit den gesetzlichen Änderungen aus § 5 Abs. 1 BKRG und zur Verwendung der erhaltenen Daten zur Prüfung auf Einheitlichkeit, Vollständigkeit und Vollzähligkeit sowie zur Erstellung eines bundesweit einheitlichen Datensatzes aus § 6 Abs.. 1, Abs. 2 Nr. 1, Nr. 2 BKRG. Das Bundeskrebsregister übermittelt den

---

Landeskrebsregistern die Ergebnisse seiner Prüfung auf Einheitlichkeit und Vollständigkeit gem. § 7 Abs. 1 BKRG. Daten können zu Forschungszwecken gem. § 8 Abs. 1 BKRG an Dritte übermittelt werden, soweit im entsprechenden Antrag nachvollziehbar dargelegt ist, dass der Umfang und die Struktur der beantragten Daten geeignet und erforderlich sind, um die zu untersuchenden Fragen zu beantworten und das im Antrag angegebene Vorhaben mit den beim Zentrum für Krebsregisterdaten vorliegenden Daten bearbeitet werden kann und eine länderübergreifende Auswertung erfordert. Allerdings unterliegt die Datenübermittlung und -bereitstellung diversen Anforderungen und Einschränkungen, u. a. der Antragstellung (Abs. 1), der Übermittlung der Daten in anonymisierter Form (Abs. 1) sowie einer Bewertung des Re-Identifizierungsrisikos und des Treffens geeigneter Gegenmaßnahmen (Abs. 5). Abweichend von Abs. 1 kann bei Erforderlichkeit und entsprechender Begründung durch den Datenempfänger eine Bereitstellung pseudonymisierter Einzeldatensätze erfolgen (Abs. 6). In diesem Rahmen gelten weitere besondere Anforderungen, um dem Persönlichkeitsschutz der betroffenen Personen gerecht zu werden: So dürfen die pseudonymisierten Datensätze nur Personen bereitgestellt werden, die einer Geheimhaltungspflicht gem. § 203 StGB unterliegen oder vor dem Zugang zur Geheimhaltung verpflichtet wurden (Abs. 7). Durch diese zusätzlichen Anforderungen werden die Möglichkeiten zur praxisgerechten Durchführung von Record Linkage erheblich erschwert oder sogar in vielen Fällen unmöglich gemacht. Eine Möglichkeit zur Überwindung dieser Limitationen könnte darin zu sehen sein, dass das Gesetz zur Zusammenführung von Krebsregisterdaten in § 10 BKRG bis zum 31.12.2024 den Auftrag zur Erarbeitung eines Konzepts zur Schaffung einer Plattform, die eine bundesweite anlassbezogene Datenzusammenführung und Analyse der Krebsregisterdaten aus den Ländern sowie eine Verknüpfung von Krebsregisterdaten mit anderen Daten ermöglicht und die klinisch-wissenschaftliche Auswertung der Krebsregisterdaten fördert, formuliert. Die Belange des Datenschutzes und der Informationssicherheit sind bei der Konzepterstellung entsprechend zu berücksichtigen.

### 5.1.3.3 Melderegisterdaten

Im Hinblick auf Meldedaten ergeben sich vor allem für Use Case 3 rechtliche Besonderheiten. Für die Erhebung der Melderegisterdaten sind die §§ 44 ff. BMG relevant. § 44 BMG (einfache Melderegisterauskunft) erlaubt auf Verlangen und für nichtkommerzielle Zwecke die Auskunft über Personen und die Übermittlung von Familiennamen, Vornamen, Doktorgrad, derzeitige Anschrift sowie die Information darüber, ob die Person verstorben ist. § 45 BMG (erweiterte Melderegisterauskunft) erlaubt bei berechtigtem Interesse zusätzlich die Übermittlung von u. a. früherem Namen, Geburtsdatum und -ort, Staatsangehörigkeit, früheren Anschriften und Sterbedatum und -ort. § 46 BMG (Gruppenauskunft) erlaubt eine Melderegisterauskunft über eine Vielzahl nicht namentlich bezeichneter Personen (Zufalls­stichprobe), wenn dies im öffentlichen Interesse (z. B. Forschung) liegt. Es können Datensätze von Personen mit Familiennamen, Vornamen, Doktorgrad, Geschlecht, Alter, derzeitige Anschriften und Staatsangehörigkeiten übermittelt werden.

### 5.1.3.4    Labordaten

In Use Case 4 mit Blick auf das Linkage von Labordaten findet sich im IfSG keine allgemeine Rechtsgrundlage für die Durchführung von Record Linkage außerhalb der Gefahrenabwehr zu reinen Forschungszwecken. Gem. § 14 Abs. 1 IfSG führt das RKI ein elektronisches Melde- und Informationssystem. Die Möglichkeit zum Record Linkage im eng verstandenen Sinne ist dabei in § 14 Abs. 4 IfSG geregelt. Demgemäß können die im elektronischen Melde- und Informationssystem verarbeiteten Daten, die zu melde- und benachrichtigungspflichtigen Tatbeständen erhoben wurden, automatisiert daraufhin überprüft werden, ob sich diese Daten auf denselben Fall beziehen. Insoweit ist für ein generelles Record Linkage von Daten aus dem Infektionsschutz außerhalb des engen Anwendungskontexts aus dem IfSG auf die allgemeinen rechtlichen Möglichkeiten gem. Art. 9 Abs. 2 lit. a DSGVO (Einwilligung) und Art. 9 Abs. 2 lit. j DSGVO i.V.m. § 27 BDSG (Forschungsprivileg) zu verweisen.

### 5.1.3.5    Versorgungsdaten

Für die Forschung mit Gesundheitsdaten sind in einem weiter gefassten Kontext zusätzlich die Vorgaben aus den Landeskrankenhausgesetzen (LKHG) zu berücksichtigen[26]. Auch hier zeigt sich die besondere Komplexität der Verarbeitung von Gesundheitsdaten, die aus dem Föderalismus herrührt (für eine kommentierte Übersicht über die aktuelle landesrechtlichen Besonderheiten sei verwiesen auf Pollmann (2021). Auch hier führt die schwere Handhab­barkeit der Regelungen häufig zu einem erheblichen Mehraufwand sowie einer Einschränkung bei der Datenerhebung, die zu Qualitätsverlusten führt. Als Beispiel mag das Projekt INDEED

---

[26]    Baden-Württemberg: LKHG gilt für Forschung nicht, vgl. § 43 Abs. 3; nur für krankenhausinterne Forschung, vgl. § 46 Abs. 1 Nr. 2a; Bayern: Art. 27 Abs. 1 S. 2, Abs. 5 schränken § 27 BDSG (Datenverarbeitung zu wis­senschaftlichen Forschungszwecken) nicht ein; Berlin: § 25; Brandenburg: § 31; Bremen: § 39; Hamburg § 12; Hessen: Keine Vorschrift zum forschungsbezogenen Datenschutz; Mecklenburg-Vorpommern: § 37; Niedersachsen: Keine Vorschrift zum forschungsbezogenen Datenschutz; Nordrhein-Westfalen: Keine Vor­schrift zum forschungsspezifischen Datenschutz, zu beachten ist jedoch § 6 GDSG NRW; Rheinland-Pfalz: § 37 Abs. 3; Saarland: § 14 Abs. 2; Sachsen: § 34 Abs. 2, Abs. 3; Sachsen-Anhalt: § 17 Abs. 2; Schleswig-Hol­stein: § 38; Thüringen: § 27a

gelten, bei dem Routinedaten aus 16 Notaufnahmen mit Daten der kassenärztlichen Vereinigung verknüpft wurden:

*„Die heterogenen datenschutzrechtlichen Rahmenbedingungen in den einzelnen Bundesländern führten zu erheblichen Verzögerungen bei der Abstimmung mit Kliniken und Behörden. Die bundeslandspezifische Gesetzeslage verhinderte zudem eine deutschlandweite Rekrutierung von Kliniken. Eine deutschlandweite und trägerübergreifende einheitliche Verfahrensweise wäre wünschenswert, würde aber einen einheitlichen Rechtsrahmen voraussetzen. Durch die Einbindung des Konsortialpartners TMF mit Expertise auf dem Gebiet des Datenschutzes in der medizinischen Forschung und speziell Großprojekten konnten die Verzögerungen eingegrenzt werden."* (Fischer-Rosinský et al., 2022).

## 5.2 Rechtliche Hürden des Record Linkage und daraus abzuleitende Forderungen

Deutlich ist dies auch in der vorangehenden Darstellung geworden, indem für die Legitimation der Datenverarbeitung innerhalb der einzelnen Use Cases unterschiedlichste Regelungen aus dem Europa-, Bundes- und Landesrecht heranzuziehen sind und überdies zwischen allgemeinen und bereichsspezifischen Vorgaben zu unterscheiden ist, die teils durch Forschungsklauseln in unterschiedlichem Umfang privilegiert werden.

Verschärft wird die allgemeine rechtliche Problematik durch die Tatsache, dass Forschungsprojekte in der Regel in der Form von Forschungskooperationen bzw. Verbundvorhaben organisiert sind und sich hieraus unterschiedlichste Rollen und Verantwortlichkeiten ergeben, so beispielsweise bezogen auf Dateneigner, Datenverarbeiter, Auftragsverarbeiter/ gemeinsame Verantwortliche, Treuhandstellen, Datenschutzbeauftragte und Aufsichtsbehörden. Nicht selten sind die Verbundpartner unterschiedlichen Rechtsräumen mit unterschiedlichen (datenschutz-)rechtlichen Anforderungen zuzuordnen und es ist zwischen öffentlichen Stellen (z. B. Universitäten und Hochschulen) und privaten Einrichtungen wie Unternehmen und Vereinen zu unterscheiden. Die aufsichts- und fachbehördlichen Kontrollstrukturen haben weitere Unklarheiten zur Folge und bedeuten einen erheblichen Mehraufwand in der Verwaltung von Forschungsdaten. In diesem Zusammenhang wäre es wünschenswert, verteilte behördliche Zuständigkeiten durch federführende Verantwortlichkeiten zu ersetzen, um Forschungsdatenprozesse zu prüfen und ggf. zu genehmigen sowie

Verwaltungsaufwände zur Durchführung von Forschungsvorhaben mit personenbezogenen Daten zu reduzieren.

Umso wichtiger erscheint es, dort, wo ein einheitlicher Rechtsrahmen besteht, namentlich im SGB, die Datenzugänge so zu gestalten, dass sich hier nicht ein zusätzlicher vermeidbarer Aufwand ergibt. Gerade im Bereich der Zugänglichkeit und Verknüpfbarkeit von Datenbeständen unter administrativer Kontrolle wäre in Deutschland allerdings ein Streamlining von Antrags- und Genehmigungsverfahren sinnvoll. Das Verfahren nach §75 SGB X z. B. ist nach wie vor recht sperrig, weil ein ganzes Bündel von Genehmigungen einzuholen ist, was allzu oft sehr zeit- und ressourcenaufwendig ist und Forschungsprojekte nicht unerheblich erschwert. Hier ließe sich beispielsweise von Schottland lernen, das zu den Ländern gehört, die über ein etabliertes System der Verknüpfung von Bevölkerungsdaten aus verschiedenen Bereichen für epidemiologische Gesundheitsstudien verfügen (Administrative Data Research UK, 2021). Zu diesen Bevölkerungsdaten gehören auch die Daten des staatlichen Gesundheitssystems NHS, die über diesen Service relativ einfach mit Daten aus den Bereichen Bildung, Arbeit und Wirtschaft, Kriminalität und Bevölkerungsstatistik verknüpft werden können. Die zuständige Einrichtung Administrative Data Research UK verfolgt hierbei eine aktive Rolle in der Datenverknüpfung und versteht sich als Serviceeinrichtung für die Forschung und nicht lediglich als genehmigende Stelle.

Die Verknüpfung von und mit Registerdaten sollte ebenfalls einheitlich in den Registergesetzen geregelt werden. Zu begrüßen wäre hier ein übergeordnetes Registergesetz, wie es derzeit im Gespräch ist.

Für eine reibungslose Verknüpfung von Daten aus verschiedenen Quellen, die ihrerseits pseudonymisiert sind, wäre ein gemeinsamer Patientenidentifikator ungemein förderlich. Zwar gibt es auch andere Methoden, Daten zu abzugleichen, sie sind aber aufwendig und führen zu Verknüpfungsfehlern.

Die Zulässigkeit der Verwendung der Krankenversichertennummer für die eindeutige Identifizierung von Patient:innen bzw. Proband:innen im Rahmen von Forschungsvorhaben wurde bereits eingehend von Hornung & Roßnagel (2015) untersucht. Die dort festgestellten Grundsätze dürften weiterhin zutreffen, denn die zugrunde liegende Rechtsprechung, insbesondere des Bundesverfassungsgerichts[27] hat sich nicht geändert. Im Ergebnis hängt die Zulässigkeit der gesetzlichen Einführung der Verwendung bestimmter Identifikatoren entscheidend davon ab, wie diese verwendet werden, d. h. im Wesentlichen wer die Nummer zu sehen bekommt. „Im Ergebnis ist deshalb die Nutzung der neuen Krankenversicherten-nummer – in dem beschriebenen Szenario der informationstechnischen und organisa-torischen Absicherung – zur Identifizierung der Versicherten bei der Erzeugung von PIDs [Patientenidentifikatoren – Anm. d. Verf.] zwar gegenwärtig datenschutzrechtlich nicht zulässig. Eine entsprechende Regelung, die eine Einwilligung der betroffenen Person ermöglichen würde, wäre jedoch verfassungsrechtlich zulässig und datenschutzpolitisch vertretbar." Die gegenwärtige Unzulässigkeit beruht auf einer Entscheidung des Bundes-sozialgerichts [28], die inzwischen eventuell anders gelesen wird, die aber in jedem Fall den Gesetzgebenden selbst nicht bindet. Der Gesetzgebende ist vielmehr lediglich durch den grundrechtlichen Rahmen begrenzt.

---

27       BVerfGE 27, 1 (6); 65, 1, 53 (57).

Der grundrechtliche Rahmen steht nach wie vor nach gängiger Auffassung der Einführung einer übergreifenden Identifikationsnummer etwa nach dänischem Muster entgegen, die in den unterschiedlichsten Zusammenhängen von Bankgeschäften bis zu Gesundheitsdiensten verwendet wird und praktisch öffentlich zugänglich ist. So soll verhindert werden, dass die Identifikationsnummer zur Persönlichkeitsprofilbildung verwendet wird, indem Daten aus allen Betätigungsbereichen eines Individuums zusammengezogen und zu Handlungs-mustern oder Vorlieben verdichtet werden. Insofern ist die Verwendung der Steuer-ID für nicht-steuerliche Zwecke nicht gänzlich unproblematisch. Die Schaffung einer Gesundheits-ID, die lediglich im Forschungszusammenhang verwendet wird, trägt demgegenüber den verfassungsrechtlichen Bedenken Rechnung, weil sie gerade bereichsspezifisch bleibt.

---

[28]     BSGE 102, 134ff. zur krankenversicherungsrechtlichen Verwendung von Patientendaten und der abschlie-ßenden Regelung des SGB im Umgang mit diesen Daten.

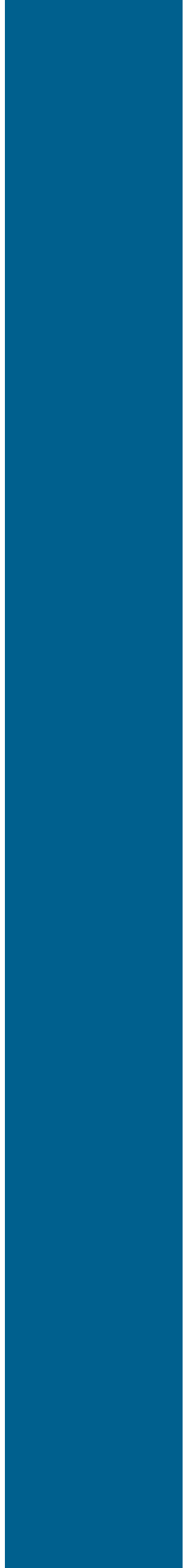

# 6 Schlussfolgerung und Ausblick: Effizientes Record Linkage zur Verbesserung der Gesundheitsforschung in Deutschland

Unter Record Linkage von personenbezogenen Gesundheitsdaten versteht man die Zusammenführung von für die Gesundheitsforschung relevanten Datensätzen auf Personenebene. Bei einer solchen Zusammenführung ist die möglichst fehlerfreie und vollständige Verknüpfung der digitalen Identitäten aus den zu verknüpfenden Datenquellen unerlässlich, um mit diesen Daten aussagekräftige Forschungsergebnisse erzielen zu können. Zur Ableitung der digitalen Identitäten werden Identifikatoren verwendet, die in den jeweils zusammenzuführenden Datensätzen enthalten sein müssen. In der Praxis der Gesundheitsforschung werden dafür häufig direkt personenidentifizierende Daten (direkte Identifikatoren) verwendet, die entweder im Klartext oder kodiert für den Linkage-Prozess zur Verfügung stehen.

Kodierte Identifikatoren werden benötigt, um eine Re-Identifizierung der betroffenen Personen anhand von personenidentifizierenden Daten weitgehend auszuschließen (Privacy-Preserving Record Linkage, PPRL). Dafür werden diese Daten in sogenannte Kontrollnummern (z. B. bei Linkage-Prozessen mit Krebsregisterdaten) oder Bloomfilter (z. B. bei Linkage-Prozessen in der Medizininformatik-Initiative oder dem Netzwerk Universitätsmedizin (NUM-RDP)) transformiert, die die datenschutzfreundliche Verknüpfung ohne den Austausch der Identitätsdaten im Klartext erlauben. Kann mit einem einzelnen Identifikator jede Person eindeutig identifiziert werden, spricht man von einem Unique Identifier. Als solcher bietet sich beispielsweise für gesetzlich Versicherte die Krankenversichertennummer (KVNR) an, die beim exakten Linkage Verwendung finden kann. Dabei erfolgt die Zuordnung der Daten zu einer Person bei exakter Übereinstimmung der KVNR. Die KVNR steht allerdings häufig nicht zur Verfügung. Sie kann z. B. grundsätzlich nicht bei bestimmten Personengruppen (z. B. Privatversicherten) oder bestimmten Datentypen (z. B. SARS-CoV-2-Sequenzdaten oder Meldeamtsdaten) für die Verknüpfung genutzt werden. In der Gesundheitsforschung wird deswegen häufig das fehlertolerante probabilistische Record Linkage unter der Verwendung von mehreren Identifikatoren, wie z. B. Geschlecht, Geburtsdatum, Vor- und Nachname, verwendet. Diese Verfahren bergen jedoch die Gefahr von falsch positiven (Homonymfehler) oder falsch negativen Zuordnungen (Synonymfehler). Die Gefahr von Fehlern beim Record Linkage wird zusätzlich erhöht durch:

1. die Größe der zu verknüpfenden Datensätze und die daraus resultierende Häufung von gleichen Namen oder Geburtsdaten sowie

2. Fehler in den Identifikatoren (z. B. durch Schreibfehler in Namen).

Demzufolge ist häufig – falls möglich – eine zusätzliche manuelle Überprüfung der Verknüpfungen erforderlich, was einen hohen Personalaufwand erfordert. Beim Record Linkage auf Basis von kodierten Identifikatoren kann zudem die temporäre Einbeziehung von unkodierten identifizierenden Patienten- oder Probandendaten (Klartext-IDAT) im manuellen Überprüfungsprozess vonnöten sein.

Eine zentrale Datenschutzmaßnahme beim Record Linkage stellt die informationelle Gewaltenteilung dar. Ihr Prinzip beruht auf der Aufteilung der für das Linkage nötigen Forschungsinfrastrukturen in unabhängige Teilstrukturen bzw. Module. Der Zugang zu den Daten ist dabei strikten Regularien unterworfen. So werden die medizinischen Daten und die personenidentifizierenden Daten getrennt und unabhängig voneinander verwaltet. Die medizinischen Daten werden ausschließlich pseudonymisiert abgespeichert. Das hierfür benötigte Identitätsmanagement wird in der Regel von einer zentralen Treuhandstelle oder Vertrauensstelle koordiniert, die vertrauenswürdig ist und weisungsfrei von den einzelnen Institutionen des Forschungsvorhabens agieren kann. Für diesen Zweck führt diese Stelle eine Patienten-/Probandenliste, in der der Bezug der IDAT zu Pseudonymen hinterlegt ist. Diese Liste ist die Grundlage für den Abgleich im Record Linkage-Prozess.

In der Praxis müssen die Details des Record Linkage-Vorhabens im Vorhinein festgelegt werden, um die notwendigen Genehmigungen von Datenhaltern und Aufsichtsgremien zu erhalten. In Box 7 ist eine Übersicht über die realen Record Linkage-Möglichkeiten nach Art der Daten angegeben (für Details sei auf die Use Cases in Kapitel 3 verwiesen). Die dazugehörigen rechtlichen Regelungen und Auslegungen sowie die Prozesse sind nicht standardisiert und können sich in Abhängigkeit von den verwendeten Identifikatoren sowie von Bundesland zu Bundesland und Datenhalter zu Datenhalter unterscheiden (siehe Box 8 und Box 9 für eine Zusammenstellung der rechtlichen und datenspezifischen Herausforderungen). Gleiches gilt für etwaige nachträgliche Anpassungen und die finale Zustimmung bzw. Ablehnung des Vorhabens. Zudem muss bedacht werden, dass bei manchen Vorhaben (z. B. beim Linkage mit epidemiologischen Primärdaten) ggf. auch noch die Zustimmung der Studienteilnehmenden, d. h. ihre informierte Einwilligung, eingeholt werden muss.

Insgesamt ist festzuhalten, dass das Record Linkage für die Gesundheitsforschung in Deutschland ein aufwendiges und rechtlich unsicheres Unterfangen darstellt. So müssen für jeden Anwendungsfall und jedes Record Linkage-Projekt individuelle Lösungen entwickelt, geprüft, ggf. modifiziert und – falls positiv beschieden – umgesetzt werden. Aufgrund des Fehlens eines Unique Identifiers (siehe Box 10) und des hohen administrativen, rechtlichen und organisatorischen Aufwands bei der Umsetzung kommt es zu Verknüpfungsfehlern und geringer Ausnutzung der verfügbaren Daten, wenn beispielsweise nur Daten einzelner weniger Datenhalter (z. B. einzelner Krankenhäuser, Krankenversicherungen oder Krebsregister) genutzt werden können oder dürfen. Dies führt zu eingeschränkter Verallgemeinerbarkeit und zu unpräzisen und verzerrten Forschungsergebnissen. Rechtliche Unsicherheiten führen sogar dazu, dass Record Linkage-Vorhaben nicht umgesetzt werden, sodass wichtige medizinische Fragen unbearbeitet bleiben.

## Box 7: Übersicht zur Verknüpfung von Daten aus unterschiedlichen Quellen

### Was geht – was geht nicht?

Übersicht über die Möglichkeiten zur Verknüpfung zweier Datensätze auf Personenebene nach Art der Daten hinsichtl. Einheitlichkeit, Aufwand, Bevölkerungsabdeckung, Fehlerquote und Datentiefe.

### Epidemiologische Primärdaten – Krankenkassendaten (siehe Use Case 1)

Für das RL zwischen epidemiologischen Primärdaten und Krankenkassendaten werden verschiedene Verfahren eingesetzt. Wenn die Krankenversichertennummer als Identifikator für gesetzlich Versicherte verwendet wird, kann im Allgemeinen von einer sehr kleinen Fehlerquote ausgegangen werden. Aufgrund der Vielzahl der verschiedenen Krankenkassen und dem damit verbundenen Aufwand für Einholung der Genehmigungen und Abstimmung der Prozesse wird häufig nur ein geringer Teil der Krankenkassendaten für ein RL genutzt. Die Krankenkassendaten gehalten vom Zentralinstitut für die kassenärztliche Versorgung oder den Kassenärztlichen Vereinigungen weisen eine geringere Datentiefe auf und/oder decken nur einzelne Bundesländer ab. Daten von Privatversicherten werden im Allgemeinen (noch) nicht genutzt.

## Epidemiologische Primärdaten – Krebsregisterdaten
(siehe „Kohortenabgleich" aus Use Case 3)

Die Verfahren beim sogenannten Kohortenabgleich unterscheiden sich von Bundesland zu Bundesland und Anwendungsfall zu Anwendungsfall. Der administrative Aufwand ist abhängig von der Anzahl der eingeschlossenen Landeskrebsregister. Die Krebsregisterdaten decken die gesamte Bevölkerung der jeweiligen Bundesländer ab. Die Fehlerquote der Verknüpfung hängt wiederum davon ab, welche Identifikatoren eingesetzt werden (können) und ob eine zusätzliche manuelle Prüfung möglich ist.

## Krankenkassendaten – Krebsregisterdaten (siehe Use Case 2)

In diesem Fall unterscheiden sich die Verfahren von Anwendungsfall zu Anwendungsfall, insbesondere in Abhängigkeit der berücksichtigten Krankenkassen und Krebsregistern. Aufgrund der hohen Zahl an Kassen und Krebsregistern sowie unterschiedlicher Datenschutzbedenken ist ein Linkage nur mit einer kleinen Anzahl Kassen und Krebsregistern möglich. Die Fehlerquote des Linkage hängt dabei maßgeblich von den zur Verfügung stehenden Identifikatoren ab.

## Krankenkassendaten – Klinische Versorgungsdaten (Ohlmeier et al., 2015)

Grundsätzlich sind unterschiedliche Verfahren für diesen Anwendungsfall denkbar. Die Abdeckung der Bevölkerung ist durch die Schnittmenge der beteiligten Krankenkassen und Krankenhäuser begrenzt. Falls die Versichertennummer Verwendung findet, ist von einer geringen Fehlerquote auszugehen.

## Krebsregisterdaten – Klinische Versorgungsdaten
(siehe „Routine-Linkage" aus Use Case 3)

Im Routinebetrieb der Krebsregister werden unterschiedliche Verfahren für die klinische und epidemiologische Krebsregistrierung eingesetzt. Dabei werden ausschließlich Daten, die für das Krebsregister relevant sind, berücksichtigt. Die Verfahren schließen eine manuelle Prüfung ein. Die Fehlerquote ist sehr gering. Aufgrund des gesetzlichen Auftrags decken die Krebsregister mit ihrer Arbeit die gesamte Bevölkerung ab.

## Klinische Versorgungsdaten – Klinischen Versorgungs- oder Studiendaten
(Standortintern: siehe Use Case 5)

Innerhalb eines Standorts kann für ein Record Linkage aufgrund des Patientenkontakts im Behandlungskontext auf Klarnamen und die Versichertennummer zugriffen werden, was zu einer sehr geringen Fehlerquote führt.

**Klinische Versorgungsdaten – Klinische Versorgungs- oder Studiendaten**
(Standortübergreifend: siehe Use Case 6 & 7)

Innerhalb von eingerichteten Konsortien und Netzwerken wurden einheitliche Record Linkage-Verfahren etabliert, die häufig eine manuelle Prüfung ermöglichen, was zu einer geringen Fehlerquote führt. Die Abdeckung der Bevölkerung hängt von der Beteiligung der Standorte in den jeweiligen Netzwerken und Konsortien ab.

## Box 8: Rechtliche Herausforderungen

- Je nach Verarbeitungssituation und Datenkategorie unterschiedliche Rechtsgrundlagen
- Fragmentierte aufsichts- und fachbehördliche Kontrollstrukturen, und damit erheblicher Mehraufwand bei Forschungsprojekten auf der Grundlage von Daten aus verschiedenen Quellen, insbesondere:
  - Rechtliche Unterschiede bei Record Linkage mit Daten gesetzlicher und privater Krankenkassen
  - Fehlendes Konzept für bundesweites Record Linkage von Krebsregisterdaten mit anderen Datenquellen (in Entwicklung, § 10 BKRG)
  - Länderspezifische Regelungen und Auslegungen bei Record Linkage mit Daten von gesetzlichen Krankenkassen, Krebsregistern oder klinischen Routinedaten
  - Fehlen von rechtlichen Rahmenbedingungen zum Aufbau einer Treuhandstelle für Sars-CoV-2-Sequenz- und Meldedaten

→ **Insgesamt führen diese Beschränkungen und Unsicherheiten dazu, dass nur (Bruch-)Teile der bestehenden Daten verknüpft werden (können), was die Bearbeitung wichtiger gesundheitlicher Fragestellungen verhindert.**

## Box 9: Datenspezifische Herausforderungen

- Auffindbarkeit der Daten
- Datenzugangsbeschränkungen
- Fehlende Metadaten (z. B. Datensatzbeschreibungen)
- Mangelnde Interoperabilität der Daten

## Box 10: Herausforderungen bedingt durch das Fehlen eines Unique Identifiers

- Krankenversichertennummer als Unique Identifier nur eingeschränkt nutzbar, da weder für alle Bevölkerungsgruppen noch für alle Datenquellen verfügbar
- Fehlerhafte und indirekte Identifikatoren (mit denen nur in Kombination eine Identifizierung möglich ist, wie z. B. über Geschlecht, Diagnose und Gemeindekennziffer) erhöhen die Zahl der Verknüpfungsfehler; dies gilt insbesondere bei
  – manueller Eingabe von Identifikatoren
  – geringer Überschneidung der vorhandenen Identifikatoren der zu verknüpfenden Datenquellen
  – sehr groben indirekten Identifikatoren (wie z. B. bei Geschlecht oder Gemeindekennziffer)
  – geringer Datenqualität der Identifikatoren
  – mangelnder Standardisierung von Namensschreibweisen auf Basis der Aussprache für die Bevölkerung in Deutschland
  – großen zu verknüpfenden Datenquellen
- Aufwendige manuelle Nachbearbeitung zur Korrektur fehlerhafter Verknüpfungen
- Hoher Anteil von auszuschließenden Verknüpfungen aufgrund uneindeutiger Zuordnung
- → **Datenverlust und falsche Verknüpfungen führen zu verzerrten Forschungsergebnissen**

## 6.1    Forderungen an Stakeholder „Wie kann man Defizite beheben?"

Wie in den Kapiteln 3-5 gezeigt wurde, gibt es viele Herausforderungen, die ein effektives Record Linkage unverhältnismäßig erschweren und damit die Gesundheitsforschung behindern und teilweise sogar unmöglich machen. In Folgenden sollen einige Vorschläge zur Lösung der offenen Fragen vorgestellt werden. Die Umsetzung dieser Vorschläge richtet sich an unterschiedliche Interessensgruppen: die Community selbst kann zur Verbesserung ebenso beitragen wie die Verwaltung und letztlich der Gesetzgeber. Dabei ist ein gemeinsames Vorgehen notwendig, denn am Ende hängt eine forschungs- und evidenzbasierte Verbesserung der gesundheitlichen Versorgung der Bevölkerung davon ab.

### 6.1.1  Unique Identifier

Record Linkage-Verfahren ohne einen Unique Identifier als Identifikator erhöhen das Risiko für Verknüpfungsfehler, sodass der Ruf nach einem solchen Unique Identifier für das Record Linkage berechtigt und politisch aktuell bleibt (Interdisziplinäre Kommission für Pandemieforschung der Deutschen Forschungsgemeinschaft (DFG), 2021; Nationale Forschungsdateninfrastruktur (NFDI) e.V., 2022; Niemeyer et al., 2021; Rat für Sozial- und Wirtschaftsdaten, 2023). Wünschenswerte Eigenschaften eines solchen Unique Identifiers wurden bereits von Hillestad et al. (2008) für die eindeutige Identifizierung von Patient:innen im US-amerikanischen Gesundheitssystem aufgestellt. Danach soll ein solcher Identifier die folgenden Eigenschaften haben:

- Eineindeutig und universell: Jede Person soll genau einem eigenen Identifier zugeordnet sein.

- Nicht offenlegend („Non-disclosing"): Es sollen keine persönlichen Daten enthalten sein (also z. B. nicht Geburtsdatum oder Name).

- Unveränderlich: Der Identifier soll sich während des gesamten Lebens nicht ändern (im Unterschied zu Adresse oder Namen).

- Verifizierbar: Integrierte Prüfnummern sollen Validität der Nummer überprüfbar machen können, um Eingabefehler zu vermeiden.

Von den existierenden IDs in Deutschland kommen für einen Unique Identifier mehrere in Frage, die unterschiedliche Stärken und Schwächen haben: Naheliegend wäre zunächst die Verwendung von Sozial- bzw. Rentenversicherungsnummer (SVNR/RVNR) oder

Krankenversichertennummer (KVNR), wobei letztere aus der ersten abgeleitet ist. Vorteilhaft wäre hier, dass die Krankenversichertennummer schon lange im Gesundheitsbereich verwendet wird, so dass sie sich auch für Kohortenstudien wie die NAKO Gesundheitsstudie gut eignet. Allerdings ist nicht jeder Person eine solche Nummer zugewiesen. Privatversicherten ist in der Regel keine KVNR zugeordnet (auch wenn sich dieser Anteil durch die verpflichtende Nutzung der KVNR durch die private Krankenversicherungen im Rahmen der Nutzung der elektronischen Patientenakte (ePA) in den kommenden Jahren verringern dürfte). Die Rentenversicherungsnummer wird erst seit 2005 bei Geburt flächendeckend vergeben und ist somit auch noch nicht für alle Personen vorhanden.

Demgegenüber werden durch die Steuer-ID (Identifikationsnummer nach § 139b Abgabeordnung) schon heute nahezu alle Personen abgedeckt. Negativ fällt hier ins Gewicht, dass eine bereichsfremde Nutzung von Identifikatoren bislang im Hinblick auf den Datenschutz eher kritisch gesehen wird. Gleichwohl ist die Nutzung der Steuer-ID in den vergangenen Jahren schon „bereichsfremd" erweitert worden. So ist mit dem Registermodernisierungsgesetz 2021 zu Zwecken der Verbesserung, Entbürokratisierung und nutzerfreundlichen Digitalisierung von Verwaltungsleistungen vorgesehen, die einheitliche Steuer-ID in zahlreichen Verwaltungsbereichen als Identifikator zu nutzen. Auf die Steuer-ID können dann 51 verschiedene Behörden, Datenbanken und Verzeichnisse (vom Melderegister im Einwohnermeldeamt, Personenstandsregister im Standesamt und Personalausweis- und Passregister über Fahrzeugregister, Luftfahrerdatei, nationalem Waffenregister und Ausländerbehörde bis hin zu Stammsatzdateien der Renten- und Sozialversicherungen und dem Versichertenverzeichnis der Krankenkassen) zugreifen und diese als Zuordnungskriterium speichern (gem. Anlage zu § 1 Identifikationsnummerngesetz [IDNrG]). Der Steuer-ID kann hiermit faktisch in der Zukunft die Rolle einer „Bürger-ID" zukommen. Es stellt sich somit die Frage, ob die Steuer-ID auch für die Datenauswertung im Gesundheitswesen gesetzlich genutzt werden kann, zumal eine Speicherung bei den Krankenkassen – als kassenunabhängiges Merkmal – in den Versichertenstammdaten ohnehin mit der o. a. Gesetzgebung vorgesehen ist. Nicht ausgeblendet werden kann hierbei der Umstand, dass die Verknüpfung verschiedener Bereiche über eine einheitliche „Bürger-ID" politisch, gesellschaftlich und datenschutzrechtlich umstritten ist. Dies dürfte auch auf die Verknüpfung mit sensiblen Gesundheitsdaten zutreffen.

Intensiv zu prüfen wäre daher auch die Option, aus den vorhandenen gesetzlich geregelten Identifikatoren (KVNR, SVNR / RVNR, Steuer-ID und elektronische Identität auf dem Personalausweis (eID)) ein geschütztes bereichsspezifisches Pseudonym („Forschungspseudonym") zu erzeugen. Eine solche neu zu schaffende versorgungsnahe Gesundheits-ID könnte das Problem lösen, ohne dass zugleich eine ungewollte Verknüpfung mit persönlichen Daten aus anderen Bereichen erfolgen kann. Diese Gesundheits-ID müsste zwingend in allen Dokumentations- und Abrechnungsprozessen im Gesundheitswesen mitgeführt werden. An den Stellen, an denen dies nicht möglich ist, wie in historischen Datenbeständen, müsste wiederum aus einer der o. a. primären IDs durch Umschlüsselung die Gesundheits-ID erzeugt werden können. Erforderliche Schutzmaßnahmen sowie eine zentrale Vertrauensstelle zur sichereren Verwahrung der ID-Zuordnung wären zu schaffen. Diese Stelle müsste so aufgesetzt werden, dass sie als Serviceeinrichtung zur kontrollierten Datenverknüpfung für die Forschung und nicht lediglich als genehmigende Stelle fungiert (siehe Beispiel Schottland, Kapitel 5).

Wir schlagen die Einrichtung einer **interdisziplinären Kommission** unter Führung des BMG vor, die eine konkrete Lösung erarbeitet. Dabei ist in jedem Fall eine enge Verzahnung mit der EU sinnvoll (insbesondere zum konzeptuellen Abgleich mit der europäischen eID der EU-Bürger:innen gemäß der sogenannter eIDAS-Verordnung[29]); auch vergleichende Betrachtungen aus anderen europäischen Ländern mit Unique Identifiern sollten direkt in die Arbeiten der Kommission einfließen. Verwaltungswissenschaftliche Expertise, Sicherheitsexpertise und Domänenexpert:innen der Sozialversicherungen und der Gesundheitsforschung und Versorgung sollten ebenfalls einbezogen werden. Die Kommission sollte ein durchgängiges Konzept entwerfen, das auch die Konsequenzen für Dokumentationsprozesse und Anforderungen an eine Sicherheitsarchitektur benennt, und eine anschließende rechtliche und politische Betrachtung zur Verfügung stellen. Die wissenschaftliche Community sollte eingebunden werden, diesen politischen Prozess mit Fachexpertise und Darlegungen des Nutzens von Datenverknüpfung sowie mit Einschätzungen zur Praktikabilität möglicher Lösungsvorschläge der Kommission konstruktiv zu unterstützen (Details zur Zusammensetzung und Arbeit einer solchen Kommission siehe Niemeyer et al. (2021) (S. 247 ff.)).

---

[29]  Verordnung (EU) Nr. 910/2014 des Europäischen Parlaments und des Rates vom 23. Juli 2014 über elektronische Identifizierung und Vertrauensdienste für elektronische Transaktionen im Binnenmarkt und zur Aufhebung der Richtlinie 1999/93/EG

Auf europäischer Ebene gibt es bereits Initiativen, die bemüht sind, ein europaweites Record Linkage zu ermöglichen (z.B. mit dem Pseudonymisierungstool Spider[30]). Diese sollten ebenfalls Berücksichtigung finden.

---

**Box 11: Kernaussagen und Empfehlungen zum Unique Identifier**

→ Verwendete Record Linkage-Verfahren ohne Unique Identifier aufwendig und fehleranfällig

→ Zur Verbesserung des Record Linkage ist Unique Identifier erforderlich

→ Mit vorhandenen Identifikationsnummern (wie Krankenversichertennummer oder Steuer-ID) Unique Identifier bereits verfügbar

→ Umschlüsselung einer vorhandenen primären Identifikationsnummer zu einer sekundären bereichsspezifischen Gesundheits-ID („Forschungspseudonym") für Record Linkage möglich

→ Debatte über geeigneten Unique Identifier für die Gesundheitsforschung notwendig

→ Einrichtung einer interdisziplinären Kommission zur Erarbeitung einer Lösung zur Etablierung eines kontrollierten Record Linkage-Verfahrens mit einem Unique Identifier empfehlenswert

---

## 6.1.2 Zentrale Genehmigungsbehörde und Rechtssicherheit

Bei Forschungsvorhaben besteht bezüglich des Datenschutzes häufig eine erhebliche Rechtsunsicherheit, weil die deutschen Datenschutzaufsichtsbehörden keine Genehmigungen erteilen, sondern lediglich bei angenommenen Verstößen einschreiten. Das führt dazu, dass von Ausnahmeklauseln für die Forschung nur selten Gebrauch gemacht wird. Insbesondere wenn Forschungsvorhaben einen großen Aufwand erfordern, wie beispielsweise Vorhaben zur Sekundärdatennutzung oder zum Aufbau von Daten-plattformen, bedarf es einer sicheren rechtlichen Basis. Die Anwendung von gesetzlichen Ausnahmeregelungen setzt aber eine Interessenabwägung voraus, die von der verantwortlichen Stelle, die die Daten zur Verfügung stellt, selbst zu treffen ist. Dabei entsteht jedoch das Risiko, dass ggf. eine solche Entscheidung im Nachhinein von der verantwortlichen Datenschutzaufsichtsbehörde nicht geteilt wird – mit allen daraus sich ergebenen Konsequenzen. Insbesondere die stärkere Betonung der persönlichen Verantwortung von internen Datenschutzbeauftragten durch die DSGVO macht es unattraktiv, von Ausnahme-regelungen Gebrauch zu machen. Selbst wenn Ausnahmeregelungen im Einzelfall ein

---

30  https://eu-rd-platform.jrc.ec.europa.eu/spider/, Zugriff am 09.05.2022

Forschungsprojekt ermöglichen könnte, wird diese Möglichkeit nicht genutzt, weil die DSGVO eine persönliche Haftung und Verantwortung vorsieht. Ein solch erhebliches persönliches Risiko wird verständlicherweise im Allgemeinen gescheut. Außerdem ist es gerade die Aufgabe und Erwartung an Datenschutzbeauftragte und Rechtsabteilungen, Compliance sicherzustellen. Die sichere Lösung ist daher in den meisten Fällen die Nichtanwendung einer Ausnahmeregelung, weil man bei einer Interessenabwägung immer zu unterschiedlichen Ergebnissen kommen kann.

Abhilfe könnte hier geschaffen werden, indem die Entscheidung über die Einschlägigkeit und konkrete Anwendung von Ausnahmeregelungen zum Einwilligungserfordernis von einer staatlichen Stelle geprüft und klagefähig entschieden wird. So wäre sichergestellt, dass hier eine Kasuistik entsteht, aus der sich eine größere Rechts- und damit Planungssicherheit ergibt. Auch gegenüber Betroffenen wäre diese Lösung vorzugswürdig, weil sie Transparenz herstellt und Betroffenenrechte durch ein Anfechtungsrecht der behördlichen Entscheidung gesichert werden können.

Die Folge der derzeit bestehenden Rechtsunsicherheit ist zudem, dass ein notwendiger Diskurs zur rechtspolitischen Überarbeitung der gesetzlichen Ausnahmeregelungen nicht entstehen kann, weil die jeweiligen Forschungsprojekte gar nicht erst zustande kommen.

In bisherigen Verhandlungen mit den Datenschutzaufsichtsbehörden, z. B. im Rahmen der Medizininformatik-Initiative (MII) oder der NAKO Gesundheitsstudie sowie insbesondere in den seit mehr als zwei Jahrzehnten erfolgten Abstimmungen mit Datenschutzaufsichtsbehörden zu Datenschutzfragen in der medizinischen Forschung über die TMF, konnte viel erreicht werden, um eine übergreifende Datennutzung für die Gesundheitsforschung zu ermöglichen, wie unter anderem die Einigung auf den Broad Consent. Dieses Verhandlungsverfahren stößt aber zuweilen auch an seine Grenzen. Wie das Beispiel der MII zeigt, kann dieses Vorgehen sehr langwierig sein und zu nicht unerheblichen Verzögerungen führen. Darüber hinaus werden häufig Einigungen im Nachhinein wieder in Frage gestellt, was durch personelle Wechsel in einer Behörde begünstigt wird. Eine solche Verhandlungslösung hat somit keine Bestandskraft.

Zudem wird die Debatte mit vielen unterschiedlichen Behörden dadurch erschwert, dass nicht nur Landesgesetze einen unterschiedlichen Wortlaut aufweisen, sondern auch bundesgesetzliche Grundlagen unterschiedlich ausgelegt und gehandhabt werden. Diese

Verflechtung von Bundes- und Landesebene zeigt sich auch beim Verfahren nach § 75 SGB X, nach dem trotz bundeseinheitlicher Rechtsgrundlage verschiedene Landesaufsichtsbehörden zu beteiligen sind (s.o. Kapitel 5). Eine Straffung und Vereinheitlichung zersplitterter Verfahren sowie die Einführung einer federführenden Aufsichtsbehörde in Anlehnung an die federführende Ethikkommission nach der *Clinical Trial Regulation* (CTR) wären hier empfehlenswert und würden eine deutliche Verbesserung darstellen.

Der Vorteil am SGB-Verfahren ist, dass die Genehmigung eine gewisse Rechtssicherheit bietet, die sonst im allgemeinen Datenschutzrecht fehlt. In Frankreich besitzt die *Commission Nationale de l'Informatique et des Libertés* (CNIL) eine solche Genehmigungskompetenz. Es wäre wünschenswert, wenn eine solche Institution für den Bereich der Gesundheitsforschung auch in Deutschland eingerichtet werden könnte. Diese Stelle könnte dann z. B. in Anspruch genommen werden, um Rechtssicherheit bei der Nutzung von Forschungsklauseln zur einwilligungsfreien Datennutzung zu erzielen und haftungsrechtliche Risiken zu vermeiden. Ansätze hierzu finden sich auf EU-Ebene im kürzlich veröffentlichten Vorschlag der EU-Kommission zur Verordnung für die Einführung des Europäischen Gesundheitsdatenraumes[31]. Praktikable und auf die deutsche Situation passende Vorschläge sind hier notwendig, um Deutschland angemessen zu repräsentieren.

> ## Box 12: Empfehlungen zur Einführung einer zentralen Genehmigungsbehörde
>
> → Straffung administrativer Verfahren
>
> → Einführung einer federführenden Aufsichtsbehörde, z. B. in Anlehnung an die französische Ethikkommission, inkl. behördlicher Genehmigungskompetenz zur Schaffung von Rechtssicherheit

### 6.1.3 Regelungsbedarf bei Record Linkage ohne Informed Consent

Nach wie vor wird bei vielen Forschungsvorhaben die informierte Einwilligung für das Record Linkage eingeholt, bevor ein Linkage durchgeführt wird, wie beispielsweise bei der NAKO Gesundheitsstudie oder der Medizininformatik-Initiative. Die spezifische persönliche Einwilligung legt bestimmte Record Linkage-Möglichkeiten fest, die sich später nicht mehr ändern lassen, ohne die Teilnehmenden erneut zu kontaktieren – wovon in der Regel aufgrund

---

31  https://health.ec.europa.eu/ehealth-digital-health-and-care/european-health-data-space_de, Zugriff: 17.04.2023

des Aufwands für Forschende und Teilnehmende abgesehen wird. Da das Datenschutzrecht es bislang verbietet, Einwilligungsdokumente so zu verfassen, dass möglichst umfassende Record Linkage-Szenarien abgedeckt wären, treten hierdurch notwendigerweise Lücken im Hinblick auf zukünftige Forschungsprojekte auf.

Wenn die Verlinkung von Datensätzen, deren Erhebung nicht auf einer Einwilligung beruht und somit kein Kontakt zu den Personen hergestellt wurde, mit Datensätzen, die der Einwilligung bedürfen, ermöglicht werden soll, müsste in der entsprechenden gesetzlichen Grundlage ein zukünftiges Record Linkage mitgedacht werden, die über die initiale gesetzliche Zweckgebundenheit hinausgeht. So sind zurzeit etwa Registerdaten nicht mit Biobankdaten verknüpfbar, wenn weder die Einwilligung der Bioprobenspender:innen noch das entsprechende Registergesetz eine solche Verlinkung vorsieht. Falls die Verlinkung gesetzlich erlaubt würde, müsste im Einwilligungsprozess darüber aufgeklärt werden.

Überdies sind gesetzliche Erlaubnistatbestände nicht forschungsfördernd, wenn die abstrakten gesetzlichen Anforderungen mit der Einholung einer Einwilligung verknüpft sind, wie dies für die Regelung in § 75 SGB X der Fall ist. Durch die Auslegung, dass die Einwilligungs-freiheit auf Ausnahmefälle beschränkt ist, wird der forschungsbezogenen Realität nicht angemessen Rechnung getragen. So ist beispielsweise die Durchführung von Großforschungs-vorhaben mit Krankenkassendaten mit einer zusätzlichen Einwilligung weder zumutbar noch praktikabel. Zu beachten ist aber, dass sich die DSGVO auch so verstehen lässt, dass für die Forschungsdatenverarbeitung gerade nicht das Primat der Einwilligung besteht.

### 6.1.4  Ausbau und Schaffung von Registern und Bereitstellung von Sekundärdaten zur wissenschaftlichen Nutzung

Eine Voraussetzung für die Durchführung von Record Linkage-Projekten der Gesundheits-forschung ist, dass personenbezogene Gesundheitsdaten für eine Verknüpfung zur Verfügung stehen. Dazu müssen einerseits vorhandene Sekundär- und Registerdaten im Sinne der FAIR-Prinzipien (Wilkinson et al., 2016) aufbereitet werden, das heißt, die personenbezogenen Daten müssen auffindbar („Findable"), zugänglich („Accessible"), interoperabel („Interoperable") und wiederverwendbar („Reusable") sein und insbesondere ein Record Linkage mit anderen Datenquellen ermöglichen. Dies betrifft unter anderem die Datenquellen, die in Kapitel 3 und in Kapitel 6 behandelt werden sowie medizinische Register (Niemeyer et al., 2021). Die Auffindbarkeit der Register und Sekundärdatenquellen als wichtigen ersten Schritt wird in Zukunft über die Registrierung der Datenquellen in dem NFDI4Health German Central Health Study Hub[32] gewährleistet werden können.

Andererseits sollten – neben der systematischen Erschließung der vorhandenen Datenquellen – weitere nützliche personenbezogene Register aufgebaut werden, um das volle Potenzial von Record Linkage ausschöpfen zu können. Diese Forderungen decken sich mit bereits formulierten Forderungen aus Wissenschaft und Gesundheitswesen (Interdisziplinäre Kommission für Pandemieforschung der Deutschen Forschungsgemeinschaft (DFG), 2021; Rat für Sozial- und Wirtschaftsdaten, 2023; Sachverständigenrat zur Begutachtung der Entwicklung im Gesundheitswesen, 2021). Dies betrifft insbesondere die folgenden aufzubauenden oder zu erschließenden Datenquellen:

- In einem **Mortalitätsregister oder Mortalitätsindex** werden basierend auf Todesbescheinigungen die medizinischen Informationen zu jedem Todesfall registriert. In Bremen und im Saarland sind auf Landesebene solche Register etabliert. Ein bundesweites Register für die wissenschaftliche Forschung gibt es bislang allerdings nicht, obwohl es seit langem empfohlen wird und konkrete Implementierungsvorschläge inkl. nötiger Gesetzestexte und Kosten- und Nutzenschätzungen vorliegen (German Data Forum, 2011; Luttmann et al., 2019; PROGNOS AG, 2013).

- In einem **medizinischen Geburtsregister** werden systematisch Informationen zu allen Geburten gesammelt. Dieses schließt insbesondere medizinische Informationen zur Mutter, Schwangerschaft und Geburt sowie zum Neugeborenen ein. In Deutschland gibt es kein bundesweites medizinisches Geburtsregister zur wissenschaftlichen Nutzung, obwohl dies unter anderem für die Erhebung von Fehlbildungen bei Neugeborenen von großer Bedeutung wäre (Bundesärztekammer, 2022).

- Die **elektronische Patientenakte (ePA)**[33] befindet sich im Prozess der stufenweisen Einführung (Thiel & Deimel, 2022); Wissenschaftlicher Dienst des Deutschen Bundestages, 2022). Dieser Prozess wird aktuell von der Bundesregierung noch einmal überarbeitet und soll zudem beschleunigt

---

32    https://csh.nfdi4health.de/, Zugriff: 08.12.2022

33    https://www.bundesgesundheitsministerium.de/patientendaten-schutz-gesetz.html, Zugriff: 16.03.2023

werden (Bundesministerium für Gesundheit, 2023). Es ist geplant, dass die Daten der ePA auch der Forschung zur Verfügung gestellt werden können. Zur tatsächlichen Umsetzung müssen allerdings zunächst drei Bedingungen erfüllt sein: (1) Patient:innen müssen zunächst eine ePA beantragen bzw. im Rahmen einer möglichen Einführung einer Opt-out-Lösung dieser nicht widersprechen; (2) medizinische Einrichtungen müssen die nötige IT-Infrastruktur bereitstellen, damit medizinische Daten in der ePA abgelegt werden können; und (3) Patient:innen müssen in die Freigabe der jeweiligen Daten zu Forschungszwecken informiert einwilligen („Datenspende"). Aus Forschungssicht ist es zudem wichtig, dass die Daten die FAIR-Kriterien erfüllen.

- **Impfungen** werden in Deutschland über die Krankenkassen abgerechnet und sind somit in den Krankenkassendaten enthalten. Allerdings trifft dies nur auf Impfungen zu, die eine Kassenleistung darstellen. Dies ist beispielsweise bei COVID-19-Impfungen in Deutschland nicht der Fall. Eine umfassende wissenschaftliche Untersuchung der durchgeführten COVID-19-Impfungen hinsichtlich Wirksamkeit und Sicherheit ist somit in Deutschland nicht möglich, was von wissenschaftlichen Fachgesellschaften frühzeitig kritisiert wurde (DGEpi & GMDS, 2020). Für zukünftige Pandemien ist es aus diesem Grund ratsam, Impfungen zu erfassen und die personenbezogenen Daten anschließend zeitnah für Forschungszwecke zur Verfügung zu stellen.

Zum wissenschaftlichen Nutzen von nationalen Gesundheitsregistern sei auf die Übersichtsarbeiten von Sund et al. (2014) und Furu et al. (2010) verwiesen, die die besondere Bedeutung der oben genannten Register für die Sozial-, Pharmako- sowie Umweltepidemiologie in den Nordischen Ländern (Dänemark, Finnland, Island, Norwegen und Schweden) herausstellen und eine Vielzahl an Beispielen von Studien zu gesundheitlichen Berufsrisiken, sozialer Ungleichheit, Verwendung und Wirkungen von Arzneimitteln sowie die Beurteilung von Umwelteinflüssen aufführen.

Das hohe Potenzial von Sekundärdaten für die Gesundheitsforschung, das sich in Zukunft auch in Deutschland durch die Nutzung der ePA-Daten entfalten kann, zeigt sich zudem anhand der Clinical Practice Research Datalink-Datenbank aus dem Vereinigten Königreich. Diese Datenbank wird mittlerweile von mehr als 1750 Allgemeinmedizinern gespeist und deckt 24% der Bevölkerung des Vereinigten Königreichs ab. Auf Basis dieser Daten wurden bereits mehr als 3000 Publikationen mit Peer-Review-Verfahren veröffentlicht, zuletzt beispielweise zu Krebssterblichkeit von Personen mit Typ-2-Diabetes oder zu Komorbiditäten bei Personen mit Arthrose (Gallagher et al., 2021)[34].

### 6.1.5 Infrastruktur und Servicestruktur für Record Linkage mit zentralen Komponenten

Mithilfe des geplanten und bereits im Koalitionsvertrag der Bundesregierung an- gekündigten Gesundheitsdatennutzungsgesetzes soll die wissenschaftliche Datennutzung verbessert werden. Hierfür soll eine *dezentrale* Forschungsdateninfrastruktur aufgebaut werden[35]. Diese folgt der Struktur in der Gesundheitsversorgung für Datenerfassung und -speicherung, die ebenfalls dezentral erfolgt (z. B. in Arztpraxen, medizinische Versorgungszentren, Universitätskliniken). Gleiches gilt auch für Forschungsdaten aus klinischen oder epidemiologischen Studien, die dort entstehen, wo die Behandlung stattfindet bzw. wo Proband:innen untersucht oder befragt werden.

Um alle relevanten Gesundheitsdaten sicher, umfassend und effektiv für die Forschung nutzen zu können, bedarf es idealerweise einer Forschungsdateninfrastruktur, die ein fehlerfreies Record Linkage und ein effektives Identitätsmanagement ermöglicht, das mittels moderner Pseudonymisierungstechniken höchste Datenschutzanforderungen erfüllt. Zurzeit wird die jeweilige Forschungsdateninfrastruktur in der Regel vorhabenbezogen aufgebaut. Diese folgt dabei der dezentralen Struktur der Datenquellen (siehe Use Cases, Kapitel 3). Dies hat den Vorteil, dass an eben diesen Stellen aufgrund der vorhandenen Fachexpertise Daten am einfachsten gepflegt und verwaltet werden können und gleichzeitig die notwendige Interoperabilität hergestellt werden kann. In Deutschland fehlen allerdings interoperable Lösungen, die ein sicheres Record Linkage über verschiedene Studien und Datenkörper hinweg ermöglichen. Um diese Situation zu verbessern, sollen nicht etwa Parallelstrukturen in der Datenhaltung geschaffen werden, zum Beispiel durch eine zusätzliche zentrale Datenhaltung, sondern es soll ermöglicht werden, dass die dezentralen Datenhalter die Daten zur Verfügung stellen, sodass die Daten anlassbezogen und bedarfsgerecht für die Forschung genutzt und verknüpft werden können, wozu Vorhaben wie NFDI4Health beitragen wollen.

---

34  https://cprd.com/bibliography, Zugriff 09.03.2023

35  https://www.bundesregierung.de/breg-de/service/gesetzesvorhaben/koalitionsvertrag-2021-1990800, Zugriff: 25.11.2022

Hierfür erscheint es sinnvoll, bestehende Strukturen für die sekundäre Datennutzung entsprechend auszubauen oder neue zu schaffen. Beispiele für relevante Datenhalter und Forschungsdateninfrastrukturen, die bei der Anpassung Berücksichtigung finden sollten, wären unter anderem:

- die Datenintegrationszentren (DIZ) der Universitätsmedizin für die Medizininformatik-Initiative (MII) und das Netzwerk Universitätsmedizin (NUM),

- die Local Data Hubs der NFDI4Health [36],

- das Forschungsdatenzentrum Gesundheit [37] beim Bundesinstitut für Arzneimittel und Medizinprodukte (BfArM) für Krankenversicherungsdaten und ePA-Daten,

- die vom RatSWD akkreditierten Forschungsdatenzentren (FDZ) [38], u. a. mit der pharmakoepidemiologischen Forschungsdatenbank (GePaRD) und dem FDZ der Rentenversicherung,

- das Zentrum für Krebsregisterdaten am RKI [39],

- alle regionalen Krebsregister,

- die Krankenversicherungen mit den vertragsärztliche Gesundheitsdaten,

- die kommende genomDE-Plattform mit den Genomdaten in der Trägerschaft des BfArM,

- die Clinical Communication Platform (CCP) des DKTK für die translationale Krebsforschung sowie

- das Krebs-Forschungsdatenzentrum (onkoFDZ) zur Verknüpfung von Daten klinischer Krebsregistern, der Zentren der Deutschen Krebsgesellschaft, onkologischer Spitzenzentren und gesetzlicher Krankenkassen [40].

---

36  https://www.nfdi4health.de/en/service/local-data-hub.html, Zugriff 04.08.2023

37  https://www.forschungsdatenzentrum-gesundheit.de/, Zugriff: 30.11.2022

38  https://www.konsortswd.de/datenzentren/alle-datenzentren/, Zugriff: 30.11.2022

39  https://www.krebsdaten.de/, Zugriff: 30.11.2022

40  https://www.bundesgesundheitsministerium.de/ministerium/ressortforschung/handlungsfelder/forschungsschwerpunkte/krebsregisterdaten/onkofdz.html

Bei der Entwicklung kompatibler Strukturen sollten u.a. die Herausforderungen beim Datenzugang, die Interoperabilität und die rechtliche Grundlage der Datennutzung außerhalb des jeweiligen eigenen Aufgabenfeldes Beachtung finden. Zudem wäre es empfehlenswert, eine Forschungsdateninfrastruktur in den jeweiligen Einrichtungen aufzubauen, die unabhängig von dem jeweiligen Vorhaben vorgehalten wird und darüber hinaus kompatibel mit allen für die Gesundheitsforschung relevanten nationalen Forschungsdaten-infrastrukturen ist.

Die Datenintegrationszentren (DIZ) an den Universitätskliniken, die im Rahmen der MII eingerichtet wurden, stellen bereits eine dezentrale Infrastruktur für die sekundäre Nutzung für die Daten der Universitätskliniken dar. Diese dienen der Zusammenführung und Aufbereitung von Gesundheitsdaten an dem jeweiligen Standort sowie der Sicherstellung von Daten-qualität und Datenschutz. Die aufbereiteten Daten werden dann zur Nutzung für die medi-zinische Forschung zur Verfügung gestellt. Die DIZ dienen somit als lokale Data Hubs in einer föderierten Forschungsdateninfrastruktur. Die gleiche Idee verfolgt die NFDI4Health mit der Einrichtung von sogenannten *Local Data Hubs* für alle forschungsrelevanten personen-bezogenen Gesundheitsdaten, weswegen eine tiefere Zusammenarbeit der Initiativen vonnöten ist, um Doppelstrukturen zu vermeiden.

Um eine effektive dezentrale Forschungsdateninfrastruktur zu etablieren, mit der ein umfassendes Record Linkage möglich ist, sind zentralisierte Regelungen in Form von Governance und Koordination sowie spezifische Komponenten wie Services unverzichtbar. Da eine Zusammenführung von dezentral vorliegenden Forschungsdaten notwendig ist, um eine national nutzbare Infrastruktur zu schaffen, ist es demzufolge passender, von einer *dezentral-föderierten* Forschungsdateninfrastruktur zu sprechen, angesichts der für eben diese Zusammenführung erforderlichen Komponenten. Diese zentral organisierten Services werden im Folgenden vorgestellt. Für die DIZ fungiert das Forschungsdatenportal für Gesundheit (FDPG) als zentraler Zugang und für die Local Data Hubs der German Central Health Study Hub.

Um der Notwendigkeit eines Unique Identifiers (siehe Abschnitt 6.1.1) beim Record Linkage von personenbezogenen Gesundheitsdaten Rechnung zu tragen, sollten zentrale Instanzen die Aufgabe zur Vergabe und Verwaltung von Identifikatoren übernehmen (vgl. Abbildung 11). Diese zentralen Identitätsverwalter bilden den Zugang zu den möglicherweise auf verschiedene Organisationseinheiten verteilten Services und müssen mit den o.g. Strukturen der sekundären Datennutzung kompatibel sein.

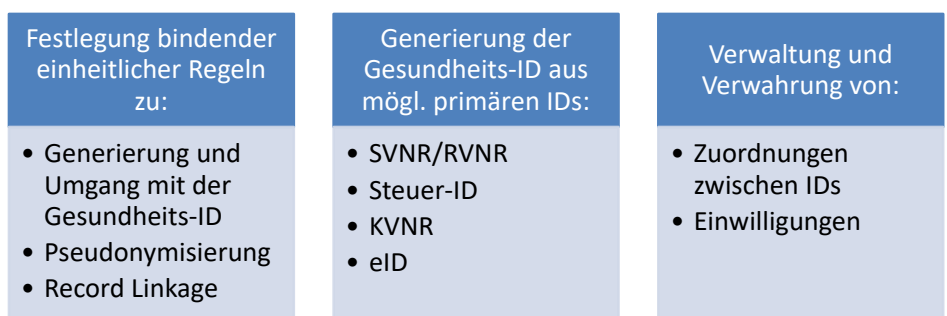

Abbildung 11: Aufgaben von zentralen Instanzen (z. B. Vertrauensstellen) zur Ermöglichung eines Record Linkage mit einem Unique Identifier (Forschungspseudonym/Gesundheits-ID)

Aus den primären Identifikatoren kann eine bereichsspezifische Gesundheits-ID („Forschungspseudonym") für das Record Linkage generiert werden, die von zentraler Stelle verwaltet wird. In diesen Zusammenhang bedarf es einer Vereinheitlichung von Regeln für das Record Linkage und die Pseudonymisierung, um höchsten Ansprüchen der Linkage-Qualität und des Datenschutzes zu genügen. So kann – anders als in den geschilderten Use Cases (siehe Kapitel 3) – in der zu schaffenden Forschungsdateninfrastruktur ein exaktes Record Linkage unter Verwendung eines Unique Identifiers (d. h. des dazugehörigen umgeschlüsselten Forschungspseudonym bzw. der Gesundheits-ID) erfolgen und so fehlerfrei verknüpfte Auswertungsdatensätze erstellt werden. Ein vereinfachter beispielhafter möglicher Datenfluss, der den Regeln einer zentralen Instanz folgt und ein Record Linkage von Gesundheitsdaten zweier Datenhaltern unter Verwendung einer solchen Gesundheits-ID ermöglicht, ist in Abbildung 12 dargestellt. Da die Verknüpfung der Identitäten und die Pseudonymzuweisung eng miteinander verzahnt sind, muss eine entsprechende zentrale Instanz für beide Vorgänge grundsätzliche Regularien formulieren.

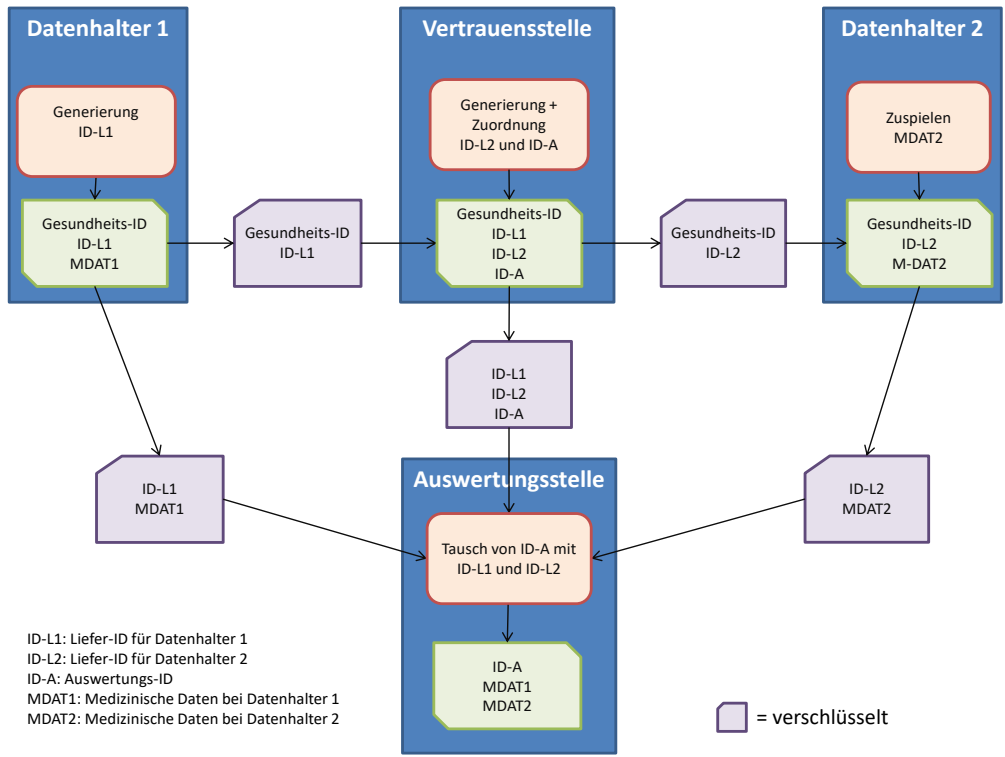

Abbildung 12: Vereinfachter beispielhafter möglicher Datenfluss bei einem Record Linkage von Gesundheitsdaten zweier Datenhalter, die die Gesundheits-ID führen, über eine Vertrauensstelle.
Diese verwaltet alle IDs und sorgt für die Übermittlung der Gesundheits-ID von Datenhalter 1 zu Datenhalter 2. Die Gesundheitsdaten der Datenhalter werden ausschließlich an die Auswertungsstelle übermittelt, von der ein Auswertungsdatensatz erstellt werden kann.

Über die elektronische Patientenakte (ePA) sollte ein Zugang zu institutionsübergreifenden Versorgungsdaten für Forschende gemäß § 363 SGB V geschaffen werden. Die ePA dient dann nicht nur als zentraler Zugriffspunkt für Patient:innen auf ihre Daten, sondern liefert auch Informationen über erteilte Einwilligungen zur Datennutzung. Mit der ePA können Patient:innen ihren Datenzugriff und ihre Einwilligungen zur Nutzung ihrer Daten jederzeit einsehen und verwalten. Um die Nutzung von Sekundärdaten für Gesundheitsforschung zu ermöglichen, sollte eine zentrale Instanz zur Verwaltung von Einwilligungen etabliert werden. Dies ermöglicht es Patient:innen, sich einfach und transparent über die Verwendung ihrer Daten zu informieren und ihre Einwilligungen jederzeit zu ändern oder zu widerrufen. Diese Optionen und die Einführung eines zentralen Tranparenzportals, mit dem Bürger:innen über die Datennutzung informiert werden, können zur Akzeptanz auf Seiten der Bevölkerung beitragen und die Bereitschaft erhöhen, eigene Gesundheitsdaten für die Forschung zur Verfügung zu stellen.

Die Daten, die für das Record Linkage zur Verfügung stehen und dezentral gehalten werden, sollten über ein zentrales Anfrageportal zugänglich gemacht werden. Ein solches Portal ermöglicht Forschenden, Daten und Datenzugänge zu überblicken und Record Linkage-Anfragen zu stellen. Dieses Portal sollte einen einfachen und intuitiven Zugang zu allen relevanten Informationen (inkl. Übersicht über Datenbestände für die standortübergreifende Forschung, Machbarkeitsanfragen, vertragliche Regelungen zur unkomplizierten Datennutzung) bieten. Mit dem Forschungsdatenportal für Gesundheit (FDPG) der MII existieren hierzu bereits Vorarbeiten, das als zentrales Element dienen kann[41]. Darüber hinaus sollen über das German Central Health Study Hub der NFDI4Health Daten aus epidemiologischen und klinischen Studien, Sekundär- und Registerdaten sowie Public Health-Daten an zentraler Stelle auffindbar gemacht werden[42]. Ein kurzfristiges Ziel muss daher eine funktionable Verzahnung des German Central Health Study Hubs der NFDI4Health mit dem FDPG der MII sein, um Parallelstrukturen zu vermeiden.

Auf europäischer Ebene hat die Europäische Kommission für den Zeitraum von 2019 bis 2025 die Schaffung eines europäischen Raums für Gesundheitsdaten (European Health Data Space, EHDS) in Form einer derzeit als Entwurf vorliegenden Verordnung avisiert. Der EHDS *„unterstützt Einzelpersonen dabei, die Kontrolle über ihre eigenen Gesundheitsdaten zu bewahren, fördert die Nutzung von Gesundheitsdaten für eine bessere medizinische Versorgung, für Forschung, Innovation und Politikgestaltung, und ermöglicht es der EU, das Potenzial von Austausch, Nutzung und Weiterverwendung von Gesundheitsdaten unter gesicherten Bedingungen voll auszuschöpfen"*[43]. Auch wenn Details zur Ausgestaltung des EHDS noch zu klären sind, kann schon heute eine Grundannahme getroffen werden: Um den Anschluss der nationalen Forschungsdateninfrastruktur an den EHDS zu ermöglichen, ist es wichtig, eine Hub-Funktion zu etablieren, die eine einfache und sichere Möglichkeit bietet, Daten mit anderen Institutionen des EHDS zu teilen. Für weitere Ausführungen zu den aufgeführten Forderungen sei auf das Gutachten „Gesundheitsdaten heterogener Datenhalter" (TMF 2022, unveröffentlicht) verwiesen, das für das BMG erstellt wurde.

---

**Box 13: Empfehlungen zur Infrastruktur**

→ Schaffung von geeigneten Schnittstellen zur Einbindung der etablierten dezentralen Struktu­ren der Datenquellen und Datenhalter (u. a. Datenintegrationszentren der Universitätskliniken, Krebsregister und Forschungsdatenzentren)

→ Etablierung einer dezentral-föderierten Forschungsdateninfrastruktur

• mit folgenden zentralen Komponenten:

  – Zentrale Instanz zur Vergabe von bundeseinheitlichen direkten Identifikatoren und zur Formulierung grundsätzlicher, bindender Regeln für das Record Linkage

  – Zentrale Vertrauensstelle zur Verwaltung der Zuordnung von bereichsspezifi­schen Pseudonymen zum primären Identifikator einer Person

  – Zentrale Instanz zur Verwaltung von Einwilligungen

• und mit folgenden Anknüpfungspunkten:

  – Elektronische Patientenakte (ePA) zur Sicherstellung des Zugangs gemäß § 363 SGB V zu den Daten der ePA und zur Einsicht über erteilte Einwilligungen

  – NFDI4Health German Central Health Study Hub

  – Forschungsdatenportal für Gesundheit (FDPG) der MII

  – Europäischer Raum für Gesundheitsdaten (EHDS)

## 6.1.6 Kompetenzförderung und Verfahrenssicherheit bei Pseudonymisierung und Linkage

Eine fehlerfreie und sichere Methode zur Verknüpfung von Daten ist durch die Verwendung verschlüsselter Unique Identifier gegeben, sofern dieser Identifier für alle Personen in den zu verknüpfenden Datenbanken vorliegt. Da dies in Deutschland voraussichtlich in naher Zukunft nicht der Fall sein wird, erfordert die Verknüpfung von Datensätzen von mehreren Datenhaltern zunächst fehlertolerante Record Linkage-Verfahren basierend auf indirekten Identifikatoren. Die gegenwärtige Forschungsdatenlandschaft ist zudem durch eine Vielzahl von Gesundheitsdatenquellen in diversen datenhaltenden Institutionen gekennzeichnet, was zu einer erheblichen Heterogenität der Daten führt. Die Verknüpfung unterschiedlicher Datenquellen, wie Primär- und Sekundärdaten, stellt sowohl technisch als auch organisa­torisch und datenschutzrechtlich eine Herausforderung dar. Dabei gelten je nach Datenhalter unterschiedliche gesetzliche Regelungen und Zuständigkeiten von Aufsichtsbehörden. Pflege und Speicherung personenbezogener Daten sind Aufgaben der datenhaltenden

Institution, für die derzeit (noch) keine einheitlichen Standards oder Vorgaben existieren. Die Erfassung und Speicherung von personenidentifizierenden Daten als Identifikatoren für ein Record Linkage hängen stark von den spezifischen Rahmenbedingungen der datenhaltenden Institution ab. Folglich ist es ratsam, angemessenes strategisches und technisches Know-how bereitzustellen, um der Komplexität der Datenverknüpfungsprozesse gerecht zu werden.

Zur Förderung des Wissenstransfers und Erfahrungsaustauschs im Bereich der Pseudonymisierungs- und Record Linkage-Verfahren empfiehlt es sich, dass Wissenschafts-förderer in Zusammenarbeit mit dem BMG einen Fachworkshop oder eine Reihe von Workshops durchführen. Dabei sollten die praktischen Verfahrensweisen, verfügbare Technologien und Werkzeuge sowie kontinuierliche Weiterentwicklungen wie neuartige Verfahren (z. B. Secure Multi-Party Computation) zusammengestellt und vermittelt werden, unter Ein-beziehung aktueller Initiativen zum Aufbau der Forschungsdateninfrastruktur wie MII, NFDI, Netzwerk Universitätsmedizin (NUM), Deutsche Zentren der Gesundheitsforschung (DZG), Nationales-Centrum für Tumorerkrankungen, Nationale Dekade gegen Krebs und genomDE[44]. Das Register-Gutachten gibt hierzu weitere Details (Niemeyer et al., 2021).

## 6.1.7 Gesellschaftliche Akzeptanz

Verknüpfte Gesundheitsdaten bieten aufgrund ihrer Informationstiefe eine wertvolle Grundlage für die Gesundheitsforschung, deren Potenzial für die Verbesserung der Gesundheit der Bevölkerung bislang noch nicht ausgeschöpft wird. Daher ist es erforderlich, eine umfassende gesellschaftliche Diskussion über die Zukunft der datenintensiven Gesundheits-forschung zu führen, die neue Modelle der Einwilligung und Partizipation mit einschließt. Die Förderung eines gesellschaftlichen Vertrauens in die Institutionen, Personen und Prozesse, die für die Erhebung, Verarbeitung und Auswertung von Gesundheitsdaten ver-antwortlich sind, stellt eine zentrale Voraussetzung für den Erfolg der datenbasierten Medizin dar. Das Vertrauen der Bürger:innen in die Gesundheitsforschung ist entscheidend dafür, dass Gesundheitsdaten in der Gesundheitsforschung genutzt und mit anderen Daten verknüpft werden können. In Deutschland gibt es bundeslandspezifische gesetzliche Re-gelungen, die die Rechtslage in Bezug auf Datennutzung für die Gesundheitsforschung un-übersichtlich gestalten und die praktische Umsetzung erschweren oder sogar verhindern. Zudem besagen diese Regelungen, dass die Verwendung und Verknüpfung personen-bezogener Daten ohne die Einwilligung der Betroffenen nur in Ausnahmefällen und nach Abwägung der Interessen akzeptiert werden kann (Strech et al., 2020) (siehe auch Abschnitt 3.2, Abschnitt 3.3 sowie Abschnitt 5.1.2). Daher muss die Nutzung von Daten, wenn keine Ausnahmeregel angewendet wird, durch eine informierte Einwilligung der betroffenen Person legitimiert sein. Dies bedeutet, dass sie vorab über den Zweck, die Art der Nutzung, die Risiken und den Nutzen des Forschungsvorhabens informiert wird und die Freiwilligkeit ihrer Einwilligung garantiert ist.

---

[44]   https://www.bundesgesundheitsministerium.de/themen/gesundheitswesen/personalisierte-medizin/genomde-de.html, Zugriff 25.01.2023

In der medizinischen Forschung kann es jedoch vorkommen, dass bei der Einwilligung nicht alle möglichen Verwendungszwecke der Daten bekannt sind. Aus diesem Grund werden häufig umfassende Einwilligungen eingeholt, die als Broad Consent bezeichnet werden. Der Broad Consent der Medizininformatik-Initiative (MII) wurde als Rechtsgrundlage für die Forschung mit medizinischen Routinedaten entwickelt und mit der Konferenz der unabhängigen Datenschutzbeauftragten des Bundes und der Länder abgestimmt. Er soll es ermöglichen, dass die medizinische Forschung auf der Grundlage der DSGVO pseudonymisierte klinische Daten bundesweit nutzen und verknüpfen kann. Patient:innen haben damit die Möglichkeit, ihre informierte Einwilligung zur Verwendung ihrer Gesundheitsdaten der medizinischen Routineversorgung für Forschungszwecke zu geben. Diese Daten können für einen Zeitraum von fünf Jahren pseudonymisiert ausgewertet werden. Nach Ablauf dieser Frist muss eine erneute Einwilligung eingeholt werden. Die Organisation der Einholung der erneuten Einwilligungen muss allerdings noch diskutiert und geregelt werden. Alternativen zur informierten Einwilligung in Form von gesetzlich verankerten Ausnahmeregelungen kommen selten zur Anwendung (siehe dazu auch Abschnitt 5.1.2 und Abschnitt 6.1.2). Die informierte Einwilligung zur Nutzung und Verknüpfung der Daten für medizinische Forschung wird in der Regel im Laufe der medizinischen Behandlung erteilt. Da die medizinische Behandlung eine Stresssituation darstellt, in der das medizinische Personal häufig unter Zeitdruck steht, kann es zu Verständnisproblemen bei den Datenspender:innen kommen. Es ist daher empfehlenswert, die Entscheidung über die Datenspende von der medizinischen Behandlung und möglicherweise sogar allgemein von der krankheitsbedingten Situation der Patient:innen zu trennen und in den Alltag zu verlegen.

Im Allgemeinen führen Opt-in-Verfahren allerdings nicht zu realistischen Zustimmungsraten. Beispielsweise standen bei den Repräsentativbefragungen der Bundeszentrale für gesundheitliche Aufklärung von 2022 zu dem Thema Organ- und Gewebespende zwar 84% der Befragten diesem Thema positiv gegenüber, aber nur 44% haben ihre Entscheidung in einem Organspendeausweis oder einer Patientenverfügung festgehalten[45].

Eine Option, dem Bedarf in der digitalisierten Gesundheitsforschung gerecht zu werden und bestehende Gesundheitsdaten, die in der Regel für Versorgungszwecke erhoben wurden, rechtskonform zu nutzen, könnte darin bestehen, eine gesetzliche Erlaubnis zu nutzen und den Betroffenen dennoch die Entscheidung zu überlassen. Dies kann nicht nur durch die Einholung der Zustimmung des Betroffenen (Opt-in), sondern auch durch die Möglichkeit eines Widerspruchs der Betroffenen (Opt-out) erfolgen (Strech et al., 2020). Dies würde dem Umstand Rechnung tragen, dass die Bereitschaft der Bürger:innen grundsätzlich sehr groß ist, ihre Daten für die Forschung zur Verfügung zu stellen. Im August 2019 führte das Forschungsinstitut forsa *Politik- und Sozialforschung GmbH* im Auftrag der TMF eine Umfrage in Deutschland durch, um zu verstehen, welche Haltung die Bevölkerung gegenüber Forschungsdatenspenden einnimmt und welche Einwilligungsform bevorzugt wird (Lesch et al., 2022). Nach dieser Studie sind 79% der Befragten grundsätzlich bereit, persönliche Gesundheitsdaten für die Durchführung von medizinischen Forschungs-projekten zur Verfügung zu stellen. Unter diesen waren lediglich 27%, die eine Einholung ihrer Einwilligung für jedes einzelne Forschungsprojekt bevorzugen. Unter den Befragten, die nicht oder nur bedingt zu einer Datenspende für medizinische Forschungszwecke bereit sind, wurden als Gründe Bedenken bezüglich eines möglichen Datenmissbrauchs oder möglicher Verstöße gegen das Datenschutzrecht angegeben. Außerdem schien es für die Bereitschaft, persönliche Daten zur Verfügung zu stellen, von Bedeutung zu sein, wer die Daten nutzt. Universitäten und öffentliche Forschungseinrichtungen genießen ein großes Vertrauen, während die Nutzung der Daten für kommerzielle Forschung kritischer gesehen wurde. Darüber hinaus befürworten die Bürger:innen laut einer im August 2022 durchgeführten repräsentativen Online-Umfrage zunehmend, dass ihre Daten auch zu Opt-out-Bedingungen – d. h. ohne vorher eingeholte Einwilligung – und unter strengen Schutzmaßnahmen pseudonymisiert für Forschungszwecke genutzt werden. Damit muss die Möglichkeit bestehen, jederzeit und ohne Schwierigkeiten der Datenspende zu widersprechen. Unter einer entsprechend ausformulierten einwilligungsbefreiten Regelung konnte eine weitaus höhere Akzeptanz zur Nutzung der Daten selbst durch kommerzielle Forschungseinrichtungen beobachtet werden. Sicherlich spielte die aktuelle Aufgeschlossenheit der Öffentlichkeit

---

45    https://www.organspende-info.de/zahlen-und-fakten/einstellungen-und-wissen, Zugriff: 21.02.2023

aufgrund der Corona-Pandemie eine Rolle, entscheidend aber war der Aspekt, dass eine solche Regelung Schutz- und Kontrollmaßnahmen beinhaltet und so den Bedenken von Patient:innen (Richter et al., 2021) und der Allgemeinbevölkerung begegnet wird (Richter et al. eingereicht 2023). Vor dem Hintergrund der hohen Bereitschaft der Bevölkerung, an Forschung teilzuhaben und sich zu informieren, ist es angeraten, Maßnahmen zur Förderung der Gesundheitsdatenkompetenz zu ergreifen und das Vertrauen in die Arbeit der Gesundheitsforschung noch weiter zu stärken. Dafür muss vermittelt werden, dass bereits ein hohes Maß an Datensicherheit in der Forschung garantiert wird (u. a. durch geeignete Verschlüsselungs- und De-Identifikationsstrategien). Die Bürger:innen sollten zudem transparent über die Studien, die aus ihrer Datenspende resultieren, informiert werden. Diese Transparenz schafft somit zusätzliches Vertrauen in die digitale Gesundheitsforschung.

Dabei könnte ein nationales Transparenzportal eine geeignete Möglichkeit darstellen, über Forschungsvorhaben und Betroffenenrechte zu informieren und Informationskampagnen auf nationaler Ebene zu unterstützen. So würden der Bevölkerung nicht nur einheitliche Informationen zur Verfügung gestellt, sondern es könnte auch eine breite gesellschaftliche Diskussion über Sekundärdatennutzung und Verknüpfung von Gesundheitsdaten angestoßen werden. Hier sollte auch die forschungskompatible ePA Berücksichtigung finden. Sie kann der Bevölkerung als Zugriffspunkt auf ihre Daten dienen, einschließlich der Informationen über erteilte Einwilligung zu Datennutzung und -verknüpfung.

Über ein solches Transparenzportal könnten darüber hinaus nötige Interaktionen der Bürger:innen mit verantwortlichen Stellen ermöglicht werden. Bei einem Widerspruch zur Datennutzung oder -verknüpfung könnten so die Datenspender:innen mit den zuständigen Treuhandstellen in Kontakt treten und von ihrem Widerrufsrecht Gebrauch machen. Aber auch bei dem Wunsch zur Datenspende für bestimmte Forschungsvorhaben könnten den interessierten Bürger:innen über das Portal direkt oder indirekt erste organisatorische Vereinbarungen angeboten werden, wie z. B. Vereinbarung von Terminen zu einem Aufklärungsgespräch.

Bei der Schaffung dieser niederschwelligen Zugangs- und Informationspunkte müssen die unterschiedlichen Kompetenzen zu Gesundheit und Digitalem berücksichtigt werden. Verständlichkeit und Nutzerfreundlichkeit müssen Kernpunkte dieser Angebote sein.

Hierfür könnte das sogenannte Projektregister[46] als Transparenzportal des Forschungsdaten-portals für Gesundheit der MII als Vorbild und Ausgangspunkt dienen. Über dieses Register werden der Bevölkerung Informationen zu medizinischen Forschungsvorhaben angeboten, für die Daten via Broad Consent genutzt werden.

Die gesellschaftliche Akzeptanz kann auch durch die Partizipation von Patient:innen in der Gesundheitsforschung gefördert werden. Das bedeutet, dass die Perspektiven von Patient:innen in der Konzeption (z. B. Fragestellungen generieren), in der Organisation (z. B. Erstellung von Aufklärungs- und Einwilligungsdokumenten) sowie in der Aufbereitung und Kommunikation von Ergebnissen (z. B. Beteiligung an der Berichterstellung) Berücksichtigung finden. Dafür können Patientenvertretungen z. B. in Steering Committees von Forschungs-vorhaben und über die Mitwirkung an gemeinsamen projektbezogenen Veranstaltungen eingebunden werden. Das *European Joint Programme on Rare Diseases* begreift die Patientenpartizipation als die Gesamtheit kontinuierlicher Vorgänge, die von der Mitwirkung an der Rekrutierung von Patient:innen für Studien über die Überprüfung von Durchführ-barkeit und Relevanz von Forschungsvorhaben, die zielgruppengerechte Aufbereitung von Ergebnissen bis hin zur Einbeziehung von Patient:innen als Projektpartner – inklusive wissen-schaftlicher Mitarbeit – reichen [47]. Letzteres dient u. a. auch der Identifizierung von Patienten-bedürfnissen. Diese Vorgehensweise zur Förderung der Patientenpartizipation ist beispielsweise bei der Beforschung der seltenen Erkrankungen von höchster Relevanz, da gerade hier die Forschung von einem Pooling der Daten auf europäischer und inter-nationaler Ebene stark profitieren kann.

Die Verantwortung für die Gewährleistung einer angemessenen Beteiligung von Patient:innen sowie Studienteilnehmenden liegt bei der Forschung selbst und ist nicht per se eine Aufgabe der Gesetzgebung. Dennoch kann der Gesetzgeber Richtlinien vorgeben, die verlangen, dass entsprechende Nachweise für eine Beteiligung der Betroffenen erbracht werden und / oder die Finanzierung an die Umsetzung dieser Richtlinien knüpfen. Allerdings müssen ent-sprechende Mittel auch bei der Antragstellung eingestellt werden dürfen und durch den Drittmittelgeber im Fall einer Bewilligung auch als erstattungsfähig angesehen werden. Im Rahmen der Medizininformatik-Initiative (MII) sind beispielsweise Patientenorganisationen

---

46    https://forschen-fuer-gesundheit.de/menu_register.php, Zugriff 25.01.2023

47    https://www.ejprarediseases.org/our-actions-and-services/patients-in-research/, Zugriff 22.02.2023

strukturell in ein Dialogforum eingebunden. Das Forum fungiert als übergreifende Instanz und hat die Aufgabe, externe Partner, einschließlich Patientenorganisationen wie z. B. dem Aktionsbündnis Patientensicherheit e.V. und der Patientenvertretung im Gemeinsamen Bundesausschuss, in die strategische Planung und Entwicklung der MII einzubeziehen.

Damit Gesundheitsdaten umfassend genutzt werden können und so das Potenzial datengetriebener medizinischer Forschung erschlossen werden kann, bedarf es einer Kultur des Datenteilens, für die eine breite gesellschaftliche Akzeptanz vonnöten sein muss. Um diese zu erreichen, sind klare Regeln, ein hohes Maß an Transparenz, Aufklärung und Beteiligung sowie ein barrierefreier Service zu Datenspende und zu Widersprüchen erforderlich.

---

**Box 14: Empfehlungen zur Förderung der gesellschaftlichen Akzeptanz**

→ Informationskampagnen zur Förderung
  • der öffentlichen Gesundheitsdaten- und Digitalkompetenz
  • einer breiten gesellschaftlichen Debatte zur Sekundär- und Registerdatennutzung
→ Zentrales Transparenzportal zur Information von Patient:innen über Datennutzungen und zur Kommunikation mit Forschungsvorhaben
→ Förderung der Einbeziehung von Patientenvertretungen in Forschungsvorhaben
→ Weiterbildung in Wissenschaftskommunikation stärken

---

## 6.2 Ausblick

Sowohl die im Koalitionsvertrag anvisierte dezentral-föderierte Forschungsdateninfrastruktur als auch die von der EU-Kommission verfolgte Etablierung eines europäischen Gesundheitsdatenraums (EHDS) verlangen nach einem sicheren und effektiven Record Linkage für die Gesundheitsforschung. Die fehlerfreie, aufwandsarme und datenschutzkonforme Zuordnung von Personenidentitäten ist entscheidend für den Erfolg des Record Linkage. Die zurzeit aufgrund des Fehlens eines Unique Identifiers verwendeten Ansätze führen jedoch zu Verknüpfungsfehlern (siehe Use Cases Kapitel 3). Eine hohe Verknüpfungsqualität kann bei exaktem Record Linkage entweder über den Abgleich einer Vielzahl an identifizierenden Daten oder aber unter Verwendung eines Unique Identifiers ermöglicht werden. Jedoch ist das exakte Linkage auf Grundlage von vielen personenidentifizierenden Daten durch ein erhebliches Datenschutzrisiko gekennzeichnet. Die Verwendung von bereichsspezifischen Unique Identifiern beim exakten Record Linkage in der zu schaffenden Forschungsdateninfra-struktur trägt nicht nur zur Korrektheit der Zuordnung bei, sondern erhöht den Datenschutz durch Wahrung des Grundsatzes der Datenminimierung in Hinblick auf die Verwendung von personenidentifizierenden Daten. Die Zusammenführung von dezentral vorliegenden Daten zum Zwecke der Forschung muss dabei für eine bundesweite nutzbare Infrastruktur durch zentrale Governance-, Koordinations- und Service-Einheiten organisiert werden.

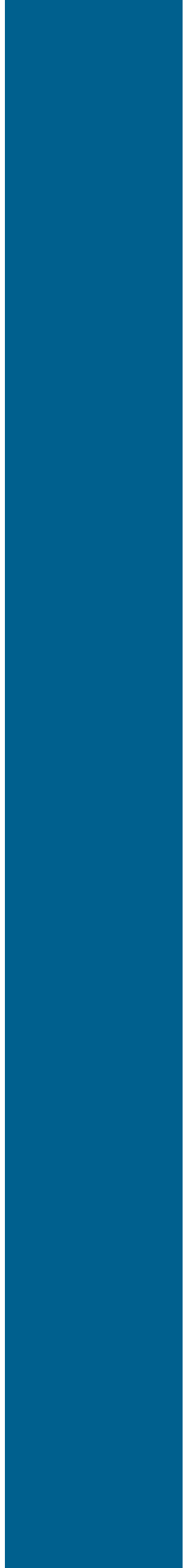

# Autorenliste

Timm Intemann[a]/Knut Kaulke[b], Dennis-Kenji Kipker[c,d], Vanessa Lettieri[c], Christoph Stallmann[e], Carsten O. Schmidt[f], Lars Geidel[f], Martin Bialke[f], Christopher Hampf[f], Dana Stahl[f], Martin Lablans[g,h,i], Florens Rohde[k], Martin Franke[k], Klaus Kraywinkel[l], Joachim Kieschke[m], Sebastian Bartholomäus[n], Anatol-Fiete Näher[o,p], Galina Tremper[g,h,i], Mohamed Lambarki[g,h,i], Stefanie March[q], Fabian Prasser[r], Anna Christine Haber[r], Johannes Drepper[b], Irene Schlünder[b], Toralf Kirsten[k,s], Iris Pigeot[a,t], Ulrich Sax[u], Benedikt Buchner[c], Wolfgang Ahrens[a]/Sebastian C. Semler[b]

a     Leibniz-Institut für Präventionsforschung und Epidemiologie – BIPS, Bremen

b     TMF – Technologie- und Methodenplattform für die vernetzte medizinische Forschung e.V., Berlin

c     Universität Augsburg, Augsburg

d     Fachbereich Rechtswissenschaft, Universität Bremen, Bremen

e     Universität Magdeburg, Magdeburg

f     Universitätsmedizin Greifswald, Greifswald

g     Deutsches Krebsforschungszentrum Heidelberg, Heidelberg

h     Deutsches Konsortium für Translationale Krebsforschung (DKTK), Heidelberg

i     Medizinische Fakultät Mannheim, Universität Heidelberg, Mannheim

k     Universität Leipzig, Leipzig

l     Robert Koch-Institut, Zentrum für Krebsregisterdaten, Berlin

m     Epidemiologisches Krebsregister Niedersachsen, Oldenburg

n     Landeskrebsregister NRW, Bochum

o     Robert Koch-Institut, Berlin

p     Institut für Medizinische Informatik, Charité Universitätsmedizin Berlin, Berlin

q     Hochschule Magdeburg-Stendal, Magdeburg

r     Berlin Institute of Health at Charité Universitätsmedizin Berlin, Berlin

s     Hochschule Mittweida, Mittweida

t     Fachbereich Mathematik und Informatik, Universität Bremen, Bremen

u     Universitätsmedizin Göttingen, Göttingen

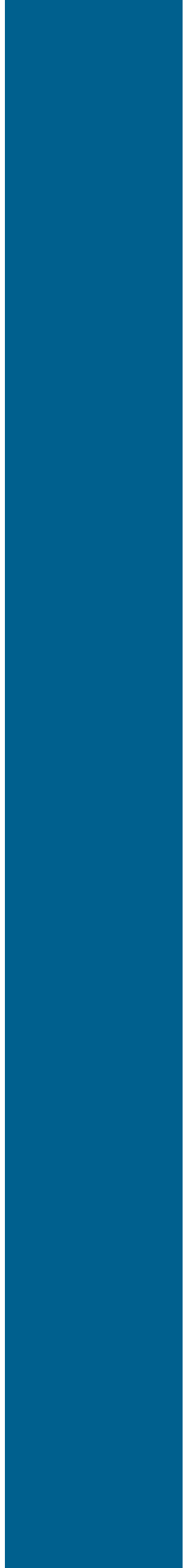

# Beiträge der Autoren und Autorinnen

*jeweils in alphabetischer Reihenfolge*

**Konzept:**

Wolfgang Ahrens, Timm Intemann, Knut Kaulke, Iris Pigeot, Sebastian C. Semler

**Erster Entwurf und Bearbeitung:**

Kapitel 1: Wolfgang Ahrens; Kapitel 2: Martin Bialke, Martin Franke, Knut Kaulke, Martin Lablans, Stefanie March, Florens Rohde; Abschnitt 3.1: Timm Intemann, Carsten O. Schmidt, Christoph Stallmann; Abschnitt 3.2: Timm Intemann, Klaus Kraywinkel; Abschnitt 3.3: Sebastian Bartholomäus, Timm Intemann, Klaus Kraywinkel, Joachim Kieschke; Abschnitt 3.4: Anatol-Fiete Näher; Abschnitt 3.5: Ulrich Sax; Abschnitt 3.6: Martin Lablans, Mohamed Lambarki; Abschnitt 3.7: Martin Lablans, Galina Tremper; Abschnitt 3.8: Martin Bialke, Lars Geidel, Christopher Hampf, Dana Stahl; Abschnitt 3.9: Martin Bialke, Lars Geidel, Christopher Hampf; Abschnitt 3.10: Martin Bialke, Lars Geidel, Christopher Hampf; Kapitel 4: Knut Kaulke, Martin Lablans, Galina Tremper; Kapitel 5: Benedikt Buchner, Dennis-Kenji Kipker, Vanessa Lettieri, Irene Schlünder, Sebastian C. Semler; Kapitel 6: Wolfgang Ahrens, Timm Intemann, Knut Kaulke, Irene Schlünder, Sebastian C. Semler

**Überprüfung und Überarbeitung:**

Wolfgang Ahrens, Johannes Drepper, Anna Christine Haber, Timm Intemann, Knut Kaulke, Toralf Kirsten, Stefanie March, Iris Pigeot, Fabian Prasser, Sebastian C. Semler

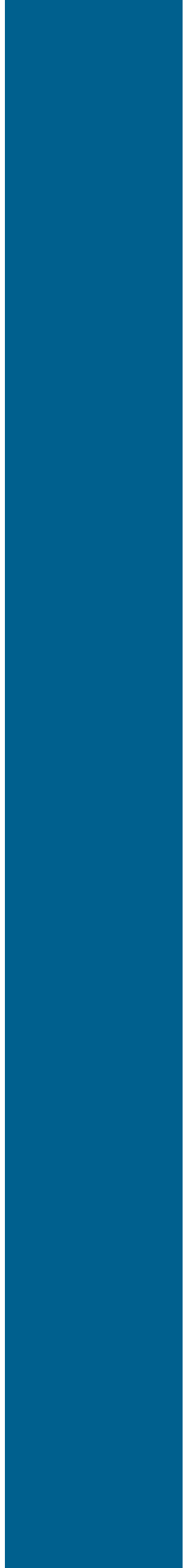

# Danksagung

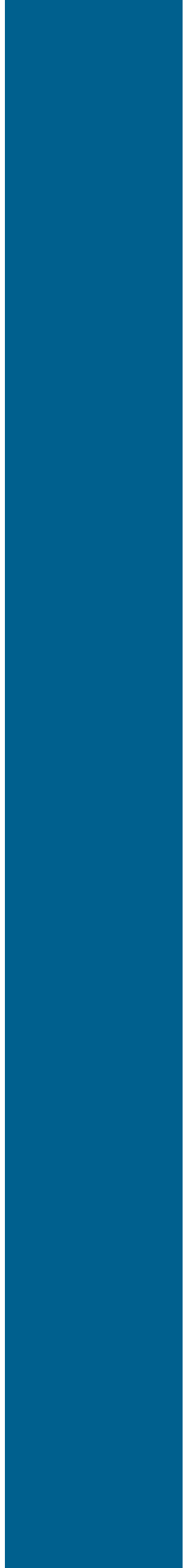

# Literaturverzeichnis

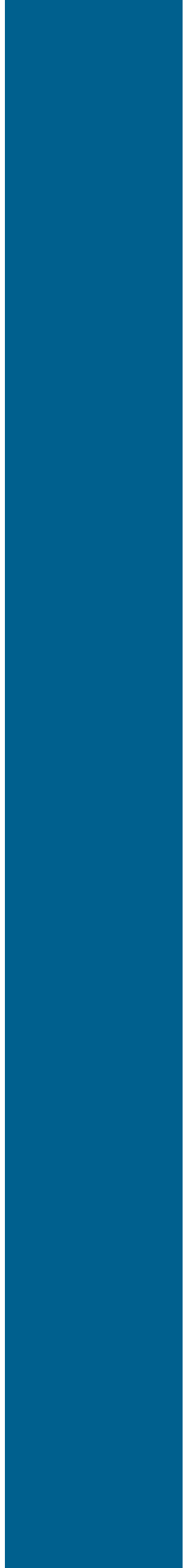

# Anhang

## Glossar

Tabelle Anhang 1: Glossar

| Begriff | Beschreibung |
|---------|--------------|
| Anonymisierung | Die personenbezogenen Daten werden in diesem Vorgang derart verändert, „dass die Einzelangaben über persönliche oder sachliche Verhältnisse nicht mehr oder nur mit einem unverhältnismäßig großen Aufwand an Zeit, Kosten und Arbeitskraft einer bestimmten oder bestimmbaren natürlichen Person zugeordnet werden können." **(BDSG a.F. § 3 Abs. 6)** |
| Blocking | Blocking, das in der Literatur alternativ auch als Filtering oder Indexing bezeichnet wird, dient einer Effizienzsteigerung des Verknüpfungsprozesses. Statt alle Beobachtungseinheiten eines Datenbestandes mit allen Beobachtungseinheiten eines anderen Datenbestandes abzugleichen, werden nur jene Datenpaare abgeglichen, die bei einem oder mehreren Identifikatoren identisch sind (klassisches Blocking) oder eine sehr hohe Ähnlichkeit aufweisen. **(Christen , 2012a, S. 69)** |
| Bloomfilter (BF) | Ein Bloomfilter ist eine von Bloom entwickelte Datenstruktur, die zur effizienten Überprüfung der Zugehörigkeit von Mengen eingesetzt werden kann. Bloomfilter können verwendet werden, um festzustellen, ob zwei Datenmengen näherungsweise übereinstimmen. Die Methode der Bloomfilter-Kodierung wird im Rahmen des Privacy-Preserving Record Linkage (PPRL) in verschiedenen praktischen Anwendungen zur Verknüpfung sensibler Daten eingesetzt. Technisch betrachtet bestehen Bloomfilter aus Ketten von Nullen und Einsen, deren Länge vorab zu definieren ist. Beim PPRL werden die Original-Identifikatoren in eine Bloomfilter-Datenstruktur überführt/kodiert. Dabei bestimmen Streuwertfunktionen (Hashfunktionen), wo Nullen in der Kette durch Einsen ersetzt werden. Diese entstandenen Kodierungen lassen keinen Rückschluss auf die Original-Identifikatoren zu, können aber für ein fehlertolerantes Record Linkage herangezogen werden (Bloom, 1970; Broder & Mitzenmacher, 2003). Darüber hinaus gibt es weitere Verfahren, die zusätzlich den Bloomfilter bearbeiten, z.B. indem Einträge vertauscht werden, sodass bestimmteAngriffsszenarien verhindert oder erschwert werden. Solche zusätzlichenVerfahren werden als Härtungen bezeichnet. **(Schnell et al., 2009a), (Christen et al., 2020)** |
| Dateneigner | „Unter diesem Begriff werden im Rahmen der Guten Praxis Sekundärdatenanalyse (GPS) diejenigen Institutionen verstanden, die die Daten (primär) erheben, speichern und nutzen. Dateneigner und Primärnutzer sind als Synonym zu verstehen. Der Begriff Dateneigner hebt jedoch zusätzlich hervor, dass der Primärnutzer auch die rechtliche Verfügungsgewalt über die Daten besitzt. Im Bereich der Gesetzlichen Sozialversicherung sind Dateneigner beispielsweise Krankenkassen oder Rentenversicherungsträger, die versichertenbezogene (medizinische) Daten für administrative Aufgaben speichern, ebenso wie (Krebs-)Registerstellen, arbeitsmedizinische Untersuchungsstellen oder epidemiologische Einrichtungen." **(Swart et al., 2015, S. 126)** |
| Datenintegrationszentrum (DIZ) | Diese Einrichtungen dienen der datenschutzgerechten Akkumulation und dem Austausch von harmonisierten Forschungs- und Versorgungsdaten an den universitätsmedizinischen Standorten. Dabei spielt die Interoperabilität zwischen den Standorten eine wesentliche Rolle. „Die aufbereiteten Daten werden dann zur Nutzung für die medizinische Forschung bereitgestellt und Forschungsergebnisse werden über die DIZ in die Versorgung zurückgeführt. In den DIZ werden somit die technischen und organisatorischen Voraussetzungen für die standortübergreifende Datennutzung zwischen Krankenversorgung und medizinischer Forschung geschaffen." [48] |

---

[48]    https://www.medizininformatik-initiative.de/de/konsortien/datenintegrationszentren, Zugriff 14.04.2022

| Begriff | Beschreibung |
|---|---|
| Datenlinkage | Datenlinkage bezeichnet die personenbezogene Verknüpfung verschiedener Datenquellen für Forschungszwecke mittels geeigneter Identifikatoren. „Im Kern wird hierbei ein umfassender Prozess betrachtet der von der Planung eines Forschungsvorhabens über die eigentliche Zusammenführung unterschiedlicher Datenquellen bis hin zur Auswertung und Nutzung durch weitere Personen inklusive der Löschung der Forschungsdaten und Anonymisierung reicht; kurz: alle Prozessschritte, die in der Anwendung von Datenlinkage zu beachten sind." **(March et al., 2019)** |
| Deterministisches Linkage | **Siehe Fehlertolerantes Linkage** |
| Distanzbasiertes Linkage | **Siehe Fehlertolerantes Linkage** |
| Exaktes Linkage | „Beim exakten Linkage führt man die Daten mehrerer Datenquellen nur bei exakter Übereinstimmung eines eindeutigen Verknüpfungsschlüssels (z. B. Sozialversicherungsnummer) oder mehrerer Linkage-Variablen zusammen (Match-Merge Linkage). Gibt es Fehler oder unterschiedliche Schreibweisen im Schlüssel oder in den Linkage-Variablen, können die Daten mit exaktem Linkage nicht zusammengebracht werden." **(March et al., 2018)** |
| Fehlertolerantes Linkage | „Ist kein exaktes Linkage möglich, sind fehlertolerante Verfahren erforderlich, welche die Zahl der verknüpften Beobachtungseinheiten erhöhen können, wie z.B. durch die Nutzung von String-Metriken (siehe „Distanzbasiertes Linkage"). Innerhalb der fehlertoleranten Linkage-Methoden können wiederum mehrere Arten unterschieden werden: regelbasiertes, distanzbasiertes und probabilistisches Linkage. Das Finden und Festlegen von Schwellenwerten für die Entscheidung, ob Datensätze zusammengeführt werden sollen, ist eine der wesentlichen Hausforderungen des Datenlinkage [Im vorliegenden White Paper wird für die Beschreibung der personenbezogenen Datenverknüpfung die Begrifflichkeit „Record Linkage" verwendet – Anm. d. Verf.], bei der auch bestimmte *Sonderfälle* zu berücksichtigen sind, die je nach verwendeten Daten unterschiedlich sein können (z. B. Zahlendreher, gleichgeschlechtliche Zwillinge, (Nicht-)Berücksichtigung zweiter Vornamen)." <br><br> **Regelbasiertes Linkage** <br><br> Bei regelbasiertem Linkage wird durch Regeln definiert, „welche Identifikatoren komplett übereinstimmen müssen, und bei welchen eine teilweise Übereinstimmung ausreichend ist. Eine Regel könnte beispielsweise lauten, dass der Nachname und beim Vornamen die ersten drei Buchstaben übereinstimmen müssen sowie, dass das Geburtsjahr höchstens um drei Jahre abweichen darf. Eine solche Regel ist leicht zu evaluieren und beinhaltet eine gewisse Fehlertoleranz, sie birgt jedoch eine erhöhte Gefahr einer falsch positiven Klassifikation." … „Eine spezielle und in der Forschungspraxis sehr gängige Variante des regelbasierten Linkage wird als deterministisches Linkage bezeichnet. <br><br> **Deterministisches Linkage** <br><br> Dabei werden Identifikatoren oder Transformationen von diesen (bspw. durch Anwendung des phonetischen Codes) weiterhin auf exakte Übereinstimmung hin überprüft. Die Entscheidungsregel kann jedoch vorsehen, dass nur ein bestimmter Anteil davon übereinstimmen muss (z. B. „5 von 7 Identifikatoren müssen übereinstimmen, darunter mind. Nachname und Geburtsjahr"), um ein Datenpaar als einen Match zu klassifizieren." |

| Begriff | Beschreibung |
|---|---|
| | **Distanzbasiertes Linkage** |
| | Bei diesem Linkageverfahren wird die Ähnlichkeit zweier Identifikatoren berechnet. Dabei wird z.B. überprüft, wie viele Änderungen in der Schreibweise eines Namens (Identifikators) nötig sind, um ihn in den zu vergleichenden Namen umzuwandeln. Dabei gilt, je weniger Änderungen dafür nötig sind, um so ähnlicher sind sich diese Namen. So ist für das Vergleichspaar „Maier" und „Mayer" nur die Änderung eines Buchstaben nötig, um die Namen ineinander umzuformen. |
| | **Probabilistisches Linkage** |
| | Der in der medizinischen Forschung hauptsächlich genutzte Ansatz ist das probabilistische Record Linkage-Verfahren."..."Wesentlich an diesem Modell ist die Annahme von Wahrscheinlichkeiten, also einem probabilistischen Modell, für die Funktionswerte von Vergleichs von Nachnamen in zwei Datensätzen) jeweils unter der Bedingung, dass die dem zu vergleichenden Datenpaar zugrunde liegenden Personen identisch/nicht-identisch sind. Dies nutzt die Tatsache, dass die Übereinstimmung mancher Identifikatoren mehr Aussagekraft hinsichtlich der Zusammengehörigkeit zweier Beobachtungseinheiten [ oder Datensätze – Anm. d. Verf.] aufweisen als die Übereinstimmung anderer. Konkret kann z.B. der Nachname deutlich mehr verschiedene Ausprägungen annehmen als das Geschlecht." **(March et al., 2018)** |
| Hashfunktion | Eine Hashfunktion ist eine Funktion, die Zeichenketten von beliebiger Länge (z.B. Namen) auf Zeichenketten gleicher Länge abbildet (z.B. alle 10-stelligen ganzen Zahlen). **(Christen et al., 2020)** |
| Identifikator / Identifier | Ein Identifikator oder Identifier ist ein personenbezogenes Merkmal, das der Identifizierung einer Person dient. Häufig wird dabei zwischen direkten und indirekten Identifikatoren unterschieden. Als direkte Identifikatoren werden in der Regel eindeutige Einzelmerkmale bezeichnet (z.B. Sozialversicherungsnummer). Als indirekte Identifikatoren werden im Allgemeinen Merkmale bezeichnet, die nur in Kombination zur Identifizierung führen (z.B. die Kombination von Geburtsdatum, Geschlecht, Gemeindekennziffer und Erkrankung). Potenziell kann jedes personenbezogene Merkmal als indirekter Identifikator herangezogen werden. Siehe **March et al. (2018)** und **Christen et al. (2020)** für Details. |
| Identifizierend | Siehe **personenidentifizierend** |
| Informierte Einwilligung / Informed Consent (IC) | „Die Einwilligungserklärung eines Patienten oder Probanden ist nur wirksam, wenn sie auf der freien Entscheidung des Betroffenen beruht. Er ist zuvor über den vorgesehenen Zweck der Erhebung, Verarbeitung oder Nutzung seiner Daten und Proben aufzuklären („Informed Consent"). Die Wirksamkeit der Einwilligungserklärung erfordert deren Schriftform. Soll sie zusammen mit anderen Erklärungen schriftlich erteilt werden, ist sie besonders hervorzuheben. Materiellrechtlich setzt die Einwilligungserklärung die Einsichtsfähigkeit des Erklärenden voraus." **(Pommerening et al., 2014)** |
| | Die wesentlichen Kriterien für eine informierte Einwilligung sind, dass die betroffene Person sowohl über Wissen als auch über Verständnis verfügt, dass die Einwilligung aus freien Stücken und ohne Zwang oder unzulässige Beeinflussung gegeben wird und dass die betroffene Person klar und deutlich auf das Recht hingewiesen wird, ihre Einwilligung jederzeit zurückzuziehen. |

| Begriff | Beschreibung |
|---------|--------------|
| | Weitere Aspekte der Einwilligung nach Aufklärung im Zusammenhang mit epidemiologischer und biomedizinischer Forschung sowie die Kriterien, die bei der Einholung der Einwilligung erfüllt werden müssen, sind in der International *Ethical Guidelines for Epidemiological Studies* (Council for International Organizations of Medical Sciences (CIOMS) in collaboration with the World Health Organization (WHO), 2009) und den *International Ethical Guidelines for Biomedical Research Involving Human Subjects* (Council for International Organizations of Medical Sciences, 2002) festgelegt. |
| | Die informierte Einwilligung (gelegentlich auch als „datenschutzrechtliche Einwilligung" bezeichnet) ist i.d.R. die Voraussetzung zur Durchführung des Record Linkage personenbezogener Daten. Gesetzlich ergibt sich die Möglichkeit zur Einholung der Einwilligung der betroffenen Person allgemein aus **Art. 6 Abs. 1 S. 1 lit. a DSGVO** und für besondere Daten-kategorien (z. B. Gesundheitsdaten, genetische Daten, biometrische Daten) aus **Art. 9 Abs. 2 lit. a DSGVO**, der aufgrund der größeren Schutzwürdigkeit dieser Daten zusätzliche Voraussetzungen zur Einwilligungserteilung bestimmt. Im Forschungskontext ist die Besonderheit des sogenannte „Broad Consent" zu beachten, der den Anwendungsbereich der Ein-willigung ausdehnt. Unter formalen Gesichtspunkten ist die Einwilligungs-erteilung von der Entbindung von der ärztlichen Schweigepflicht gem. § 203 StGB zu trennen. Im Kontext von medizinischen Forschungs-vorhaben wird die Verwendung des Begriffs „Einverständniserklärung" nicht empfohlen, „da er die Manifestation des Willens des Be-troffenen nicht deutlich ausdrückt." **(Pommerening et al., 2014, S. 214)** |
| Personenbezogene Daten | Personenbezogene Daten sind "alle Informationen, die sich auf eine iden-tifizierteoder identifizierbare natürliche Person (im Folgenden „betroffene Person") beziehen". **(DSGVO, Art. 4, Abs. 1)** |
| Personenidentifizierend | Merkmale sind personenidentifizierend, falls mit ihnen eine Person an-gesehen, die direkt oder indirekt, insbesondere mittels Zuordnung zu einer Kennung wie einem Namen, zu einer Kennnummer, […] oder zu einemoder mehreren besonderen Merkmalen, die Ausdruck der […] Identitätdieser natürlichen Person sind, identifiziert werden kann;" **(DSGVO, Art. 4, Abs. 1)** |
| Primärdaten | „Primärdaten sind Daten, die im Rahmen ihres originär vorgesehenen Verwendungszwecks aufbereitet und analysiert werden." **(Swart et al., 2014, S. 125)** |
| Privacy-Preserving Record Linkage (PPRL) | Unter dem Überbegriff Privacy-Preserving Record Linkage lassen sich Verfahren zusammenfassen, die eine Verknüpfung von Datensätzen unter-schiedlicher Dateneigner ermöglichen, ohne dass zwischen den Daten-eignern personenidentifizierende Daten ausgetauscht werden. Zu diesem Zweck kann eine Vertrauensstelle / Treuhandstelle herangezogen werden. In der Regel werden beim PPRL für die Verknüpfung verschlüsselte Identifi-katoren (z. B. Kontrollnummern, Bloomfilter) herangezogen, die keinen oder nur mit unverhältnismäßig großem Aufwand herstellbaren Rückschluss auf die Identität der Personen erlauben. **(Christen, 2012a), (Christen et al., 2020)** |
| Probabilistisches Linkage | Siehe **Fehlertolerantes Linkage** |

| Begriff | Beschreibung |
|---|---|
| Pseudonymisierung | Pseudonymisierung bezeichnet „die Verarbeitung personenbezogener Daten in einer Weise, dass die personenbezogenen Daten ohne Hinzuziehung zusätzlicher Informationen nicht mehr einer spezifischen betroffenen Person zugeordnet werden können, sofern diese zusätzlichen Informationen gesondert aufbewahrt werden und technischen und organisatorischen Maßnahmen unterliegen, die gewährleisten, dass die personenbezogenen Daten nicht einer identifizierten oder identifizierbaren natürlichen Person zugewiesen werden;" **(DSGVO, Art. 4, Abs. 5)** |
| | „Dies kann beispielsweise durch die Ersetzung des Probanden-Namens durch eine Kenn-Nummer geschehen. [...] Ziel der Pseudonymisierung ist es aber nicht, den Personenbezug irreversibel abzutrennen, sondern lediglich durch ein eindeutiges Kennzeichen – ein Pseudonym – zu ersetzen, das für sich genommen die Identifikation der dahinterstehenden Person ausschließt oder aber wesentlich erschwert. Grundsätzlich bleiben pseudonymisierte Daten allerdings personenbeziehbar: Es existiert ein „Geheimnisträger", der die Zuordnung von Person zu Pseudonym kennt oder wiederherstellen kann und entsprechend vertrauenswürdig und geschützt sein muss." **(Pommerening et al., 2014, S. 231–232)** |
| | Diese Aufgabe kann z. B. eine Treuhandstelle wahrnehmen (siehe Treuhandstelle). |
| | **Dierks & Roßnagel (2020)** ergänzen hierzu, dass grundsätzlich zwei Arten von pseudonymen Daten verzeichnet werden. Da wären zum einen die Daten, deren Zuordnung zu einer bestimmten Person für den Verantwortlichen ausgeschlossen werden kann, wenn ihm die Zuordnungsregel nicht bekannt ist. Somit handelt sich in dieser Konstellation nicht um personenbezogene Daten. Das wäre beispielsweise in den medizinischen Forschungsvorhaben der Fall, wo eine Treuhandstelle in den Datenfluss mit einbezogen wird. D. h. die IDAT werden sicher und von den MDAT getrennt und durch ein Pseudonym ersetzt. Die Zuordnungsregel wird dann an unabhängiger Stelle (Treuhandstelle) unzugänglich für Forschende hinterlegt. |
| | Zum anderen werden bei der zweiten Art pseudonymer Daten zwar der Verarbeitungsprozess der zu verarbeitenden Daten und die Zuordnungsregel für Identifizierung der Personen auch separiert, die Forschenden, die eigentlich nur die Pseudonyme kennen, können aber in Ausnahmefällen die Aussetzung der Pseudonymität beantragen, um z. B. Diskrepanzen zu klären. In dem Fall wären für sie diese Daten personenbezogen. |
| | „Diese Unterscheidung kennen auch die Definitionen des § 3 Abs. 6a BDSG a.F. und § 67 Abs. 8a SGB X a.F., wenn sie Pseudonymisieren definieren als „das Ersetzen des Namens und anderer Identifikationsmerkmale durch ein Kennzeichen zu dem Zweck, die Bestimmung des Betroffenen auszuschließen oder wesentlich zu erschweren". „Die Bestimmung des Betroffenen auszuschließen", entspricht der anonymisierenden Wirkung der Pseudonymisierung für den Verantwortlichen, der die Zuordnungsregel nicht kennen kann. „Die Bestimmung des Betroffenen zu erschweren", entspricht der zweiten, der risikomindernden Art pseudonymer Daten. Diese pseudonymen Daten sind aber weiterhin personenbezogene Daten." **(Dierks & Roßnagel, 2020, S. 272)** |

| Begriff | Beschreibung |
|---|---|
| Record Linkage von personenbezogenen Gesundheitsdaten | **Record Linkage von personenbezogenen Gesundheitsdaten** bezeichnet den Prozess des Vergleichens von Einträgen aus zwei oder mehreren Datenbanken mit dem Ziel, Einträge zu identifizieren, die sich auf dieselbe Person beziehen, um die Datenbanken für die Gesundheitsforschung entsprechend zusammenzuführen. Für die Umsetzung von **Record Linkage von personenbezogenen Gesundheitsdaten** in der Praxis werden in diesem Dokument sowohl datenschutzrechtliche als auch organisatorische und technische Aspekte berücksichtigt. |
| Regelbasiertes Linkage | Siehe **Fehlertolerantes Linkage** |
| Sekundärdaten | „Sekundärdaten sind Daten, die einer Auswertung über ihren originären, vorrangigen Verwendungszweck hinaus zugeführt werden. Maßgeblich für die Einstufung als Sekundärdaten sind Unterschiede zwischen dem primären Erhebungsanlass und der nachfolgenden Nutzung. Für die Einstufung ist es unerheblich, ob die weitergehende Nutzung durch den Dateneigner selbst oder durch Dritte erfolgt. Demnach sind beispielsweise Routinedaten einer Krankenkasse nicht nur Sekundärdaten, wenn sie für wissenschaftliche Fragestellungen genutzt werden, sondern z.B. auch dann, wenn sie durch die Krankenkasse für Zwecke der Versorgungsplanung herangezogen werden." **(Swart et al., 2015, S.126)** |
| | Der Begriff der Sekundärdaten wird oftmals umgangssprachlich synonym mit anderen Begriffen wie „claims data", administrative Daten, Abrechnungs- oder Routinedaten verwendet. Diese Daten gehören zwar zu den Sekundärdaten, beziehen sich aber jeweils nur auf eine Teilmenge aller möglichen Sekundärdaten. **(March et al., 2018)** |
| Sozialdaten | „Sozialdaten sind personenbezogene Daten […], die von einer in §35 des Ersten Buches genannten Stelle [z.B. Rentenversicherung oder Versicherungsämter – Anm. d. Verf.] […] verarbeitet werden." **(SGB X § 67 Abs. 2)** |
| | Sozialdaten unterliegen besonderen datenschutzrechtlichen Auflagen, die nach der Anonymisierung auf diese nicht mehr zutreffen. **(March et al., 2018)** |
| Symmetrische Verschlüsselung | Als symmetrische Verschlüsselung wird eine Verschlüsselung bezeichnet, bei der Sender und Empfänger den gleichen Schlüssel zum Ver- und Entschlüsseln verwenden. **(Christen et al., 2020)** |
| Treuhandstelle (THS) | Eine Treuhandstelle agiert zwischen Forschungsdatengebern und der Nachfrageseite von Forschungsdaten als eine unabhängige und neutrale Vertrauensinstanz, die idealerweise einer besonderen Geheimhaltungspflicht unterliegt, z.B. ein Notar oder ein externer Arzt." **(Pommerening et al., 2014, S. 214)** |
| | Eine Treuhandstelle umfasst „technische und organisatorische Maßnahmen zur Gewährleistung grundlegender Anforderungen an Datenschutz und IT-Sicherheit" **(Bialke et al., 2015)** |

| Begriff | Beschreibung |
|---------|--------------|
| | In ihrer unabhängigen und weisungsfreien Position zwischen der Forschungsdaten liefernden Stelle und den Forschenden stellt sie die Wahrung der Betroffenenrechte von Patient:innen und Proband:innen sicher. Ein wichtiger Bestandteil dieser Strategie ist die elektronisch geführte Patientenliste, die die Verknüpfung identifizierender Patientendaten (IDAT) zu Pseudonymen speichert und deren Verwaltung der Treuhandstelle obliegt. Sie erfüllt in dieser Weise in Funktion eines „vertrauenswürdigen Dritten" ein breites Aufgabensektrum. Beispielsweise „anonymisiert oder pseudonymisiert die Treuhandstelle die von der Daten besitzenden Stelle übermittelten personenbezogenen Daten und übermittelt nur die anonymisierten bzw. pseudonymisierten Daten an den Forschenden weiter. Auf diese Weise bleibt der Kreis derjenigen, die Kenntnis von personenbezogenen Daten erhalten, eng begrenzt, und die Datensicherheit kann effektiv gewährleistet werden." **(Pommerening et al., 2014)** |
| | Folgende Aufgabenbereiche werden in medizinischen Forschungsvorhaben in einer Treuhandstelle behandelt: |
| | • Verwaltung der IDAT (Patientenliste) inkl. direkter Identifikatoren für Record Linkage |
| | • Einwilligungsmanagement |
| | • Widerrufsmanagement von Einwilligungen |
| | • Pseudonymisierung / Anonymisierung von personenbezogenen Daten |
| | • Re-Identifikation bei Vorliegen entsprechender Voraussetzungen (z. B. bei einer datenschutz-rechtlichen Anfrage). |
| | **Föderierte Treuhandstelle / federated Trusted Third Party (fTTP)** |
| | Eine föderierte Treuhandstelle ermöglicht den Betrieb von Infrastrukturen, um Machbarkeitsanfragen auf Patienten-/Probandeneinwilligungsbasis in dezentral organisierten Datenbeständen zu bearbeiten, Teildatenkörper zu selektieren und diese einrichtungsübergreifend datenschutzkonform zusammenzuführen sowie mit weiteren Datenquellen zu verknüpfen. Die föderierte Treuhandstelle setzt ein Privcy-Preserving Record Linkage (PPRL) praktisch um und dient dabei der Qualitätsbildung und der konsequenten Wahrung der Betroffenenrechte in diesen Infrastrukturen. |
| Vertrauensstelle | Der Begriff Vertrauensstelle wird heutzutage in medizinischen Forschungsvorhaben oftmals – so auch in diesem Dokument – synonym mit dem Begriff Treuhandstelle verwendet. |

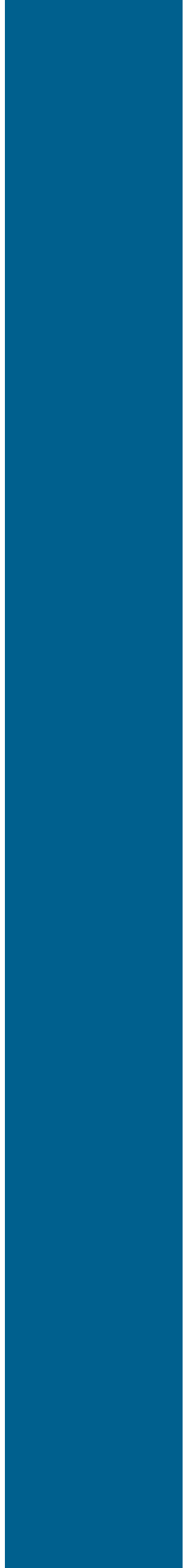

# Record Linkage-Tools

Tabelle Anhang 2: Aktuelle, kostenfreie Record Linkage-Tools, die sowohl deterministisches als auch probabilistisches Linkage erlauben

| Tools | Links | Letztes Up-date | Unterstützte Linkage-Verfahren | Art der Unterstützung bei der Auflösung von Synonymfehlern / Dubletten / Überein-stimmungen | Unterstützte Webschnitt-stellen-Standards | Grafische Oberfläche? | Ist die Software eingebettet in ein PIMS*? | Entwickelnde Einrichtungen |
|---|---|---|---|---|---|---|---|---|
| E-PIX ® | https://www.toolpool-gesundheitsforschung.de/produkte/e-pix | 2023 | Fellegi & Sunter (1969), PPRL, unterstützt diverse Vergleichsalgorithmen (basierend beispielsweise auf der Levenshtein-Distanz oder der Phonetik) | Web-Schnittstelle, grafische Oberfläche | SOAP, FHIR | Ja | Ja | Treuhandstelle der Universitätsmedizin Greifswald |
| Mainzelliste | https://www.toolpool-gesundheitsforschung.de/produkte/main-zelliste | 2021 | EpiLink, PPRL über Bloomfilter, PPRL über Secure Multi Party Com-putation (über MainSEL: https://bio.tools/mainsel) | REST-Schnittstelle, grafische Oberfläche | REST | Ja | | Community aus zzt. 8 Einrichtungen |
| PRIMAT | https://www.toolpool-gesundheitsforschung.de/produkte/primat-private-matching-toolbox | 2021 | Fellegi & Sunter (1969), PPRL | | | | Nein | Uni Leipzig |
| SOEMPI | https://github.com/MrCsabaToth/SOEMPI | 2014 | Fellegi & Sunter (1969), PPRL | | | Ja | | |

*PIMS: Patienten-Identifikatoren-Management-System